\documentclass{article}
\usepackage[T1]{fontenc}
\usepackage[utf8]{inputenc}
\usepackage[fontsize=11pt]{fontsize}
\usepackage[a4paper, total={6.5in, 9.5in}]{geometry}
\usepackage{amsmath}
\usepackage{amssymb}
\usepackage{float}
\usepackage{graphicx}
\usepackage{booktabs}
\usepackage{multirow}
\usepackage[hidelinks]{hyperref}
\usepackage{caption}
\usepackage{subcaption}
\usepackage{breakcites}
\usepackage{comment}
\usepackage{tabularx}
\usepackage{lscape}
\usepackage{rotating}
\usepackage{longtable}
\usepackage[flushleft]{threeparttable}
\usepackage[dvipsnames, table]{xcolor}
\usepackage{xcolor-solarized}
\usepackage{natbib}
\usepackage{soul}
\usepackage{setspace}
\usepackage[section]{placeins}
\title{Self-Selection, University Courses and \\ 
Returns to Advanced Degrees}
\author{Eleonora Brandimarti\footnote{University of Glasgow. Any comments can be directed to \url{eleonora.brandimarti@glasgow.ac.uk}. I thank Michele Pellizzari, Giacomo De Giorgi, and Peter Arcidiacono for their guidance and support. Aleksey Tetenov, Edwin Leuven and Arnaud Maurel provided valuable feedback, along with the seminar participants at Harvard Business School, NBER Productivity, University of Geneva, Duke University, Gerzensee Alumni Conference, SasCa PhD Conference, Trans-Atlantic Doctoral Conference, and Rare Voices in Economics conference. I thank Silvia Ghiselli and the AlmaLaurea research team for their expertise and hospitality throughout this project and their help accessing their resources. All mistakes are my own.}}
\date{February 12, 2025}
\begin{document}
	
	\maketitle

\begin{abstract}
Higher education often requires choosing a bachelor's and a master's degree, yet the returns of these combined choices and the role of courses in different disciplines remain understudied. This paper addresses this gap using detailed data on Italian graduates and university programs. I study the labor market returns to combinations of bachelor's and master's degrees and investigate how curriculum characteristics affect outcomes. I exploit exogenous variation in access to bachelor's and master's degrees to causally estimate the returns to 43 combinations of degrees. I organize the data in a nested model with exogenous variation in admission requirements and explore the preference profile of the sample through policy simulations that shift these requirements. I then relate the estimated returns to the academic curriculum of degrees, focusing on the role of quantitative education and timing of courses. I contribute to the literature on returns to advanced degrees by incorporating master's degrees in the discussion on how higher education affects outcomes and providing evidence on the characteristics of curricula that are positively related to labor market returns. The findings reveal substantial variation in returns to degree combinations, even among combinations with the same bachelor's degree, indicating the need to consider both types of programs. Combinations of degrees in different disciplines positively impact economic outcomes, while those in the same field perform worse. Successful combinations feature more non-quantitative education in the bachelor's, and quantitative courses alone do not explain higher returns. \\

	
\end{abstract}



\newpage
	\section{Introduction}

The literature on the returns to education is currently active on the issue of university degrees.  Recent evidence suggests that the choice of degree can significantly impact labor market outcomes \citep{altonji2018costs, hastings2013some,kirkeboen2016field,altonji2016analysis,altonji2012heterogeneity}. A critical yet underexplored aspect of this debate is the substantial heterogeneity in the amount of instruction across different disciplines within the same degree. For example, a typical degree in economics requires a considerable number of classes in law, statistics, and mathematics in addition to courses in economics.
In this paper, I investigate the labor market value of university degrees by combining administrative data---covering almost the entire population of university graduates in Italy---with detailed, purposely collected information on the disciplinary content of all university programs. The data includes information on the number of compulsory classes required in each program, with each class associated with a specific discipline. I develop a methodology to causally estimate the labor market returns to each university program and analyze the disciplinary content of programs with both high and low returns.

I carry out the empirical exercise in the context of Italy, where most students enroll in a 2-year master's program after a 3-year bachelor's. Since the early 2000s, this is the harmonized structure of university programs across the European Union. Compared to other studies estimating the returns to degrees, this setting poses the additional empirical challenge of modeling the sequential choice of bachelor's and master's, both of which can span several disciplines. I develop a novel methodology to causally estimate the returns to any combination of bachelor's and master's programs leveraging information on the strictness of entry requirements at both levels. Specifically, for master's degrees, I have collected data about the credit requirements needed to enroll in any master's based on the student's bachelor's degree. For example, accessing an engineering master's from a literature bachelor's requires acquiring additional credits in math. I exploit this information to generate variation in the choices of bachelor's and master's that is plausibly exogenous to labor market outcomes. This is organized in a nested model where students first choose a bachelor's program and then, conditional on the bachelor's, choose a master's program, while also allowing for the option of not pursuing a master's degree.


Several findings emerge from the analysis of labor market returns to 43 combinations of bachelor's and master's degrees. First, the choice of master's degree significantly impacts outcomes. Returns vary substantially even when the same bachelor's degree is paired with different master's degrees. Second, combining degrees from different disciplines can improve outcomes compared to specializing in the same field for both bachelor's and master's degrees. All the combinations of degrees associated with the best labor market returns involve master's degrees in different fields from the bachelor's degrees, while not pursuing a master's degree is generally associated with worse labor market outcomes. I then investigate two features of the combinations of degrees to understand the characteristics that relate to higher payoffs. First, I measure the amount of quantitative education in each combination of degrees and find that the relationship between labor market returns and quantitative courses is slightly U-shaped. In fact, both low- and high-earning combinations of degrees exhibit high shares of quantitative education. This finding challenges the widespread belief that degrees with more STEM (Science, Technology, Engineering, and Mathematics) education benefit students and indicates one dimension to consider when analyzing policies that incentivize enrollment in STEM. Finally, I observe that high-return combinations of degrees exhibit low shares of non-quantitative education (such as humanities, law, and education) in the master's program and relatively higher shares of non-quantitative courses in the bachelor's program. This breakdown by degree level (bachelor's or master's) underscores the importance of the timing of courses, further corroborating the centrality of master's degrees in the analysis of returns to higher education.

My findings help us better understand how university program design affects outcomes. In particular, they contribute to the policy discussion on STEM degrees by highlighting the potential pitfalls of degrees that do not appropriately balance quantitative and non-quantitative education.
Crucially, my analysis establishes the importance of advanced degrees in relation to labor market outcomes and informs on their connection to undergraduate degrees. The share of the population worldwide with a master's degree has increased steadily over the past few decades. In the U.S., the number of adults with a master's degree has more than doubled since 2000, and approximately 42\% of European students and 27\% of U.S. students embark on a master's degree every year \citep{CollegeEnrollmentEU, CollegeEnrollmentUS, uscensus2019}. Furthermore, as the U.S. higher education system allows more flexibility in the choice of classes than in Europe, the central feature of this paper -- that students cover a wide range of knowledge at university -- is likely to be even more relevant in the U.S. 
Unlike Europe, where students enroll in degrees with little flexibility, students in the U.S. can wait up to two years before declaring a major.

I contribute to the literature on returns to higher education in four directions. Reviews of this literature can be found in  \cite{altonji2012heterogeneity,altonji2016analysis,oreopoulos2013making}, and \cite{patnaik2020college}.
First, I propose an identification strategy that incorporates information about the sequential structure of the choice of degrees to causally estimate labor market returns to combinations of bachelor's and master's. Recent advancements have highlighted the limitations of using OLS and assuming selection on observables (selection on observables remains widely used in this field, e.g. \cite{andrews2022returns} on returns to college majors in the U.S.). For instance, \cite{kirkeboen2016field} use information on applications to higher education in Norway to account for individual partial rankings of degree choices and estimate ex-post local heterogeneous returns to undergraduate degrees. Similarly, \cite{hastings2013some} employ a research discontinuity design that leverages threshold-crossing admissions in Chile to compute local returns that account for university reputation. Both studies use private rankings of fields of study to identify the causal effect of bachelor's degrees at the margin. More recently, \cite{bleemer2022will} use a similar regression discontinuity approach to estimate returns to economics majors, and more selective colleges \citep{bleemer2021top}.
Structural approaches pioneered by \cite{arcidiacono2004ability} have also been proposed to estimate returns to bachelor's degrees. By imposing structure on decision-making, dynamic choice modeling methods can elicit ex-ante returns and incorporate introspective behaviors such as switching majors and non-pecuniary factors that can only be rationalized with error terms revealed in multiple stages. \cite{arcidiacono2011practical} offer an overview of the main methods.\footnote{Structural approaches have also been used to identify the effect of attending selective institutions \citep{brewer1999does} and the evolution of wage returns to education over time \citep{ashworth2021changes}. \cite{dhaultfoeuille2013inference} show that non-pecuniary factors are key ex-ante determinants of higher education attendance.} \cite{malamud2011discovering, malamud2010breadth} focuses on timing of specialization in higher education and its related probability of switching. 
He finds that early specialization in higher education is related to more costly switching. \cite{montmarquette2002young} explore how students choose their majors by incorporating idiosyncratic expected earnings and heterogeneous probabilities of success, finding that ex-ante expected earnings significantly influence choice. Conversely, \cite{beffy2012fieldofstudy} attribute most sorting to non-pecuniary factors. I contribute to this literature by proposing an identification strategy that leverages the timing of choices and exogenous variation at different stages to retrieve labor market outcomes of combinations of degrees.

Second, I add to the literature on advanced degrees by incorporating them in my analysis and shedding light on the labor market-enhancing features of degree combinations. \cite{altonji2021labor} analyze the returns to detailed types of graduate programs by comparing pre- and post-graduate earnings, accounting for preferences, ability, and previous college choices. They find considerable variations in returns that are strongly related to undergraduate choices. Similarly, \cite{arcidiacono2008economic} estimate returns to MBAs, taking advantage of the fact that admission into such programs requires previous work experience and establishing pre-MBA labor market earnings. 
\cite{altonji1993demand} estimates the returns to the highest degree obtained, including five aggregated graduate school categories, and assumes that only the highest degree matters. A few papers provide estimates of the returns to graduate degrees for specific groups of fields of study: \cite{black2003economic} for individuals with economics undergraduate majors, and \cite{bhattacharya2005specialty, chen2012women, ketel2016returns} for medical degrees. \cite{ketel2016returns} is the only paper on advanced degrees not to use US data, focusing on the Netherlands.
This article complements this body of work by focusing on returns for individuals who immediately enroll in a master's degree, which account for about 75\% of master's graduates in Italy and 15\% in the US, previously excluded from \cite{altonji2021labor}'s analysis \citep{almalaurea2020}. 
I also exploit variation in admission eligibility to master's programs to causally estimate the returns to the complete set of bachelor's and master's combinations. The additional structure and availability of exogenous variation in incentives strengthen \cite{altonji2021labor}'s results as they allow for rich counterfactual patterns and direct estimates of returns to degree combinations. 

Third, this paper relates to the growing literature on unordered treatment effects, where returns to university degrees are a compelling application \citep{bhuller20222sls, heckman2018unordered, kirkeboen2016field, mountjoy2021community}. These authors recognize that when choices are unordered, the treatment effect depends on individual preferences over the choice set, even if properly accounting for self-selection. In practice, unordered settings lead to multiple contrasting margins of treatment that grow exponentially with the choice set. The large number of combinations of degrees considered in this application makes the estimation of heterogeneous margins of treatment both intractable and difficult to interpret. \cite{bhuller20222sls} propose an IV method to obtain economically relevant treatment effects that are averages across all heterogeneous margins.\footnote{\cite{bhuller20222sls} propose an average monotonicity condition that requires instruments to increase the probability of treatment on average. Joint with a cross-effects condition that guarantees that instruments uniquely affect treatments, average monotonicity identifies properly estimated average treatment effects with multiple unordered treatments in 2SLS. In practice, their model exploits a modified first stage where each instrument affects the treatment separately.} This project additionally faces a weak instrument problem that emerges in 2SLS estimation, arising from the large number of endogenous regressors -- the combination of undergraduate and graduate degrees -- instrumented with the predicted probabilities of enrollment obtained with the nested model \citep{phillips2017structural}. While the setup is close in spirit to \cite{bhuller20222sls}, identification requires a reduced form solution to avoid using the correlation between the endogenous regressors and the instruments \citep{chernozhukov2008reduced}.

Lastly, I contribute to the literature on degree characteristics. Lastly, I contribute to the literature on degree characteristics. Despite consensus on the importance of higher education for labor market success beyond ability signaling, evidence on how degrees affect outcomes lacks a systematic approach. \cite{biasi2022education} focus on the coverage of frontier knowledge in higher education. They find that instructors play a central role in surmounting the education-innovation gap, leading to higher earnings for students with access to such knowledge. Conversely, \cite{braga2016teaching} examine the impact of instructors in college on labor market outcomes and discover a mild effect. \cite{deming2020earnings} investigate the skill decay of college graduates, discovering that earning premia decline faster for graduates in technology-intensive fields. \cite{acemoglu2022eclipse} find that CEOs in Denmark and the US with business education are responsible for less profit sharing with employees and claim that practices and values acquired in business school are responsible.
STEM degrees, characterized by quantitative and technical education, have received considerable attention. However, even within this group of degrees, there is a lack of consensus in the characteristics that are important for labor market outcomes \citep{xie2015stem}. Table \ref{tab:stem_litreview} in appendix \ref{sec:appendix_descriptives} substantiates this claim by comparing STEM definitions in the literature. By analyzing the impact of university courses by field of study on labor market returns, I contribute with the first systematic review of labor-enhancing degree characteristics across disciplines.

The rest of the paper is organized as follows. Section \ref{sec:institution} summarizes the relevant features of the Italian higher education system and discusses its similarities with the European and U.S. context. Section \ref{sec:theory} describes the theoretical framework of the analysis. It presents the stages of the model and the empirical challenges in close relation to the available data. Section \ref{sec:data} describes the main data sources on Italian graduates and university programs. Section \ref{sec:degrees} presents the results of all the stages of the model to obtain the labor market returns to 43 degree combinations. It also presents a policy simulation that shifts admission requirements to investigate how preferences affect enrollment at the intensive margin. Section \ref{sec:courses} relates the estimated returns to program characteristics such as timing, quantitativeness, and multidisciplinarity to elicit labor market enhancing characteristics. Together, these results provide the basis for the discussion on program characteristics. Section \ref{sec:conclusions} concludes.

\section{Institutional Background}\label{sec:institution}

Italy adheres to the Bologna process (initiated in 1999) that ensures comparability in higher education standards across the European Higher Education Area (EHEA), encompassing 48 European and Central Asian countries.
As a result, Italian degrees are organized as bachelor's (three years) and master's (two years) with workloads measured in credits, the unit of academic work. According to the European Credit Transfer and Accumulation System (ECTS), one credit corresponds to 25 hours of academic work, including both classes and individual study, and one academic year corresponds to 60 credits. Admission to a master's degree requires the completion of a bachelor's, and students apply to programs in specific fields of study. The Bologna Process also promotes the automatic recognition of degrees across the EHEA and encourages international student mobility.

Throughout this paper, I use the following terminology: a \textit{degree} refers to the university program that can be to either a bachelor's or a master's program, a \textit{university career} is the combined choice of a bachelor's (undergraduate) and master's (graduate) degree. A university \textit{course} is a specific subject studied within a degree, measured in \textit{credits}. Both degrees and courses come in various \textit{fields of study} (disciplines) are available, and a single course can be part of multiple degrees. The \textit{academic curriculum} describes the prescribed courses and credits that make up a degree.

For a degree to be legally valid, it must meet strict requirements regarding its curriculum, specified in terms of course content and credit amounts.
During the period analyzed (graduates from 2007 to 2014), there were 47 bachelor's degrees and 99 master's degrees.\footnote{The Italian higher education system also includes academic diplomas, one-year master's, doctoral programs, and vocational degrees. Only academic diplomas which have equal legal standing to a bachelor's degree are considered.} Some degrees are exceptionally organized as single-cycle degrees lasting five or six years, which directly confer a master's degree without a corresponding bachelor degree. These include medicine, veterinary, dentistry, architecture, law, chemical and pharmaceutical technologies, and primary education.

The academic curriculum for each degree is defined along two dimensions: the number of credits assigned to each course and the course content. Course content is standardized across degrees and universities, covering 370 possible disciplines \citep{ssd}. For example, the code MAT-5 corresponds to calculus, which can be a course in 23 bachelor's degrees and 12 master's degrees, though the credit value may vary. More than 50\% of a degree's course content and credit distribution is fixed, with students free to allocate only about 10\% of their total credits (roughly one class per year). The remaining credits are allocated to compulsory internships and thesis periods in varying proportions. Therefore, a degree is fully described by a vector of courses and credits in each discipline. Importantly, students choose degrees with predefined curricula rather than individual courses.

For statistical precision, I group bachelor's and master's degrees into ten fields of study, described in table \ref{tab:groups}. This grouping aligns with the data provider's aggregation and is adjusted for comparability across data sources (further discussed in \ref{sec:data}). A detailed list of which degrees belong to which group can be found in appendix \ref{sec:appendix_degrees}. Throughout the paper, I focus on university careers rather than individual degrees, meaning the combined choice of bachelor's and master's degrees. For instance, a career in economics involves both a bachelor's and master's in economics, whereas a career in economics and law involves a bachelor's in economics and a master's in law.


\begin{table}[htbp]
	\centering
	\caption{Fields of study description}
	\begin{tabular}{cll}
		\toprule
		\textbf{Code}  &\textbf{ Abbreviation} & \textbf{Description}\\
		\midrule
		1     & Agr.Vet.Geo.Bio. & Agriculture and veterinarian sciences, geology and biology \\
		2     & Arch.Eng. & Architecture and Engineering \\
		3     & Chem.Pharm. & Chemistry and Pharmacy \\
		4     & Econ.Mgmt. & Economics and Management \\
		5     & Educ.Psy. & Physical education, Teaching, Psychology \\
		6     & Law   & Law \\
		7     & Lit.Lang. & Literature, Languages and Humanities \\
		8     & Health & Medicine and Health-related studies \\
		9     & Pol.Soc. & Political Sciences, Sociology and Communication \\
		10    & Sci.Stat. & Math, Physics, Natural Sciences and Statistics \\
		\bottomrule
	\end{tabular}%
	\label{tab:groups}%
\end{table}%

Students with any secondary education diploma can access university.\footnote{Until the late 1960s, only students with the most academic-oriented high-school diplomas could access university. See \cite{bianchi2020expansion} and \cite{bianchi2020scientific} for the evaluation of the reforms that expanded access to higher education to all high-school graduates.} Admission into a bachelor degree can either be regulated at the national level -- as with all health-related degrees, veterinary, architecture, and primary education -- or at the university level. Since universities cannot significantly differentiate their programs in terms of content, they use selection criteria to attract students when possible. During the period of analysis, selection mostly occurred through a multiple choice exam, sometimes with additional requirements such as language tests. The presence and stringency of entry exams will be exploited for identification, as explained in sections \ref{sec:theory} and \ref{sec:data}.
Admission to a master's degree requires a bachelor's and it also typically involves meeting curricular prerequisites, conditions on the bachelor's graduation grade, and interviews. Curricular prerequisites are defined as credits in specific courses. For example, to enroll in a master's in economics, a student must have completed 53 credits in economics, statistics, and other social sciences during the bachelor's.

Tuition fees vary based on the degree, the university, and family income. About one third of students do not pay any tuition due to low family income. The average annual fee for the other students is around 1,500 euros (1,628 euros in 2019 \cite{eurydice2020fees}). Other benefits, such as housing and meal vouchers, are allocated at the regional level based on income and merit. 
Private universities charge higher tuition, usually between 10 and 15 thousand euros per year for an undergraduate program, and they govern their own merit- and need-based grants. All higher education regulations in terms of degree types, academic curricula, and admission apply to both private and public institutions. In 2011 and 2012, only 8.17\% of all university students were enrolled in private institutions \citep{istat2021}. 
	
\section{Theoretical Framework}\label{sec:theory}
The empirical exercise in this paper unfolds in two stages. First, I estimate labor market returns to university careers using a nested random utility model that accounts for the timing of choices and self-selection. Not accounting for the choice structure leads to biased results, as students self-select into university careers based on both observed and unobserved characteristics, and choices are unordered. 
Then, I use the information about the disciplinary content of degrees to investigate various policy-relevant questions on degree design. I ask whether the academic careers with the highest labor market returns are also the ones with the most quantitative or STEM content. Additionally, I check whether specializing early (during the bachelor's) or late (during the master's) in a given discipline is associated with high labor market returns. Finally, I also study whether multidisciplinarity -- pursuing a master's in a different discipline from one's bachelor's -- pays off in terms of outcomes.

This section focuses on the first part of the empirical exercise and illustrates how I retrieve the labor market returns to university careers. Section \ref{sec:dynamicchoicemodel} outlines the methods used to obtain the probabilities of enrollment into any university career that exploit the timing of choices and exclusion restrictions. Section \ref{sec:theory_TEs} illustrates how the probabilities of enrollment engage with a simple function of labor market outcomes (employment and wages) to obtain causal returns to university careers. The theoretical framework is set up in close relation with the available data, discussed in section \ref{sec:data}.

\subsection{Sequential Choices of Bachelor's and Master's}\label{sec:dynamicchoicemodel}

Here, I discuss the estimation procedure that leverages a nested logit model and exclusion restrictions to identify the individual probability of enrolling in any combination of bachelor's and master's. This modeling choice is grounded in its choice-theoretic foundation in dynamic discrete choice problems, where the intuition is that conditional on observed state variables, we can express future utility terms as functions of the probabilities that such choices occur \citep{hotz1993conditional}. 
Sequential choice problems with discrete unordered choices can be estimated with conditional choice probability (CCP) estimators that are brought to the data with nested logit models under the assumption of generalized extreme valued (GEV) distributed errors \citep{arcidiacono2011practical}. The model allows for unobserved determinants of the choices to be correlated across nests \citep{hoffman1988multinomial, mcfadden1974conditional, montmarquette2002young, bamberger1987occupational} and is implemented sequentially for tractability \citep{MCFADDEN19841395, amemiya1985advanced}. 

One important feature of my analysis -- contrary for example to \cite{montmarquette2002young} -- is that I do not model the alternative outcome of not choosing a bachelor's degree.  Instead, the assumption is that if a student is not admitted to their preferred degree program, they will opt for an alternative program rather than forego university education altogether. This assumption is primarily driven by the characteristics of the data and is deemed reasonable within the context of Italy's public, widely accessible, and affordable higher education system. 

Let $i\in I$ denote individuals, $j\in B$ denote a choice of bachelor's degree with $\dim (B)=L\in \mathbb{N}$, $m\in M$ denote a choice of master's degree or no master with $\dim (M)=L+1$, such that $jm\in B\times M$ denotes a university career and $\dim (B\times M)=L(L+1)$.
The timing is as follows: in the first period, the individual must choose a bachelor's degree; in the second period, they must choose a master's degree conditional on their choice of bachelor's; ultimately, the student enters the labor market where outcomes will depend on her choice of education. In the second period, students may additionally choose not to enroll in a master's degree, thus entering the labor market directly.\\

In the first period, a student $i\in I$ chooses a bachelor $j\in B$. The choice will depend on characteristics that vary with the student, as well as characteristics that vary with the choice. The probability that a student $i$ chooses a bachelor $j$ is given by

\begin{equation}\label{eq:fw_t1}
	P_{ij}=\dfrac{\exp\{X_i\beta_j+Z_{ij}\lambda_j\}}{\sum\limits_{k=1}^{B}\exp\{X_i\beta_k+Z_{ik}\lambda_k\}}
\end{equation}


\noindent where $X_i$ is a matrix of characteristics that vary with the individual (gender, family background, general ability) and $Z_{ij}$ is a matrix of characteristics that vary both with the individual and the choice of bachelor's (a composite measure of selectivity of admission requirements and distance to college for all bachelors'). The variation in $Z_{ij}$ ensures that the vector of probabilities for every counterfactual bachelor degree and individual $P_{ij} \,\,\forall \,\,j\in B$ can be computed.\footnote{For clarity, I omit additional covariates throughout this section such as cohort and geography fixed effects and other controls. Section \ref{sec:degrees} addresses them in detail.} 

The second nest of the model captures the choice of master's degree $m\in M$ conditional on a previous choice of bachelor's $j$, where $M$ also includes the choice of not enrolling in a master's and entering the labor market directly. Similar to the choice of bachelor's, the probability that a student $i$ chooses master $m$ conditional on bachelor $j$ is given by

\begin{equation}\label{eq:fw_t2}
	[P_{im}\mid j] \,\, =\, \, \dfrac{\exp\{X_i\beta_m+Z_{im}\lambda_m\}}{\sum\limits_{n=1}^{M}\exp\{X_i\beta_n+Z_{in}\lambda_n\}} 
\end{equation}

\noindent where  $X_i$ is defined as before and $Z_{im}$ is a matrix of characteristics that vary both with the individual and the choice of master (factors that determine the individual's eligibility for enrollment into each master's degree), conditional on the previous choice of bachelor's $j$. In practice, I observe enrollment constraints for each master's that vary with the previous choice of bachelor's and can be reconstructed for each $jm$ pair. 
Once again, the variation in $Z_{im}$ ensures that the probability of choosing every counterfactual master's degree can be computed, $P_{im}\mid j \,\,\forall\,\, j\in B,\,\, m\in M$. \\
Then, the probability of enrolling in any university career accounting for self-selection follows from equations \eqref{eq:fw_t1} and \eqref{eq:fw_t2} is given by
\begin{equation}\label{eq:instrument}
	P_{ijm}=P_{ij}\times[P_{im}\mid j] \quad\forall\,\,j\in B,\,\, m\in M
\end{equation}
where 
\begin{equation*}
    \sum_{j=1}^B\sum_{m=1}^M P_{ijm}=1 \quad\forall\ i.
\end{equation*}
$P_{ijm}$ is the predicted probability of enrollment into degree combination $jm$ that credibly accounts for self-selection since equations \eqref{eq:fw_t1} and \eqref{eq:fw_t2} account for general ability and family background inter alia, as well as exogenous variation in the ease of access into degrees. Importantly, the variation in matrices $Z_{ij}$ and $Z_{im}$ allows for the computation of the probability of choosing every counterfactual degree-pair, overcoming the main problem in the computation of returns to degrees, which is the lack of sufficient instrumental variables to account for all possible (endogenous) choices. In principle, any number of returns to degree-pairs can be computed with this approach, as long as there is sufficient variation in $Z_{ij}$ and $Z_{im}$. In practice, the estimation of the nonlinear equations \eqref{eq:fw_t1} and \eqref{eq:fw_t2} with maximum likelihood and the relatively high dimensionality of $X_{i}$, $Z_{ij}$ and $Z_{im}$ imposes constraints on the number of probabilities $P_{ijm}$ that can be estimated. This means that university careers which are infrequently chosen may be difficult to estimate.


\subsection{Returns to University Careers}\label{sec:theory_TEs}


I exploit probabilities $P_{ijm}$ to identify the effect of university career $(j,\, m)$ on labor market outcomes in a simple function 
\begin{equation}\label{eq:fw_t3_reducedform}
		y_i=X_i\beta+\sum_{j=1}^B\sum_{m=1}^M P_{ijm}\alpha_{jm}+\epsilon_i
\end{equation}
where $y_i$ is the labor market outcome of interest (log wages, employment), $X_i$ is a vector of individual characteristics and controls (gender, family background, high school grades), and $\alpha_{jm}$ denotes the effect of the potential treatment (careers) on outcomes. I interpret $\alpha_{jm}$ as the effect of university career $jm$ on the labor market outcome $y_i$. These coefficients represent my object of interest as they will then be used to investigate the relationship between degree characteristics and economic outcomes in section \ref{sec:courses}.
The empirical specification will additionally include rich sets of fixed effects (cohort, geography), detailed in section \ref{sec:degrees}. 
I resort to this functional form to address three challenges to identification: self-selection on unobserved characteristics, the unordered nature of university careers, and the considerable number of choices.

To best understand the implications of these three challenges, I compare equation \eqref{eq:fw_t3_reducedform} with the extreme case of no-self selection into university careers on unobserved characteristics. In this case, the simple OLS regression 
\begin{equation}\label{eq:fw_t3}
	y_i=X_i\beta+\sum_{j=1}^B\sum_{m=1}^M D_{ijm}\gamma_{jm}+u_i
\end{equation}
would return the effect $\gamma_{jm}$ of career (treatment) $D_{ijm}$ on outcome $y_i$ relative to some excluded category $D_{i0}$, conditional on observed individual characteristics $X_i$, and $\gamma_{jm}$ and $\alpha_{jm}$ would coincide. 
Clearly, any attempt to estimate equation \eqref{eq:fw_t3} directly will result in strongly biased results as we expect students to enroll in careers based on unobserved characteristics. I address self-selection in equation \eqref{eq:fw_t3_reducedform} by leveraging exclusion restrictions $Z_{ij}$ and $Z_{im}$ in equations \eqref{eq:fw_t1}-\eqref{eq:instrument} to compute $P_{ijm}$.\footnote{As equations \eqref{eq:fw_t1}-\eqref{eq:fw_t3_reducedform} are estimated sequentially, I obtain the standard errors of $\alpha_{jm}$ through pairwise bootstrapping, further discussed in section \ref{sec:degrees}.}

The second -- more nuanced -- challenge stems from the unordered nature of university careers. This equally affects equations \eqref{eq:fw_t3_reducedform} and \eqref{eq:fw_t3} as it concerns the identification of counterfactuals, that is, the benchmark (omitted) choice against which I measure the effect of each career. Importantly, when choices are unordered, the omitted category is non-neutral and should represent at least the second preferred option or lack of treatment \citep{kirkeboen2016field,bhuller20222sls,heckman2018unordered}. To illustrate this point, consider a simplified setting with only three choices -- math (M), humanities (H), and economics (E) -- and two observationally identical students who enroll in economics. In this case, the effect of studying economics may not be identifiable without further information on partial rankings if absent the choice of economics, the two students choose to enroll in different degrees. To address this issue, I assume that the excluded category $D_{i0}$ (and consequently $P_{i0}$) is a good proxy of lack of treatment. Section \ref{sec:degrees} describes the omitted category and its implications. 
In the example, the effect of studying economics may be heterogeneous or even contrasting depending on the choices the students would make if their preferred option were not available. A student who alternatively chooses humanities might benefit from studying economics if $y_E>y_H$, ceteris paribus, while a student who alternatively chooses math might suffer if $y_E<y_M$.\footnote{See \cite{mountjoy2021community} for a thorough discussion on contrasting margins of treatment.} This is the case in all unordered settings, with the number of heterogeneous margins of treatment increasing with the number of options. Given the high number of combinations of bachelor's and master's degrees, this setting allows for up to $L^4+2L^3-L$ margins of treatment, which are unlikely to be of economic relevance.\footnote{$\dim(B\times M)=L(L+1)$. Then, the number of possible margins of treatment is equal to $L(L+1)\cdot(L(L+1)-1)=L^4+2L^3-L$. In comparison, \cite{mountjoy2021community} focuses on three possible treatments and six contrasting margins. Similarly, a practical application of \cite{heckman2018unordered} who also focuses on unordered treatments identifies a subset of interesting margins \citep{braccioli2022unordered}. \cite{heckman2006understanding,heckman2010comparing} also investigate the constraints imposed by settings with unordered treatments.} 
The aggregation of the numerous heterogeneous margins to obtain meaningful effects requires proper weighting, which relies on two conditions: that the instruments affect choices monotonically on average, and that they do not cross-contaminate choices \citep{bhuller20222sls}. Lack of cross-contamination implies that given a university career $jm'$, any instrument $P_{jm'}$ is uniquely relevant for treatment $D_{jm'}$. This means that if instrument $P_{jm'}$ does not induce agent $i$ into treatment $jm'$, it cannot impact treatment $jm''\neq jm'$ in any way that changes behavior.\footnote{Lack of proper weighting due to cross-contamination of instruments may lead to severe misrepresentation of the treatment effects. In extreme cases, cross-contamination of instruments may result in a negative average treatment effect of career $jm$ even if all heterogeneous margins of treatment are positive \citep{bhuller20222sls}.}    
The stepwise estimation of $P_{ijm}$ with equations \eqref{eq:fw_t1}-\eqref{eq:instrument} allows for rich substitution patterns within which it is reasonable to assume average monotonicity of $Z_{ij}$ and $Z_{im}$ with respect to choices $j$ and $m$ (i.e. marginally shifting the admission requirements to one degree $j$ affects choices monotonically on average within each career $jm$). The nested setup also reduces the chances of cross-contamination between $P_{ijm}$ and $D_{ijm}$ as variation in admission requirements is allowed to simultaneously affect many outcomes.
Taken together, these conditions are necessary to ensure that instruments induce changes in treatment uptake in a single, threshold-crossing  manner even in an unordered setting \citep{vytlacil2002independence, heckman2018unordered}.

The third challenge addressed by equation \eqref{eq:fw_t3_reducedform} pertains to the number of career effects $\alpha_{jm}$ of interest which can be as high as $L(L+1)$. By exploiting the reduced form, I do not need to leverage the correlation between $P_{ijm}$ and $D_{ijm}$ for identification, as would be the case in a two-staged least squares setting where $P_{ijm}$ serves as an instrument for treatment $D_{ijm}$ \citep{chernozhukov2008reduced}. To understand why the dimension of $\alpha_{jm}$ can be an issue, consider the following modified 2SLS with a simplified first stage regression proposed by \cite{bhuller20222sls} to ensure the proper weighting of heterogeneous margins

\begin{equation}\label{eq:fw_t3_firststage}
	D_{ijm'}=X_i\beta_{jm'}+P_{ijm'}\varphi_{jm'}+v_{ijm'}
\end{equation}

\noindent for any arbitrary treatment $jm'\in B\times M$, such that treatment effects $\psi_{jm}$ are calculated as
\begin{equation*}
    	y_i=X_i\beta+\sum_{j=1}^B\sum_{m=1}^M \hat{D}_{ijm}\psi_{jm}+u_i.
\end{equation*}

Equation \ref{eq:fw_t3_firststage} differs from the first-stage equation in a standard 2SLS framework because only the instrument pertaining to the treatment on the left-hand side is included, i.e., $\varphi_{jm'}$ is a scalar.\footnote{Standard 2SLS requires the estimation of $B\times M$ first stage equations for every career $(j, m)$:
$$
D_{ijm}=X_i\beta_{jm}+\sum_{k=1}^B\sum_{n=1}^M P_{ikn}\varphi_{kn}+\nu_{ijm}.
$$
}
As the number of endogenous choices increases, it becomes increasingly plausible that at least some instrument $P_{ijm}$ is not sufficiently correlated with treatment $D_{ijm}$ even when it is relevant, thus incurring a weak instrument problem. When probabilities $P_{ijm}$ are jointly strongly relevant, the reduced form coefficients $\alpha_{jm}$ asymptotically identify treatment effects $\psi_{jm}$ \citep{chernozhukov2008reduced, phillips2017structural, crudu2021inference, mikusheva2022inference}. I discuss the implications of this assumption in section \ref{sec:simulations}.

By addressing these three empirical challenges, I can interpret $\alpha_{jm}$ as the average treatment effect of enrolling in university career $jm$. One alternative interpretation of $\alpha_{jm}$ that does not require the IV-equivalence assumptions on single threshold-crossing to hold relies on the structural interpretation of the nested model in section \ref{sec:dynamicchoicemodel} as a dynamic discrete choice model \citep{arcidiacono2011practical}. In this case, $\alpha_{jm}$ is the future utility term of a particular choice or the ex-ante treatment effect. The assumptions that support this interpretation require us to believe equations \eqref{eq:fw_t1} and \eqref{eq:fw_t2} accurately incorporate the determinants of the decision-making process of university career. Indeed, a wealth of sophisticated structural models has exploited this type of information to understand how students make schooling decisions \citep{arcidiacono2004ability, ashworth2021changes, dhaultfoeuille2013inference}.
Lastly, $\alpha_{jm}$ can always be interpreted as the labor market effect of shifts in the potential treatment driven by changes in the admission requirements $Z_{ij}$ and $Z_{im}$. In this setting, all instruments are jointly strongly relevant, increases in instruments $P_{ijm}$ increase the probability of treatment $D_{ijm}$ for all careers $jm$, and the nested model suggests that $P_{ijm}$ should only affect $D_{ijm}$. For these reasons, I interpret $\alpha_{jm}$ as equivalent to IV estimates.

\begin{figure}[ht] 
	\caption{Model representation}
	\includegraphics[width=0.7\linewidth]{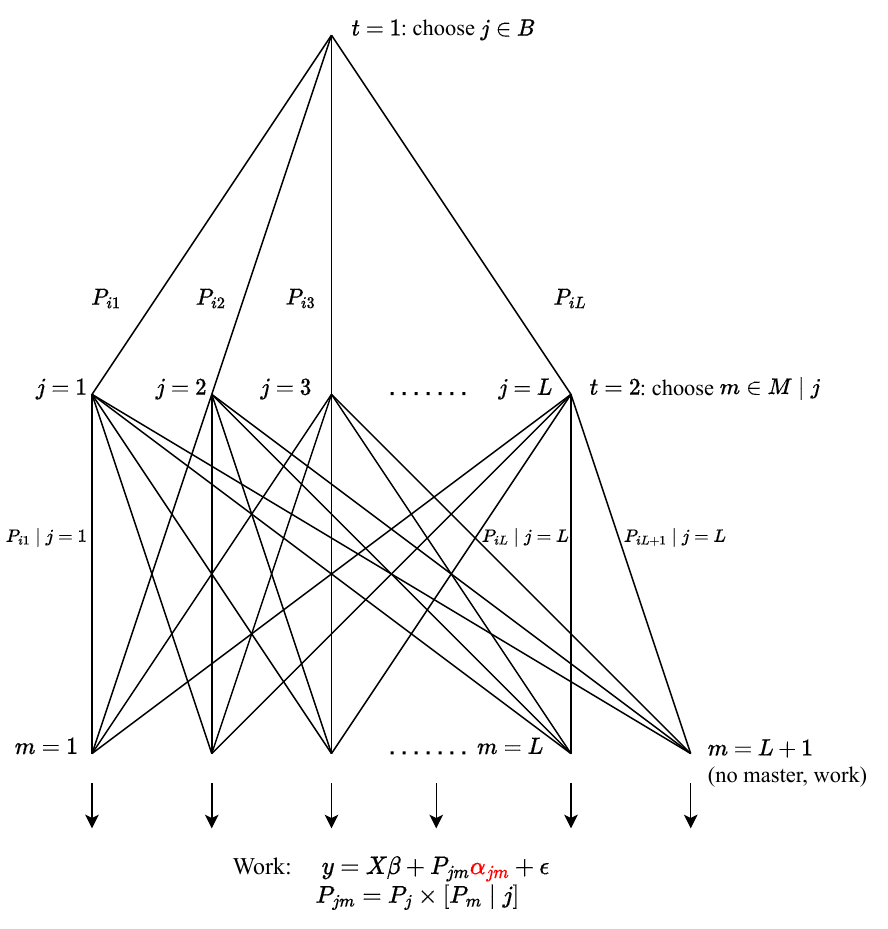}
	\centering
	\label{fig:schoolchoice_graph}
	
\end{figure}

Figure \ref{fig:schoolchoice_graph} summarizes the timing and structure of the choice of university careers and how it integrates with the estimation of labor market outcomes $\alpha_{jm}$. Exclusion restrictions that mimic admission procedures at each stage allow for the computation of counterfactual probabilities of choosing any alternative university career, partialling out the self-selection due to preferences, ability, and family background. Such counterfactual probabilities are then used as instruments for university career treatments to retrieve the causal effect of the choice of bachelor's and master's on labor market outcomes. The exploitation of timing to retrieve valid instruments allows for rich substitution patterns. 
An additional advantage of modeling the decision-making process explicitly is that, unlike standard 2SLS settings, it allows for students to be both forward-looking and introspective in their choices. In fact, by allowing for correlation between nests, the error term $\epsilon_i$ is allowed to be realized in multiple stages.
Even though the equations of the model could be jointly estimated, the lack of certain degree combinations warrants that they be estimated sequentially. This implies that all standard errors must be bootstrapped to account for the method's sequential structure.

Finally, it is worth underscoring why standard 2SLS does not produce appropriate treatment effects. Not only does it allow for instruments to cross-contaminate treatments, it also imposes the estimation of a large number of irrelevant parameters which introduce significant strain on the estimator. Including irrelevant instruments on the right-hand side of the first-stage regression will decrease the precision of the estimate of the treatment effect in the second stage because it will lead to possible collinearity between instruments and inflate the standard errors of the first-stage predictions. This is especially true if -- as it is the case -- certain probabilities $P_{ijm'}$ are close to zero for individuals who are observed to choose $jm''\neq jm'$.\footnote{Let us consider a simplified framework for presentation purposes where there are only two possible choices in each set $B=\{H,\, S\}$ and $M=\{H,\,S\}$, with $H$ denoting "humanities" and $S$ "science". Then $jm\in B\times M=\{HH,\,HS,\, SH,\, SS\}$ and the "standard" first-stage regressions of a 2SLS model become
\begin{align*}
	\begin{split}
		D_{iHH}&=  X_i\varphi_X^{HH}+P_{iHH}\varphi_{HH}^{HH}+P_{iHS}\varphi_{HS}^{HH}+P_{iSH}\varphi_{SH}^{HH}+P_{iSS}\varphi_{SS}^{HH}+u_{iHH}\\
		D_{iHS}&= X_i\varphi_X^{HS}+P_{iHH}\varphi_{HH}^{HS}+P_{iHS}\varphi_{HS}^{HS}+P_{iSH}\varphi_{SH}^{HS}+P_{iSS}\varphi_{SS}^{HS}+u_{iHS}\\
		D_{iSH}&=  X_i\varphi_X^{SH}+P_{iHH}\varphi_{HH}^{SH}+P_{iHS}\varphi_{HS}^{SH}+P_{iSH}\varphi_{SH}^{SH}+P_{iSS}\varphi_{SS}^{SH}+u_{iSH}\\
		D_{iSS}&= X_i\varphi_X^{SS}+P_{iHH}\varphi_{HH}^{SS}+P_{iHS}\varphi_{HS}^{SS}+P_{iSH}\varphi_{SH}^{SS}+P_{iSS}\varphi_{SS}^{SS}+u_{iSS}.
	\end{split}
\end{align*}
As this approach forces the estimation of $(B\times M-1)^2$ irrelevant parameters, there is a serious concern of overidentification in the first stage, which is exacerbated if some $P_{ijm}$ is small and aggravates any weak instrument bias.}


\section{Data Sources and Summary Statistics}\label{sec:data}

The empirical analysis combines three data sources. The first is an administrative student-level database covering the universe of all graduates from both bachelor's and master's programs at most Italian universities, both public and private. This database is maintained by AlmaLaurea, a consortium of universities that combines and harmonizes the original student records shared by each participating institution.
The same consortium administers surveys to all the graduates in their archives at the time of graduation and one, three, and five years later. This is my second source of data and it is individually (but anonymously) linked to the administrative records.\footnote{The AlmaLaurea Inter-University Consortium collaborates with Italian universities and the Ministry of University and Research (MUR) to monitor the labor market outcomes of Italian graduates and help match graduates with employers. Universities adhere to the consortium in different years, with 80 out of 96 universities participating in 2022. The full list of participating universities can be found in appendix \ref{sec:appendix_method}.}
The third data source is a novel archive of administrative information detailing the content of all university programs in Italy, including admission requirements for all bachelor's and master's programs.

\subsection{University Graduates}\label{sec:data_graduates}
My working sample includes all the individuals who graduated from 2007 to 2014, allowing me to observe the most recent outcomes in 2019. This dataset comprises complete information on 655 847 students. According to a comparison with the National Statistical Institute's (ISTAT) records, the raw sample covers between 62\% and 76\% of all graduates in the years of interest.\footnote{In 2007, 46 universities out of all 96 adhered to the consortium, while 64 were participating by 2014. I do not consider earlier cohorts since they only include students who graduate in July of each year, university participation was lower, and a different university system was still fading away.} 
Several analyses by the consortium suggest that the composition of their sample accurately reflects the national population of graduates over time \citep{almalaurea2019note, almalaurea2020note}. The survey data is collected online and through phone interviews. Response rates are extremely high (91\%) for the first survey administered before graduation, and remaining high also for the later ones (88\% across cohorts one year after graduation, 81\% after three years, and 75\% after five years).
The surveys provide information about socio-economic characteristics and labor market outcomes. 

Two limitations are intrinsic to the setup. First, I only observe students who complete at least a bachelor's degree. Hence, any conclusion from the empirical analysis should be interpreted at the intensive margin. 
Second, I do not observe university dropouts. 
This is relevant for master's graduates, as it is impossible to distinguish between outmigration of bachelor's graduates to institutions outside of the consortium, and master's students who drop out. To avoid confusing the two, among bachelor's graduates without a master's degree, I only keep those who report no intention of enrolling in a master's program.\footnote{The survey asks bachelor's graduates whether they intend to enroll in a master's degree abroad, enroll in a different type of program (e.g. one-year master's), or not enroll. In addition to master's graduates, I only keep bachelor's graduates who do not intend to further enroll in higher education. Fortunately, attrition due to outmigration seems low, as only 1.4\% state an intention to enroll in a master's that is not observed by the consortium.} Second, ancillary information on local labor market conditions is not available for international students who are dropped from the main analysis. They account for less than 2\% of the dataset, as most international mobility occurs through Erasmus exchanges and similar short-term exchange programs.\footnote{The employment rate for individuals 25-34 years old in the province of birth before enrollment into university summarizes local labor market conditions. The information is obtained from the National Statistical Institute (ISTAT).}


Table \ref{tab:frequency_careers} presents the distribution of individuals across university careers. Groups with fewer than 100 observations (in red) are dropped to ensure sufficient power during estimation for a total of 1 325 observations. 56 groups out of 110 contain at least 100 records. 60.8\% of graduates complete both a bachelor's and a master's degree. 24 433 (6.1\%) of master's graduates switch disciplines after the bachelor's. This value is very conservative as it depends on the grouping of degrees in broad fields. Less conservative groupings observe switching in up to 15\% of cases. Section \ref{sec:data_programs} elaborates on the grouping rule.

\begin{table}[ht]
	
	\centering
	\setlength{\tabcolsep}{3pt}
	\caption{Frequency of graduates in all university careers}
	\resizebox{\linewidth}{!}{
		\begin{tabular}{clrrrrrrrrrrrr}
			\toprule
			&       & \multicolumn{12}{c}{Master's} \\
			&       & \multicolumn{1}{c}{\begin{sideways}No Master\end{sideways}} & \multicolumn{1}{c}{\begin{sideways}AVGB\end{sideways}} & \multicolumn{1}{c}{\begin{sideways}Arc.Eng.\end{sideways}} & \multicolumn{1}{c}{\begin{sideways}Chem.Ph.\end{sideways}} & \multicolumn{1}{c}{\begin{sideways}Ec.Mg.\end{sideways}} & \multicolumn{1}{c}{\begin{sideways}Ed.Psy.\end{sideways}} & \multicolumn{1}{c}{\begin{sideways}Law\end{sideways}} & \multicolumn{1}{c}{\begin{sideways}Lit.Lan.\end{sideways}} & \multicolumn{1}{c}{\begin{sideways}Health\end{sideways}} & \multicolumn{1}{c}{\begin{sideways}Pol.Soc.\end{sideways}} & \multicolumn{1}{c}{\begin{sideways}Sci.Stat.\end{sideways}} & \multicolumn{1}{c}{\begin{sideways}Total\end{sideways}} \\
			\cmidrule{3-14}    \multirow{11}[2]{*}{\begin{sideways}Bachelor's\end{sideways}} & AVGB  & 8,387  & 26,316 & 219   & \textcolor[rgb]{ .612,  0,  .024}{59} & \textcolor[rgb]{ .612,  0,  .024}{19} & 180   & \textcolor[rgb]{ .612,  0,  .024}{*} & \textcolor[rgb]{ .612,  0,  .024}{41} & 622   & \textcolor[rgb]{ .612,  0,  .024}{29} & 932   & 36,656 \\
			& Arc.Eng. & 22,285 & \textcolor[rgb]{ .612,  0,  .024}{87} & 79,827 & 776   & \textcolor[rgb]{ .612,  0,  .024}{84} & \textcolor[rgb]{ .612,  0,  .024}{18} & \textcolor[rgb]{ .612,  0,  .024}{*} & 287   & \textcolor[rgb]{ .612,  0,  .024}{10} & \textcolor[rgb]{ .612,  0,  .024}{91} & 251   & 103,426 \\
			& Chem.Ph. & 3,902  & 118   & \textcolor[rgb]{ .612,  0,  .024}{11} & 20,643 & \textcolor[rgb]{ .612,  0,  .024}{*} & \textcolor[rgb]{ .612,  0,  .024}{*} & \textcolor[rgb]{ .612,  0,  .024}{*} & \textcolor[rgb]{ .612,  0,  .024}{*} & 260   & \textcolor[rgb]{ .612,  0,  .024}{*} & \textcolor[rgb]{ .612,  0,  .024}{18} & 24,923 \\
			& Ec.Mg. & 27,806 & \textcolor[rgb]{ .612,  0,  .024}{23} & \textcolor[rgb]{ .612,  0,  .024}{16} & \textcolor[rgb]{ .612,  0,  .024}{*} & 46244 & 123   & 208   & \textcolor[rgb]{ .612,  0,  .024}{67} & \textcolor[rgb]{ .612,  0,  .024}{31} & 1,153  & 459   & 75,993 \\
			& Ed.Psy. & 28,530 & \textcolor[rgb]{ .612,  0,  .024}{26} & \textcolor[rgb]{ .612,  0,  .024}{*} & \textcolor[rgb]{ .612,  0,  .024}{*} & \textcolor[rgb]{ .612,  0,  .024}{16} & 46,085 & \textcolor[rgb]{ .612,  0,  .024}{18} & 250   & 125   & 537   & \textcolor[rgb]{ .612,  0,  .024}{11} & 75,527 \\
			& Law   & 8,054  & \textcolor[rgb]{ .612,  0,  .024}{*} & \textcolor[rgb]{ .612,  0,  .024}{27} & \textcolor[rgb]{ .612,  0,  .024}{*} & 1,466  & 127   & 46,766 & \textcolor[rgb]{ .612,  0,  .024}{84} & \textcolor[rgb]{ .612,  0,  .024}{24} & 1,101  & \textcolor[rgb]{ .612,  0,  .024}{13} & 57,514 \\
			& Lit.Lan. & 38,343 & \textcolor[rgb]{ .612,  0,  .024}{76} & 122   & \textcolor[rgb]{ .612,  0,  .024}{27} & 693   & 595   & \textcolor[rgb]{ .612,  0,  .024}{55} & 44,974 & \textcolor[rgb]{ .612,  0,  .024}{27} & 5,788  & 166   & 90,681 \\
			& Health & 75,743 & 403   & \textcolor[rgb]{ .612,  0,  .024}{29} & \textcolor[rgb]{ .612,  0,  .024}{*} & \textcolor[rgb]{ .612,  0,  .024}{16} & 313   & \textcolor[rgb]{ .612,  0,  .024}{*} & \textcolor[rgb]{ .612,  0,  .024}{11} & 28,056 & \textcolor[rgb]{ .612,  0,  .024}{50} & \textcolor[rgb]{ .612,  0,  .024}{*} & 104,515 \\
			& Pol.Soc. & 35,003 & \textcolor[rgb]{ .612,  0,  .024}{*} & \textcolor[rgb]{ .612,  0,  .024}{65} & \textcolor[rgb]{ .612,  0,  .024}{*} & 1,562  & 599   & 1,342  & 1,949  & \textcolor[rgb]{ .612,  0,  .024}{24} & 25,324 & 112   & 65,891 \\
			& Sci.Stat. & 8,597  & 1,014  & 115   & 183   & 123   & \textcolor[rgb]{ .612,  0,  .024}{15} & \textcolor[rgb]{ .612,  0,  .024}{*} & \textcolor[rgb]{ .612,  0,  .024}{60} & \textcolor[rgb]{ .612,  0,  .024}{*} & 160   & 10,529 & 20,721 \\
			& Total & 256,650 & 27,851 & 80,283 & 21,602 & 50,088 & 48,022 & 48,316 & 47,460 & 29,063 & 34,063 & 12,449 & 655,847 \\
			\bottomrule
		\end{tabular}
	}%
	\caption*{\footnotesize Frequencies in red denote careers that are observed for less than 100 individuals. Asterisks indicate groups with fewer than 10 individuals. All groups except 3 are chosen at least once. Total amounts do not include the less frequent choices in red. AVGB -- Life Sciences, Arc.Eng. -- Architecture and Engineering, Chem.Ph. -- Chemistry and Pharmacy, Ec.Mg. -- Economics and Management, Ed.Psy. -- Education and Psychology, Lit.Lan. -- Humanities, Literature and Languages, Law -- Law, Health -- Medicine and Health, Pol.Soc. Political and Social Sciences, Sci.Stat. -- Math, Physics and Statistics.}
	\label{tab:frequency_careers}%
\end{table}%

Table \ref{tab:descr_X} presents the descriptive statistics of the main individual characteristics, summarized by bachelor's degree.
The characteristics that vary the most across fields are gender and high school type. Even though there are 62\% of women in the sample, female students are under-represented in architecture and engineering (34\%) and science and statistics (35\%), and are over-represented in education and psychology (83\%) and humanities (78\%). High school types are grouped into three main categories: sciences, humanities, and other high schools, including languages, social sciences, and vocational schools. Although no high school type precludes enrollment into any degree, more students with a humanities high school obtain bachelor's degrees in literature and languages (23\%) and law (34\%). Students from science high schools are over-represented in life sciences, engineering, chemistry and hard sciences. I include two measures of family background: parent education, measured as at least one parent with a college degree, and parent occupation, that is, at least one parent in a high-ranked profession, such as executive, entrepreneur, professional, or academic. Neither of these measures varies dramatically across fields. One exception is law degrees, where relatively more individuals have parents with college degrees (36\%) and in high-ranked occupations (30\%). I standardize high school final grades by province to account for differences in grading standards across school districts. Relatively more students with above-average high school grades enroll in engineering (62\%) and hard sciences (58\%). Below-average high school grades are observed in education (37\%), social sciences (41\%) and healthcare (42\%). 

\begin{table}[ht]
  \centering
  \setlength{\tabcolsep}{3pt}
  \caption{Description of the main individual characteristics  by bachelor's field of study.}
  \resizebox{\linewidth}{!}{
    \begin{tabular}{lccccccccccc}
 
        \toprule
        & All   & AVGB  & Arch.Eng. & Chem.Ph. & Econ.Mg. & Educ.Psy. & Law   & Lit.Lan. & Med.  & Pol.Soc. & Sci.Stat. \\
            Variables       & (1)   & (2)   & (3)   & (4)   & (5)   & (6)   & (7)   & (8)   & (9)   & (10)  & (11) \\ \midrule
          &       &       &       &       &       &       &       &       &       &       &  \\
    High School: grade (st.) & 0.00  & 0.04  & 0.30  & 0.11  & 0.04  & -0.30 & 0.08  & 0.10  & -0.19 & -0.20 & 0.22 \\
          & (1.000) & (0.969) & (0.954) & (0.955) & (0.991) & (0.954) & (0.981) & (0.984) & (1.021) & (0.979) & (0.997) \\
          &       &       &       &       &       &       &       &       &       &       &  \\
    High School: humanities & 0.15  & 0.13  & 0.08  & 0.18  & 0.07  & 0.13  & 0.34  & 0.23  & 0.14  & 0.16  & 0.06 \\
          & (0.359) & (0.337) & (0.271) & (0.380) & (0.258) & (0.342) & (0.474) & (0.423) & (0.344) & (0.365) & (0.241) \\
          &       &       &       &       &       &       &       &       &       &       &  \\
    High School: science & 0.39  & 0.52  & 0.55  & 0.57  & 0.36  & 0.27  & 0.32  & 0.26  & 0.42  & 0.29  & 0.52 \\
          & (0.487) & (0.499) & (0.498) & (0.495) & (0.481) & (0.445) & (0.468) & (0.441) & (0.494) & (0.452) & (0.500) \\
          &       &       &       &       &       &       &       &       &       &       &  \\
    Gender (1=female) & 0.62  & 0.60  & 0.34  & 0.69  & 0.54  & 0.83  & 0.63  & 0.78  & 0.68  & 0.69  & 0.36 \\
          & (0.485) & (0.490) & (0.474) & (0.463) & (0.499) & (0.378) & (0.483) & (0.414) & (0.468) & (0.464) & (0.479) \\
          &       &       &       &       &       &       &       &       &       &       &  \\
    Parents: graduate & 0.26  & 0.27  & 0.31  & 0.33  & 0.22  & 0.18  & 0.36  & 0.26  & 0.23  & 0.22  & 0.27 \\
          & (0.438) & (0.444) & (0.463) & (0.472) & (0.416) & (0.384) & (0.481) & (0.439) & (0.420) & (0.415) & (0.442) \\
          &       &       &       &       &       &       &       &       &       &       &  \\
    Parents: high-rank occ. & 0.21  & 0.21  & 0.25  & 0.26  & 0.22  & 0.16  & 0.30  & 0.21  & 0.18  & 0.20  & 0.18 \\
          & (0.410) & (0.405) & (0.431) & (0.440) & (0.411) & (0.368) & (0.459) & (0.404) & (0.388) & (0.400) & (0.383) \\
          &       &       &       &       &       &       &       &       &       &       &  \\
    Observations & 655847 & 36656 & 103426 & 24923 & 75993 & 75527 & 57514 & 90681 & 104515 & 65891 & 20721 \\ \bottomrule
    \end{tabular}%
    }%
    \vspace{10pt}\caption*{\footnotesize Column labels: AVGB -- Life Sciences, Arch.Eng. -- Architecture and Engineering, Chem.Ph. -- Chemistry and Pharmacy, Econ.Mg. -- Economics and Management, Educ.Psy. -- Education and Psychology, Lit.Lan. -- Humanities, Literature and Languages, Law -- Law, Med. -- Medicine and Health, Pol.Soc -- Political and Social Sciences, Sci.Stat. -- Math, Physics and Statistics.}
  \label{tab:descr_X}%
\end{table}%

\begin{figure}[ht]
	\centering
	\caption{Description of wages in 2015 Euros by academic career.}\label{fig:descr_y_wage}
	\includegraphics[width=\linewidth]{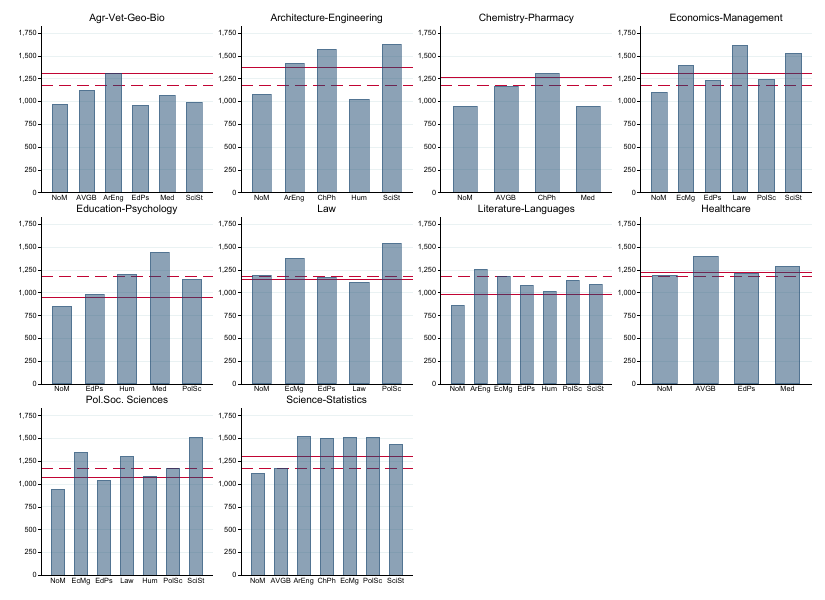}
	\caption*{\footnotesize Sub-graph titles indicate the bachelor's choice, while the fields of study on the horizontal axis refer to master's choices. The solid red line represents the average wage in 2015 Euros for the subsample of individuals who share the same bachelor's choice. The dotted red line indicates the sample average.
	NoM -- No Master, AVGB -- Life Sciences, ArEng -- Architecture and Engineering, ChPh -- Chemistry and Pharmacy, EcMg -- Economics and Management, EdPs -- Education and Psychology, Hum -- Humanities, Literature and Languages, Law -- Law, Med -- Medicine and Health, PolSc Political and Social Sciences, SciSt -- Math, Physics and Statistics.}
\end{figure}

\begin{figure}[ht]
	\centering
	\caption{Description of employment by academic career.}\label{fig:descr_y_empl}
	\includegraphics[width=\linewidth]{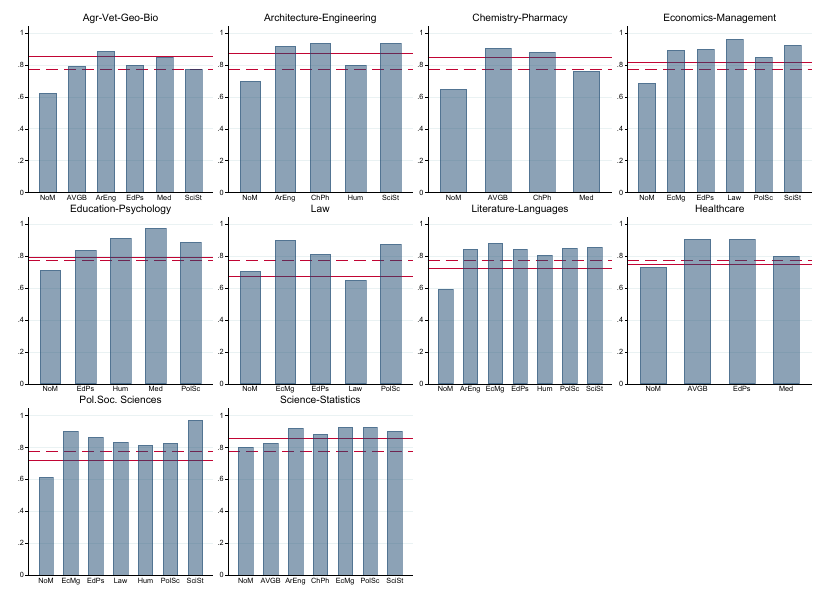}
	\caption*{\footnotesize Sub-graph titles indicate the bachelor's choice, while the fields of study on the horizontal axis refer to master's choices. The solid red line represents the average level of employment for the subsample of individuals who share the same bachelor's choice. The dotted red line indicates the sample average.
		NoM -- No Master, AVGB -- Life Sciences, ArEng -- Architecture and Engineering, ChPh -- Chemistry and Pharmacy, EcMg -- Economics and Management, EdPs -- Education and Psychology, Hum -- Humanities, Literature and Languages, Law -- Law, Med -- Medicine and Health, PolSc Political and Social Sciences, SciSt -- Math, Physics and Statistics.}
\end{figure}

The main empirical analysis focuses on two labor market outcomes: log wages and employment five years after graduation.\footnote{When the outcomes are not available five years after graduation, they are imputed using the one- and three-year survey waves. The main empirical analysis includes survey-wave fixed effects to account for these differences.} Figure \ref{fig:descr_y_wage} presents average wages in levels reported to 2015 Euros for the sample of the employed, which tallies 508 242 records (77\%), for each academic career. 
Figure \ref{fig:descr_y_empl} shows similar summary statistics for average employment levels over the whole sample of 655 847 graduates. 
Both figures \ref{fig:descr_y_wage} and \ref{fig:descr_y_empl} display differences in labor market outcomes by undergraduate choice of major by comparing the solid and dashed red lines. The figures also point to large differences in outcomes by combinations of undergraduate and graduate majors, visible by comparing the vertical bars within each subgraph. Overall, individuals without a master's degree experience worse labor market outcomes on average (first column of each subgraph). Even though these figures present unconditional means, they suggest that outcomes vary substantially across masters' choices also conditional on bachelors'. \\


\subsection{University Programs}\label{sec:data_programs}
I complement the student records with detailed information about the content and structure of all academic programs. The data on the content of programs combines various legal sources to reconstruct the compulsory features of degrees. The data on the structure of programs focuses on admission practices and results from a survey of all programs offered in Italy.

The data on the content of academic programs comes from two sources: the content requirements in terms of credits and courses of all 47 legally recognized bachelor's programs and 99 master's programs, and the official codes and description of 370 available disciplines.\footnote{Law 270/2004 provides detailed information on the legal requirements that degrees must meet. Addenda to the law have been exceptionally published over the years and are considered when relevant. The list of scientific disciplines (\textit{settori scientifico-disciplinari}) is maintained by the Italian National University Council (CUN). The total number of disciplines has increased since the years under consideration to 384.} Crucially, I observe the disciplinary content of any university course independently of the institution or the degree in which it is taught. Furthermore, for each course I observe the number of credits that must be obtained to meet the program's legal requirements. I use this information to account for different levels of specialization across degrees. For example, a course in applied economics is present in 17 bachelor's programs and 33 master's programs. However, the number of required credits varies greatly, from 4 credits in a master's program in architecture to 32 credits in a bachelor's in economics.

\begin{figure}[ht]
	\centering
	\caption{Breakdown of fields of study taught in degrees}\label{fig:degree_description}
	\includegraphics[width=\linewidth]{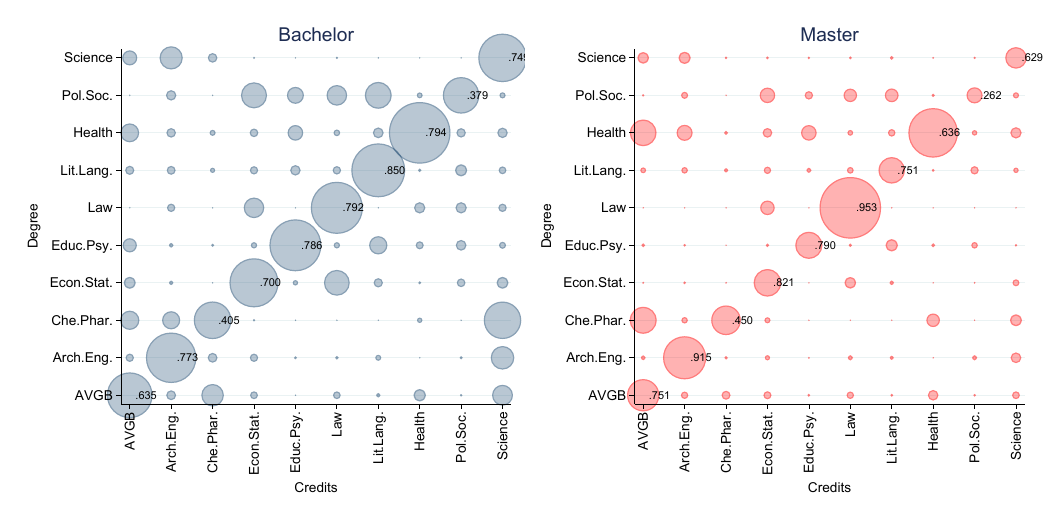}
	\caption*{\footnotesize The figure presents groups of degrees on the vertical axis plotted against the content in each degree. Larger bubbles indicate that more credits (ECTSs) in a given group of university courses are taught in a given degree. The percentages on the diagonal refer to the time spent studying the main field of study of the degree. Off-diagonal bubbles represent the credits spent studying field of study $x$ in degree $y$. A row fully describes a university degree. The left (blue) panel refers to bachelor's degrees, while the right (red) panel to master's degrees.  The groups of degrees are provided by AlmaLaurea and further aggregated for statistical precision, the full description is available in appendix \ref{sec:appendix_degrees}. The groups of university courses are provided by MIUR and further aggregated by myself. A description of the labels is summarized in table \ref{tab:groups}. The unit that defines the bubble size is one ECTS (university credit).}
\end{figure}

Figure \ref{fig:degree_description} presents a comprehensive description of the content of bachelor's (left) and master's (right) degrees at the relevant level of aggregation, by plotting them against their academic curriculum, with the total percentage of required credits in the degree's main field of study on the diagonal. Each row represents a degree by averaging the content of each program that belongs to the degree grouping.\footnote{Table \ref{tab:groups} describes the disciplines in each group.}
Significant off-diagonal variation is present, with two degrees -- chemistry and pharmacy, and political and social sciences -- requiring less than 50\% of time studying the main discipline both at the undergraduate and graduate level. While degrees specialize slightly during the master's, with a higher proportion of credits in the main domain, substantial education in off-diagonal fields remains. 
The grouping of degrees, described in table \ref{tab:groups}, serves two objectives: achieving statistical precision and yielding economically interesting results. I primarily base the grouping on that of the data provider and the Italian ministry of higher education. Infrequently chosen groups are further grouped according to the literature (table \ref{tab:stem_litreview} overviews some of the papers that were used) to maintain proximity in content. To further validate this approach, I ensure that the content of the degrees is close within group. For example, even though teaching and psychology lead to different occupations, they are grouped together for statistical precision and because a comparison of their curricula showed several similarities.  This approach is justified by the ultimate aim of this paper: understanding the role of degree content.

In addition to information about disciplinary content, I also collected information about admission requirements. I do this differently for bachelor's and master's programs to account for differences in enrollment procedures.

For bachelors' degrees, I survey the admission procedures to 2296 undergraduate programs in Italy, codifying the following information: presence of a entry exam, type of exam (standardized test, multiple choice, open-end exam, knowledge assessment), number of spots, number of applicants, and application windows.\footnote{The information on admission procedures is only widely available for the years 2018 to 2021. However, all the additional evidence that I could procure points toward high persistence in enrollment practices and admission rates.} I use this information to construct an indicator of binding admission restrictions for each bachelor's program. Specifically, I construct a dummy for each program that is equal to 1 if the bachelor's features fewer spots than applicants in the first round of admissions.  For programs lacking applicant data, I use information on the application phase dates to infer the competitiveness of the selection process. Application calls that are reopened several times or remain open well into the start of the program suggest a less stringent selection process. Hence, in the absence of applicant information, I classify programs with entry exams as not having binding admission restrictions if the call has been reopened or if the exam is a low-stakes knowledge assessment.

Admission into a master's program typically depends on a student's ability to meet eligibility requirements in terms credits acquired during the bachelor's. Additional criteria include bachelor's grades and interviews. Entry exams are rare, but may be in place for healthcare-related fields and psychology, and even then, students must meet curricular criteria. I collected information on all eligibility requirements by surveying all public university master's programs in 2020 and 2021.\footnote{Again, admission criteria are highly persistent in time such that the collected information is strongly relevant even if the years of enrollment do not match the years in which the requirements were collected.} This information is then matched with the previously collected data on academic curricula to calculate the number of credits that must be acquired beyond those already contained in the bachelor's for each pair of undergraduate and graduate degrees. For example, a student with a bachelor's degree in economics meets all the requirements for enrollment in a master's in economics. However, she must acquire 41 additional credits to be eligible for enrollment in a master's in statistics. Conversely, any student who wants to enroll in a master's in economics must have acquired 53 credits in economics, statistics, and other social sciences. The exact number of additional credits that the student must earn will depend on the content covered in her bachelor's.
When a bachelor's does not meet any eligibility criteria, the number of necessary credits is set to 180, equivalent to starting another bachelor's degree. This is the case for access into many degrees that only admit a subset of bachelor's or single-cycle master's degrees such as law or medicine which prevent students from transferring. 

\begin{table}[htbp]
  \centering
  \caption{Descriptive statistics for the exclusion restriction variables $Z_{ij}$ and $Z_{im}$}
    \begin{tabular}{lcccc}
    \toprule
    Variable & Mean  & Std. Dev. & Min   & Max \\
    \midrule
          &       &       &       &  \\
    \multicolumn{5}{l}{\hspace{1cm} A. $Z_{ij}$:\textit{ Entry Exams}} \\
    EE (AVGB) & 0.492 & 0.183 & 0.100 & 0.883 \\
    EE (Arch.Eng.) & 0.417 & 0.196 & 0     & 0.889 \\
    EE (Chem.Pharm.) & 0.634 & 0.258 & 0     & 1 \\
    EE (Econ.Mgmt.) & 0.453 & 0.377 & 0     & 1 \\
    EE (Ed.Psy.) & 0.769 & 0.234 & 0.130 & 1 \\
    EE (Law) & 0.172 & 0.235 & 0     & 0.759 \\
    EE (Lit.Lang.) & 0.165 & 0.126 & 0.001 & 0.672 \\
    EE (Health) & 0.939 & 0.068 & 0.791 & 1 \\
    EE (Pol.Soc.) & 0.299 & 0.238 & 0     & 0.852 \\
    EE (Sci.Stat.) & 0.308 & 0.273 & 0     & 1 \\
          &       &       &       &  \\
    \multicolumn{5}{l}{\hspace{1cm} B. $Z_{im}$: \textit{Constrained number of credits}} \\
    Cred. (AVGB) & 60.987 & 17.616 & 0     & 69.874 \\
    Cred. (Arch.Eng.) & 86.927 & 20.680 & 0     & 96.249 \\
    Cred. (Chem.Pharm.) & 84.485 & 22.110 & 0     & 95.891 \\
    Cred. (Econ.Mgmt.) & 50.539 & 18.686 & 0     & 58.404 \\
    Cred. (Ed.Psy.) & 65.998 & 22.539 & 0     & 82.778 \\
    Cred. (Law) & 91.487 & 38.442 & 0     & 114.910 \\
    Cred. (Lit.Lang.) & 65.866 & 5.802 & 48.790 & 69.000 \\
    Cred. (Health) & 146.272 & 33.683 & 0     & 163.571 \\
    Cred. (Pol.Soc.) & 41.140 & 20.475 & 0     & 62.066 \\
    Cred. (Sci.Stat.) & 76.325 & 9.041 & 40.040 & 86.584 \\
          &       &       &       &  \\
    \multicolumn{5}{l}{\hspace{1cm} C. $Z_{im}$:\textit{ Constrained number of credits (standardized)}} \\
    Cred. (AVGB) & -0.766 & 0.843 & -3.683 & -0.341 \\
    Cred. (Arch.Eng.) & 0.475 & 0.989 & -3.683 & 0.921 \\
    Cred. (Chem.Pharm.) & 0.358 & 1.058 & -3.683 & 0.903 \\
    Cred. (Econ.Mgmt.) & -1.266 & 0.894 & -3.683 & -0.890 \\
    Cred. (Ed.Psy.) & -0.526 & 1.078 & -3.683 & 0.276 \\
    Cred. (Law) & 0.693 & 1.839 & -3.683 & 1.813 \\
    Cred. (Lit.Lang.) & -0.533 & 0.278 & -1.349 & -0.383 \\
    Cred. (Health) & 3.313 & 1.611 & -3.683 & 4.141 \\
    Cred. (Pol.Soc.) & -1.715 & 0.979 & -3.683 & -0.714 \\
    Cred. (Sci.Stat.) & -0.032 & 0.432 & -1.768 & 0.458 \\
    \bottomrule
    \end{tabular}%
  \label{tab:descr_instruments}%
  \caption*{\footnotesize Total number of observations: 655 847; global average of constrained credits across degrees: 77.003.  AVGB -- Life Sciences, ArEn -- Architecture and Engineering, ChPh -- Chemistry and Pharmacy, EcMg -- Economics and Management, EdPs -- Education and Psychology, Hum -- Humanities, Literature and Languages, Law -- Law, Med -- Medicine and Health, PlSc Political and Social Sciences, Sci -- Math, Physics and Statistics.
}
\end{table}%

The vector of exclusion restrictions $Z_{ij}$ that regulates access into the bachelor's is built based on the previously described data on admission criteria into undergraduate programs. I build a measure of the percentage of bachelor's degrees for which the admission criterion is binding for each aggregated category of degrees as described in appendix \ref{sec:appendix_degrees} and university, and merge it with the administrative data for each individual and closest public institution.
There are thus ten variables, one for each group of bachelor's degrees, that measure the share of degrees within a group with a binding admission requirements in the institution closest to the individual's place of birth. As not all universities offer all groups of degrees and programs in different universities vary in their admission restrictions, this information will vary with the individual and the degree. Vector $Z_{ij}$ is clearly exogenous since students cannot influence the level of applicants. Panel A in table \ref{tab:descr_instruments} summarizes these ten variables, one for each bachelor's degree, that vary between 0 and 1, with 1 indicating that all degrees in a given group and institution present binding admission requirements and 0 indicating that none do. On average, the presence of binding admission requirements is lowest in humanities and highest in medicine and healthcare degrees.

The vector of exclusion restrictions $Z_{im}$ that governs admission into master's degrees includes the measures on the differences between each undergraduate's curriculum and the enrollment requirements for all master's programs. There are ten variables, one for every master's program, that vary at the individual and program level. Panels B and C in table \ref{tab:descr_instruments} summarize these ten variables, one for each master's degree, where panel B presents the average values in terms of credits, and panel C transforms the variables in panel B by standardizing them. On average, students must acquire 77 constrained credits to enter a master's program. Once again, there is substantial variation across fields of study.\footnote{These variables are standardized in the empirical analysis to improve model fit.} Average admission requirements are highest for healthcare degrees and lowest for political and social sciences. I additionally include the log distance to the closest public university to instrument the choice not to enroll in a graduate program.


\section{Returns to University Careers}\label{sec:degrees}

This section discusses the implementation of the model outlined in section \ref{sec:theory} to obtain labor market returns to combinations of undergraduate and graduate degrees. 
The relevant steps of the estimation procedure are discussed sequentially to highlight the information available at each stage, as summarized in figure \ref{fig:schoolchoice_graph}.

\subsection{Choice of Bachelor's and Master's Degrees}\label{sec:estimation_P}
Equations \eqref{eq:fw_t1} and \eqref{eq:fw_t2} are brought to the data sequentially. Although it is theoretically possible to estimate them simultaneously using a nested logit model, several data-related considerations—primarily empty cell problems due to the absence of certain combinations and significant differences in the size of degree combinations—make it more practical to estimate the equations separately. Therefore, the equations are estimated in the order presented in section \ref{sec:dynamicchoicemodel} as multinomial logit models (equations \ref{eq:fw_t1} and \ref{eq:fw_t2}). 



Throughout this section, the vector of observed individual characteristics $X_i$ will include high school grade, standardized at the province level to account for regional differences in grading standards; high school type (humanities, scientific, or other -- baseline category), gender; parents' education (at least one parent with a college degree); and parents' occupation (at least one parent in a high-ranked occupation: academics, liberal professionals, entrepreneurs, executives). Summary statistics for these variables were reported in section \ref{sec:data_graduates}. Additional controls include information on local labor markets (employment rate for 25-34 year olds in the province of birth at the time of enrollment) and an index of university quality from Censis, an independent research center, standardized to improve model fit. The battery of fixed effects $\Theta$ includes fixed effects for the year of graduation $\theta^{\text{year}}$, macro-region $\theta^{\text{geo}}$, and years since graduation $\theta^{\text{exper}}$.\footnote{I use the standard definition of macro-regions from the National Statistical Institute (ISTAT): North-East, North-West, Center, South, Islands.} 
The choice set of bachelors' $B$ is described in table \ref{tab:groups} and includes ten aggregated fields of study.
The variables belonging to vector $Z_{ij}$ are the share of binding entry exams for each group of degrees in the public university closest to the student's province of birth, previously described in section \ref{sec:data_programs} and summarized in panel A of table \ref{tab:descr_instruments}. 

\begin{figure}[htbp]
	\caption{Comparison of model and data - choice of bachelor}
	\includegraphics[width=0.7\linewidth]{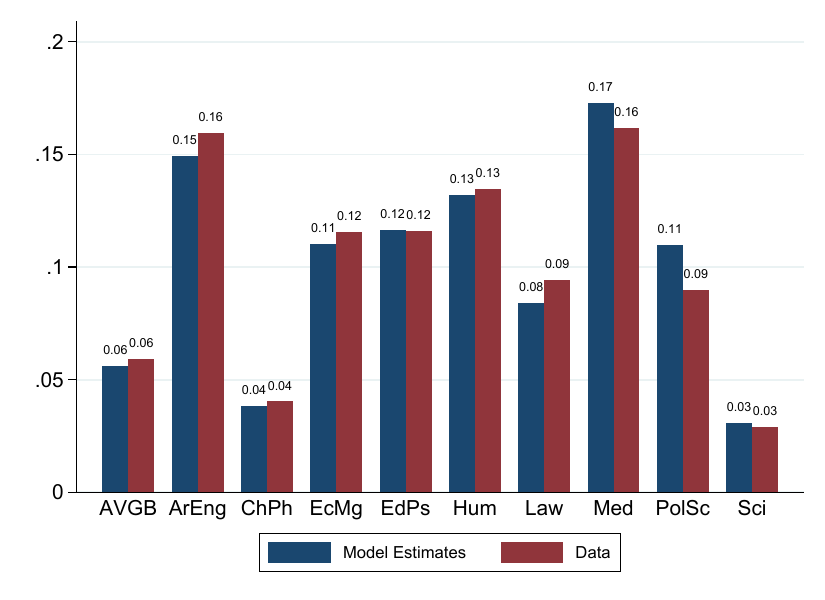}
	\centering
	\label{fig:modelfit_t1}
	\caption*{\footnotesize Model fitted on cohorts 2007-2011, predictions and data plotted for 2012-2014. Description of titles: the title refers to the previous bachelor choice on which the model is fitted. AVGB -- Life Sciences, ArEng -- Architecture and Engineering, ChPh -- Chemistry and Pharmacy, EcMg -- Economics and Management, EdPs -- Education and Psychology, Hum -- Humanities, Literature and Languages, Law -- Law, Med -- Medicine and Health, PolSc -- Political and Social Sciences, Sci -- Math, Physics and Statistics.}
\end{figure}

Table \ref{tab:t1} presents the results for equation \eqref{eq:fw_t1}. The excluded category is the choice of a bachelor's degree in humanities as it has the lowest average value of the instrument on the share of binding entry exams and it is the most geographically widespread. The exclusion restrictions are jointly strongly significant, with $\chi^2(90)=46572.60$.\footnote{Each element of $Z_j$ is also individually strongly significant with $p=0$. AVGB: $\chi^2(9)=3557.03$, Arc.Eng.: $\chi^2(9)=2672.36$, Chem.Pharm.: $\chi^2(9)=9441.17$, Econ. Mgmt.: $\chi^2(9)=3155.64$, Educ.Psy.: $\chi^2(9)=5385.44$, Hum.: $\chi^2(9)=2613.46$, Law: $\chi^2(9)=7722.88$, Health: $\chi^2(9)=6836.68$, Pol.Soc.: $\chi^2(9)=2787.76$, Sci.Stat.: $\chi^2(9)=1857.02$.} Rich substitution patterns clearly emerge. Increasing the share of programs with binding entry exams in law and health increases the probability of enrollment in all degrees compared to the baseline category (humanities). Entry exams in other degrees have more nuanced effects. Interestingly, coefficients $\lambda_j$ are positive along the diagonal for degrees in engineering, education, law, health and political sciences, indicating that decreasing the selectivity of these degrees decreases the relative probability of enrollment. This suggests that positive signaling through selectiveness may be an attribute of these degrees. Table \ref{tab:t1_mem} in appendix \ref{sec:app_mem} additionally presents the marginal effects of coefficients $\lambda$ estimated at the mean of the right-hand variables of equation \ref{eq:fw_t1}. Shifts in the share of degree programs with binding entry exams lead to substantial changes in the probability of enrolling in different degrees, following rich substitution patterns. Just like the coefficients in table \ref{tab:t1}, the marginal effects contained in table \ref{tab:t1_mem} suggest that marginally changing the bindingness of entry exams leads to significant shifts in the probability of enrolling into different degrees at the average values of the sample. Even though on average the net shift of each instrument is close to 0, the variance of the marginal effects is highest for the entry exam variables in literature and languages and health, suggesting that students are particularly reactive to the admission policies of these degrees in their decision to enroll in higher education.
I offer an additional discussion on the magnitude of the effects of the exclusion restrictions in section \ref{sec:simulations}. 
Figure \ref{fig:modelfit_t1} shows how the model fits the data. As the estimator used to fit equation \eqref{eq:fw_t1} is based on maximum likelihood, it matched group averages. To show how accurate the predictions are, I fit the model using cohorts 2007-2011 and present the average data and predictions for cohorts 2012-2014. Indeed, the model seems to match the observed choices on average quite well when I do not require matching on group averages, with differences in enrollment being less that 2 percentage points. The coefficients of equation \eqref{eq:fw_t1} are eventually used to estimate the probability $P_{ij}$ of enrolling in any bachelor's for all individuals.

\begin{table}[htbp]
	\centering
	\setlength{\tabcolsep}{3pt}
	\caption{Period 1 -- Choice of Bachelor}
	\resizebox{\linewidth}{!}{
	\begin{tabular}{lccccccccc}
		\toprule

		VARIABLES & AVGB  & Arc.Eng. & Chem.Ph. & Econ.Mg. & Ed.Psy. & Law   & Health & Pol.Soc. & Sci.Stat. \\
				& (1)   & (2)   & (3)   & (4)   & (5)   & (6)   & (7)   & (8)   & (9) \\
		\midrule
		&       &       &       &       &       &       &       &       &  \\
		\multicolumn{2}{l}{$Z_j$:\textit{ Entry Exams}}       &       &       &       &       &       &       &       &  \\
		AVGB  & -0.527*** & -0.461*** & -0.145** & -1.331*** & 0.807*** & 0.539*** & 0.526*** & 0.300*** & 0.332*** \\
		& (0.053) & (0.038) & (0.063) & (0.039) & (0.041) & (0.045) & (0.040) & (0.041) & (0.064) \\
		Arc.Eng. & 0.320*** & 0.637*** & 1.454*** & -0.831*** & -0.629*** & -0.158*** & 1.428*** & 0.375*** & 1.141*** \\
		& (0.073) & (0.054) & (0.085) & (0.055) & (0.056) & (0.061) & (0.054) & (0.058) & (0.094) \\
		Chem.Ph. & 0.282*** & 0.501*** & -0.737*** & -0.015 & -0.890*** & -0.301*** & -2.508*** & -0.333*** & -1.444*** \\
		& (0.047) & (0.034) & (0.056) & (0.034) & (0.036) & (0.039) & (0.035) & (0.037) & (0.057) \\
		Econ.Mg. & -0.109*** & -0.186*** & -0.677*** & -0.271*** & -0.180*** & -0.308*** & -1.212*** & -0.442*** & -0.143*** \\
		& (0.035) & (0.026) & (0.041) & (0.026) & (0.028) & (0.031) & (0.025) & (0.028) & (0.044) \\
		Ed.Psy. & 0.329*** & 0.924*** & -0.145*** & 0.325*** & 0.887*** & -0.103*** & 1.853*** & 0.020 & 0.848*** \\
		& (0.043) & (0.031) & (0.050) & (0.033) & (0.032) & (0.034) & (0.032) & (0.032) & (0.055) \\
		Law   & 1.848*** & 1.378*** & 1.043*** & 1.244*** & 0.813*** & 1.244*** & 1.873*** & 0.458*** & 0.736*** \\
		& (0.059) & (0.044) & (0.066) & (0.046) & (0.047) & (0.050) & (0.044) & (0.047) & (0.071) \\
		Hum   & -4.569*** & -0.499*** & -2.914*** & 0.580*** & -3.077*** & -2.332*** & -3.874*** & -2.064*** & -1.658*** \\
		& (0.096) & (0.064) & (0.103) & (0.064) & (0.068) & (0.073) & (0.071) & (0.068) & (0.100) \\
		Health & 6.876*** & 2.693*** & 7.326*** & 4.235*** & 3.999*** & 2.987*** & 6.855*** & 1.795*** & 4.261*** \\
		& (0.138) & (0.102) & (0.163) & (0.103) & (0.105) & (0.113) & (0.101) & (0.108) & (0.175) \\
		Pol.Soc. & -1.297*** & -2.262*** & 0.893*** & 0.130* & -0.888*** & -0.188** & 0.967*** & 0.674*** & 0.373*** \\
		& (0.101) & (0.073) & (0.113) & (0.073) & (0.078) & (0.085) & (0.073) & (0.079) & (0.121) \\
		Sci.Stat. & 0.864*** & -0.060 & 0.543*** & -0.256*** & 1.245*** & 0.599*** & -1.171*** & 0.436*** & 0.045 \\
		& (0.085) & (0.062) & (0.096) & (0.062) & (0.064) & (0.072) & (0.060) & (0.066) & (0.102) \\
		&       &       &       &       &       &       &       &       &  \\
		$X$ & 	\multicolumn{9}{c}{Yes} \\
		FE & 	\multicolumn{9}{c}{Yes} \\
		&       &       &       &       &       &       &       &       &  \\
		Observations & 655,847 & 655,847 & 655,847 & 655,847 & 655,847 & 655,847 & 655,847 & 655,847 & 655,847 \\
		\midrule
		\multicolumn{10}{l}{Standard errors in parentheses. *** p<0.01, ** p<0.05, * p<0.1. Pseudo $R^2=0.103$.} \\
	\end{tabular}%
	}
	\caption*{\footnotesize Excluded category: humanities. Joint test of exclusion restrictions $Z_j$: $\chi^2(90)=46572.60$, p-value=0. $X$: gender, high school grade, high school type, parent occupation, parent education, local labor market, and university quality controls. $\Theta$: Macro-region, experience and year fixed effects. }
	\label{tab:t1}%
\end{table}%

Estimating the probability of enrolling in a master's degree is slightly more cumbersome as it is conditional on the choice of bachelor's degree. I estimate ten separate multinomial logit models (equations \ref{eq:fw_t2}) on the subsample of students in each bachelor's.\footnote{Only students who are not enrolled in a single cycle degree are used to fit the model as they have to make a choice. The prediction uses the whole sample. This should not matter as the offer of single cycle degrees is plausibly exogenous to the choice and to labor market outcomes.} I then predict the probability of choosing any master's for all conditional choices of bachelor's $P_{im}\mid j\,\,\forall\,\, j\in B,\, i\in I$.

While the possible fields of study coincide between bachelor and master, the set of choices of master's $M$ is different from $B$ as it also includes the possibility of no master at all, that is, entering directly the labor market after the bachelor's. $X$ and the fixed effects are defined as before and only vary at the individual level.\footnote{Fixed effects for years since graduation are omitted due to collinearity with other covariates or lack of variation in certain subsamples.} The omitted category is always the choice of not pursuing a master's. The choice-theoretic characterization is that not pursuing a master's is equivalent to a lack of treatment conditional on the choice of bachelor's, thus always at least the second best option. Furthermore, the option is always available.
$Z_{im\mid j}$ is a rich set of exclusion restrictions that regulate access to the master's program and vary with the previous choice of bachelor's. It includes the standardized credit requirements for enrollment into each master's that vary at the individual and program level described in table \ref{tab:descr_instruments} panel B, and log distance to the closest public university. Not all degree combinations can be estimated since some are not observed in the data (table \ref{tab:frequency_careers} summarizes the available groups). Hence, only the credit requirements relevant to the possible choices are included.

\begin{figure}[ht]
	\caption{Comparison of model and data - choice of master}
	\includegraphics[width=\linewidth]{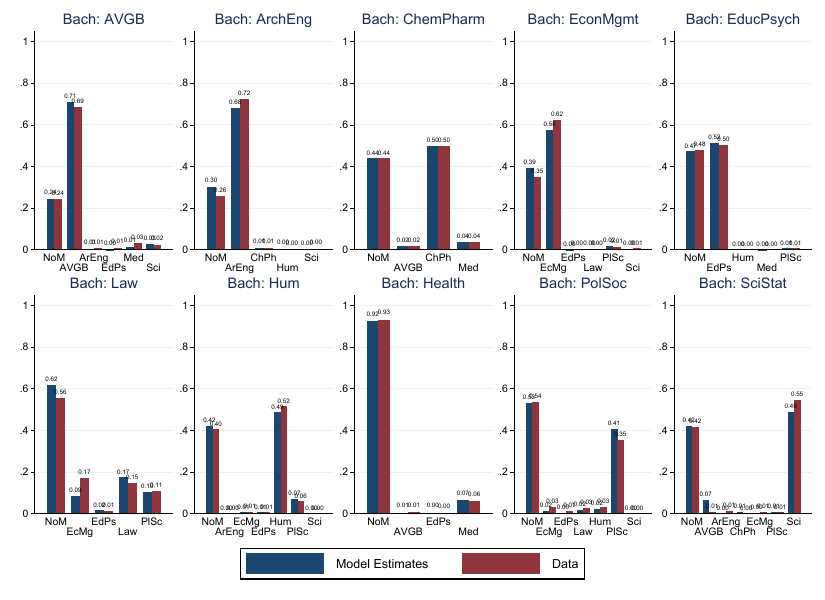}
	\centering
	\label{fig:modelfit_t2}
	\caption*{\footnotesize Model fitted on cohorts 2007-2011, predictions and data presented for cohorts 2012-2014. Students who enroll in single-cycle degrees (e.g. architecture, medicine, law) are not considered here as they do not make a schooling choice. The title of each histogram refers to the previous bachelor choice on which the model is fitted. Description of labels: AVGB -- Agriculture, Veterinary, Geology, Biology; ArEn -- Architecture and Engineering; ChPh -- Chemistry and Pharmacy; EcMg -- Economics and Management; EdPs -- P.E., Teaching and Psychology; Law -- Law; Hum -- Literature and Languages; Med -- Health; PlSc -- Political and Social Sciences; Sci -- Science and Statistics; NoM -- No Master. }
\end{figure}

Tables \ref{tab:t2_1} to \ref{tab:t2_10} in appendix \ref{sec:appendix_t2} present the results of these estimations. In all cases, the baseline category is to not enroll in a master's degree. Some exclusion restrictions on credit requirements may be dropped for collinearity or lack of variation within certain subgroups. For instance, this may occur if all students with the same bachelor face the same credit requirements for a given master's. Joint tests of the exclusion restrictions are presented in table \ref{tab:t2_exclusionrestrictions} and indicate that the exclusion restrictions are valid within each conditional choice of bachelor's. Again, rich substitution patterns emerge. In all cases except one, increasing the credit requirement in the master's with the same discipline as the bachelor's decreases the probability of enrolling in that master's. Positive coefficients indicate that the probability of enrollment increases with increases in the credit requirement with respect to the choice of not enrolling in a master's. This suggests that for certain degree combinations, the probability of enrollment increases with the additional (relative) work that the student must do. Students with graduate parents are more likely to enroll in a master's degree, with very few exceptions. Gender does not seem to systematically generate sorting into more (less) quantitative fields during the master, even though it does increase the probability of enrolling in masters' in education and psychology.\footnote{Marginal effects for the exclusion restriction variables, estimated at the means of the sample are available upon request.} The model fit is presented in figure \ref{fig:modelfit_t2} by comparing average predicted probabilities and observed enrollment. As before, equations \eqref{eq:fw_t2} are estimated on cohorts 2007-2012 and the comparison between data and estimates is presented for years 2012-2014; the model seems to predict the conditional probability of enrolling in a master well. \\


\begin{table}[htbp] 
	\centering
	\caption{Test of exclusion restrictions for equations \eqref{eq:fw_t2}}
	\begin{tabular}{lcccccc}
		\toprule
		& \multicolumn{2}{c}{All $Z_m$} & \multicolumn{2}{c}{Credit Requirements} &       &  \\
		\cmidrule(lr){2-3} \cmidrule(lr){4-5}
		Conditional Choice of Bachelor & D.f. & $\chi^2$ & D.f. & $\chi^2$ & Observations & Table \\
		\midrule
    Agr.Vet.Geo.Bio. & 25    & 3725.9 & 20    & 3691.69 & 32,494 &  \ref{sec:appendix_t2}.\ref{tab:t2_1}\\
    Architecture and Engineering & 20    & 6572.69 & 16    & 6570.72 & 79,817 & \ref{sec:appendix_t2}.\ref{tab:t2_2} \\
    Chemistry and Pharmacy & 6     & 277.14 & 3     & 273.37 & 7,398 & \ref{sec:appendix_t2}.\ref{tab:t2_3} \\
    Economics and Management & 10    & 14011.08 & 5     & 13977.24 & 75,993 & \ref{sec:appendix_t2}.\ref{tab:t2_4} \\
    P.E., Teaching and Psychology & 12    & 10142.09 & 8     & 10106.78 & 62,741 & \ref{sec:appendix_t2}.\ref{tab:t2_5} \\
    Law   & 8     & 1089.64 & 4     & 1076.1 & 10,882 & \ref{sec:appendix_t2}.\ref{tab:t2_6} \\
    Literature and Languages & 30    & 3083.15 & 24    & 3048.16 & 90,681 & \ref{sec:appendix_t2}.\ref{tab:t2_7} \\
    Healthcare and Medicine & 6     & 861.1 & 3     & 855.77 & 81,883 & \ref{sec:appendix_t2}.\ref{tab:t2_8} \\
    Political and Social Sciences & 36    & 7988.74 & 30    & 7974.45 & 65,798 & \ref{sec:appendix_t2}.\ref{tab:t2_9} \\
    Science and Statistics & 18    & 1045.78 & 12    & 1040.46 & 20,721 & \ref{sec:appendix_t2}.\ref{tab:t2_10} \\
		\bottomrule
	\end{tabular}%
	\label{tab:t2_exclusionrestrictions}%
	\caption*{\footnotesize Joint test of all exclusion restrictions for each conditional bachelor choice, d.f. denotes degrees of freedom. All reported $\chi^2$ have p-values equal to 0. $Z_m$ includes bachelor final grade (standardized), credit requirement (standardized) and distance to closest public university. Students who previously enrolled in single-cycle degrees are not used for inference.
	}
\end{table}%

Lastly, I estimate the probability of enrolling in any combination of degrees $P_{ijm}=P_{ij}\times [P_{m_i}\mid j]$ for all $i\in I$, $j\in B$ and $m\in M$. For the special case of students who end up in single-cycle degrees, $P_{ijm}=P_{ij}$ if $j=m$. 
I am left with the choice probabilities for 56 combinations of degrees.\footnote{In practice, I can only retrieve 43 returns to combinations of degrees ex post. The rationale is explained in sections \ref{sec:t3_results} and \ref{sec:courses}. A priori, all the data from 56 combinations of degrees is used.} On average, probabilities $P_{jm}$ match observed treatments $D_{jm}$. Their difference across all degree combinations is $7.14\times 10^{-9}$. Importantly, since $P_{jm}$ is the product of two probabilities, the observed maximum values are strictly lower than 1, ranging from 0.012 for (Econ.Mgmt, Educ.Psy.) to 0.748 for (Healthcare, No Master), with degree combinations chosen less frequently presenting lower ranges of probabilities of enrollment. 
Additional summary statistics for the treatments $D_{jm}$ and probabilities $P_{jm}$ can be found in table \ref{tab:diff_D_P} in appendix \ref{sec:appendix_t2}.


\subsubsection{Exclusion Restrictions \texorpdfstring{$Z_{ij}$}{Zj} and Simulations}\label{sec:simulations}

I present two policy simulations that investigate different admission policies in the bachelor's to elicit how sorting at the margin responds to shifts in entry restrictions. The focus will be on entry into bachelor degrees as it leads to remarkable shifts in the student body composition. Using the choice model set up in section \ref{sec:theory}, I shift the values of $Z_j$ in equation \eqref{eq:fw_t1} to understand how students react to entry exams. Figure \ref{fig:modelfit_t1} has previously justified the appropriateness of the model to predict the distribution of students across degrees. As the available data is not appropriate to understand the labor market outcomes of individuals who did not attend college, I am unable to assess the inbound shift that might occur if admission policies were to change substantially. For these reasons, these simulations should be interpreted as shifts in enrollment at the intensive margin. 

\begin{figure}[htbp]
	\caption{Period 1: Policy Simulations on Entry Exams}
	\includegraphics[width=\linewidth]{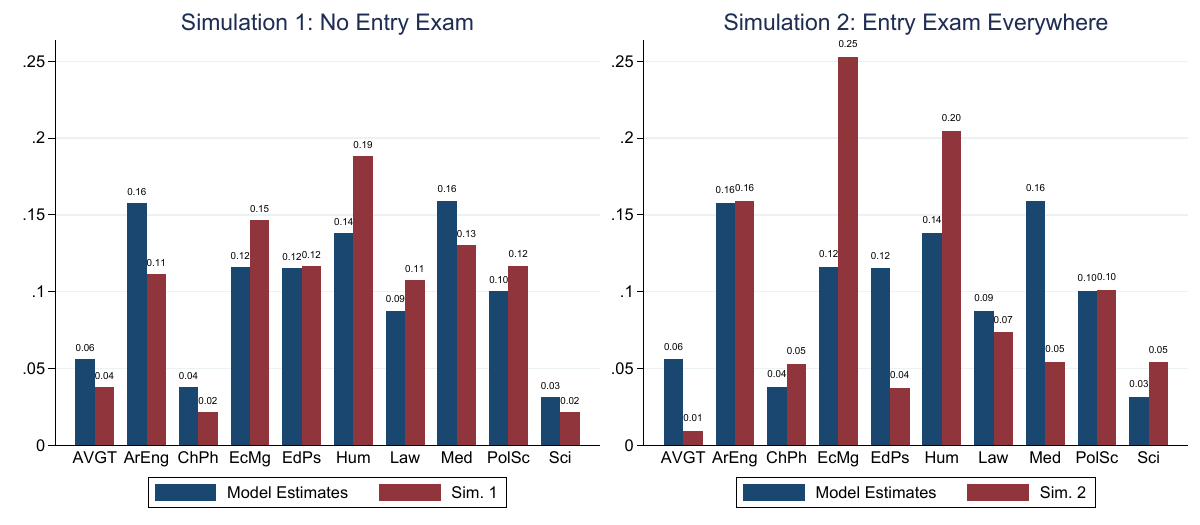}
	\centering
	\label{fig:simulations_b}
\end{figure}

In the first simulation, all variables in $Z_j$ are set to their minimum and new probabilities of enrollment in each degree are estimated using equation \eqref{eq:fw_t1}.\footnote{Values of $Z_j$ are set to their observed minimum rather than 0 for all degrees because certain degrees such as healthcare have minimum values which are very high (78\%), otherwise resulting in out-of-sample predictions.} The global effect of this policy is shown in the left panel of figure \ref{fig:simulations_b} and suggests that relaxing entry barriers would increase enrollment in economics and management, humanities (literature and languages), law, and political and social sciences, while decreasing enrollment in all the other degrees. This may be rationalized by considering that enrollment in the former is bound by entry exams, while demand for the latter may not be determined by it. This means that if there were fewer entry exams, enrollment would increase by 35.7\% in humanities (5 p.p.) and 20\% (3 p.p.) in economics and management. The largest decrease would occur in engineering, with a 31\% decrease in enrollment (5 p.p.). 
Figure \ref{fig:sim1_decomp_alldemo} in the appendix presents the results of this simulation decomposed across several individual characteristics: gender, parental occupation, education, and high school grades. While sorting into degrees varies along these dimensions, reducing entry barriers does not produce additional patterns.


An alternative simulation where entry exams are imposed everywhere is presented in the right panel of figure \ref{fig:simulations_b}. Here, all variables in $Z_j$ are set to 1 (i.e., all bachelor's programs have binding admission requirements) and new probabilities of enrollment in each degree are estimated using equation \eqref{eq:fw_t1}. Once again, enrollment in economics and humanities increases, as well as enrollment in chemistry and science. The comparison of the two simulations in figure \ref{fig:simulations_b} showcases the nonlinear substitution patterns that are possible due to the rich set of information on selective entry admissions $Z_{j}$. 

Simulation 1 in figure \ref{fig:simulations_b} suggests that the existing entry exams mostly serve the purpose of managing excess demand into less quantitative fields such as economics or humanities. In fact, if students have lower preferences for quantitative studies even after controlling for rich individual characteristics \citep{rask2010attrition, mann2013, fricke2018}, it is not surprising that removing entry barriers does not increase enrollment into such degrees. On the other hand, simulation 2 indicates the degrees where selectiveness at the margin is positively related to enrollment. One interpretation of these results is that students derive a net benefit at the margin of increasing selectiveness in economics, humanities, chemistry, and science. I rationalize the decrease in enrollment in medicine in simulation 2 by noting that entry exams are so ubiquitously present that the signal of selectiveness is saturated at the margin. Jointly, these simulations illustrate the richness of the substitution patterns allowed by the model and suggest that settings where admission requirements are assumed to relate monotonically with preferences on enrollment do not fit real world situations.\footnote{Importantly, the assumption that the instrument $P_{ijm}$ monotonically increases the take up of the treatment $D_{ijm}$ stands.}
Both of the proposed policies (elimination and imposition of binding entry exams) will reasonably induce reactions at the extensive margin as well as the intensive margin. Since individuals with no college are not observed, these results should not be interpreted as informative of global shifts in enrollment. However, they underline that when faced with multiple choices, several contrasting margins matter for sorting.
In both cases, varying the values of the exclusion restrictions induces substantial shifts in enrollment across degrees. This suggests that one of the necessary conditions for identification in the reduced form presented in section \ref{sec:theory} -- that the exclusion restrictions be strongly relevant -- is satisfied.

\let\saveFloatBarrier\FloatBarrier
\let\FloatBarrier\relax
\subsection{Returns to university careers}\label{sec:t3_results}
\let\FloatBarrier\saveFloatBarrier

The probabilities $P_{ijm}$ estimated in the previous section enter the reduced form equation \eqref{eq:fw_t3_reducedform} which is estimated with the previously described vector $x$ and fixed effects, where the labor market outcomes of interest are log wages and employment, and $jm$ only refers to combinations that are observed in the data. 

To ensure that the coefficients $\alpha_{jm}$ can be interpreted as causal effects, I choose the combination of degrees (Lit.Lang., No Master) as the excluded category to proxy lack of treatment. Undergraduate degrees in humanities exhibit the lowest levels of binding entry exams and are available in 54 out of 67 public universities. Combined with "No Master", this university career serves as the most credible benchmark.

The results for the vector of coefficients $\beta$ are presented in table \ref{tab:t3_ols_reducedform}.\footnote{Table \ref{tab:diff_empl_unempl} in section \ref{sec:appendix_descriptives} reports the differences in observed characteristics $X$ between the sample of employed and unemployed to assist the interpretation of the results on log wages conditional on employment.} All of the equations' standard errors are bootstrapped using full iterations of the entire model to account for the probabilities being predicted (equations \eqref{eq:fw_t1}-\eqref{eq:fw_t3_reducedform}). For comparison, I also present OLS results where treatments $D_{jm}$ substitute probabilities $P_{jm}$, thus not controlling for self-selection (equation \eqref{eq:fw_t3}). 

The reduced form coefficients in columns (2) and (4) of table \ref{tab:t3_ols_reducedform} follow the sign and significance level of the OLS coefficients (columns 1 and 3) for almost all the main explanatory variables, where the magnitude of the effects increases. This is likely driven by the correction for endogeneity in the observed choices of university careers. Higher grades are strongly positively related to higher chances of being employed, whereby they do not improve wages (conditional on employment). Similarly, having a science high school degree improves outcomes in terms of employment, but not wages conditional on working. Surprisingly, once we control for university careers, women are more likely to be employed than men, even though they experience lower wages. This is likely due to selection on gender into different university careers.

Coefficients $\alpha$ cannot be interpreted as causal treatment effects without taking into account that the probabilities $P_{jm}$ vary along a scale that is strictly smaller than one, as discussed in section \ref{sec:estimation_P}. By rescaling the coefficient by the maximum observed probability of choosing a given career $(j, m)$, the effect becomes
\begin{equation}\label{eq:TE}
	\tilde{\alpha}_{jm}=\alpha_{jm}\cdot\max_I(P_{jm})
\end{equation}
which can be interpreted as a shift in labor market outcomes induced by an increase in the probability of choosing said career from 0 to the sample's maximum, ceteris paribus.\footnote{In this setting, the causal effect of university careers $(j, m)$ is driven by several potentially small subsamples which may display different observed characteristics, both in $X$ and in covariate patterns of $P_{jm}$. Hence, when treatment effects are abnormally large (or small), it is difficult to distinguish between non-credible estimates which are not estimated precisely and credible estimates with large magnitudes due to strong self-selection. I introduce a regulating criterion to rule out treatment effects with excessive magnitudes. For employment, I ensure that all treatment effects, summed with the average predicted probability of the baseline are constrained between 0 and 1. I obtain the boundaries $\tilde{\alpha}(\text{empl})\in[-0.62, 0.38]$ and disregard treatment effects that exhibit larger magnitudes. For log(wages), I compare the treatment effect obtained in \eqref{eq:TE} with the maximum (minimum) deviations from the baseline predicted in the sample. Similarly, I disregard treatment effects beyond boundaries $\tilde{\alpha}(\text{ln(wage)})\in[-1.04, 2.27]$, in levels, this corresponds to monthly salaries between 187 and 7186 Euros. I further correct the out of sample estimated treatment effects by weighting them by the 95\% percentile of $P_{jm}$ and drop the rest.}
 In the end, I obtain 43 credible TEs for both log wages and employment. 

\begin{table}[ht]
	\centering
	\caption{$\beta$ coefficients for labor market outcomes.}
	\begin{tabular}{lcccc}
		\toprule
		& \multicolumn{2}{c}{log(wage)| employed} & \multicolumn{2}{c}{employment} \\
		\cmidrule(lr){2-3} \cmidrule(lr){4-5}
		VARIABLES &   OLS     & Red. Form     &   OLS    & Red. Form \\
		& (1)   & (2)   & (3)   & (4) \\
		\midrule
		&       &       &       &  \\
		$X$ &       &       &       &  \\
		High School: grade (st.) & -0.018*** & -0.937*** & 0.004*** & 3.670*** \\
		& (0.001) & (0.178) & (0.001) & (0.210) \\
		High School: humanities & -0.079*** & -1.067*** & -0.032*** & -0.128 \\
		& (0.003) & (0.376) & (0.002) & (0.197) \\
		High School: science & -0.048*** & -2.744*** & -0.020*** & 15.052*** \\
		& (0.002) & (0.647) & (0.001) & (0.888) \\
		Gender (1=female) & -0.154*** & -1.956*** & 0.009*** & 3.257*** \\
		& (0.003) & (0.644) & (0.001) & (0.373) \\
		Parents: graduate & -0.042*** & -1.016*** & -0.027*** & 4.431*** \\
		& (0.003) & (0.184) & (0.001) & (0.285) \\
		Parents: high-ranked occup. & 0.004 & -0.072 & 0.002 & 1.584*** \\
		& (0.003) & (0.196) & (0.001) & (0.108) \\
		&       &       &       &  \\
		Additional controls & Yes   & Yes   & Yes   & Yes \\
		FE & Yes   & Yes   & Yes   & Yes \\
		$D_{jm}$ & Yes   &       & Yes   &  \\
		$P_{jm}$ &       & Yes   &       & Yes \\
		&       &       &       &  \\
		Observations & 508,242 & 508,242 & 655,847 & 655,847 \\
		R-squared & 0.101 &       & 0.125 &  \\
		Mean $y$ & 6.887 & 6.887 & 0.775 & 0.775 \\
		\bottomrule
	\end{tabular}%
	\label{tab:t3_ols_reducedform}%
	\caption*{\footnotesize Reduced form results from equation \eqref{eq:fw_t3_reducedform}, OLS results from equation \eqref{eq:fw_t3}. Columns (2) and (4) feature bootstrapped standard errors with 104 iterations. Additional controls for local labor markets and university quality.}
\end{table}%

\begin{figure}[ht]
	\caption{Comparison the distributions of OLS coefficients $\gamma_{jm}$ and reduced form treatment effects $\alpha_{jm}$ }
	\includegraphics[width=\linewidth]{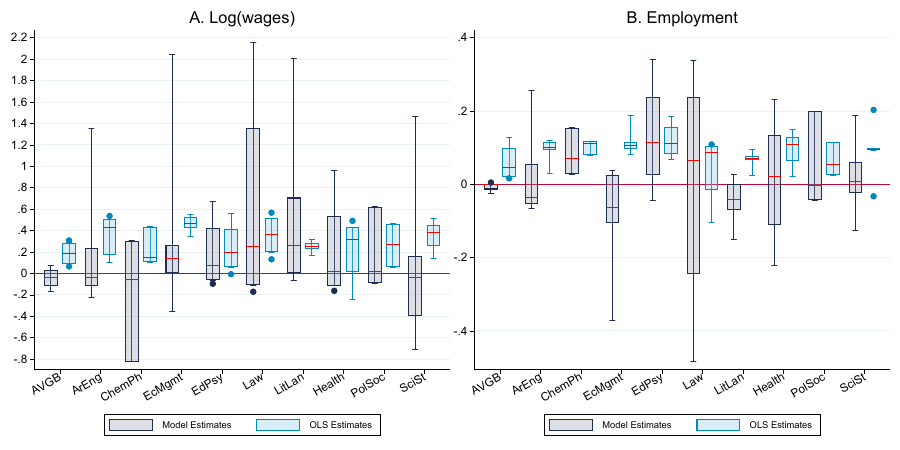}
	\centering
	\label{fig:t3_te_box}
	\vspace{-.5cm}
	\caption*{\footnotesize Generalized box plots for the distribution of returns by bachelor's. Dark blue markers denote reduced form (RF) coefficients $\alpha_{jm}$ \eqref{eq:fw_t3_reducedform}, light blue markers denote OLS coefficients $\gamma_{jm}$ \eqref{eq:fw_t3}. Red markers denote medians. The baseline is (Lit.Lang., No Master). }
\end{figure}

Figure \ref{fig:t3_te_box} compares the distributions of treatment effects $\alpha_{jm}$ and OLS coefficients $\gamma_{jm}$ for university careers and both labor market outcomes and emphasizes three main findings. Notably, this comparison makes use of the strong assumptions discussed in section \ref{sec:theory} that justify the IV-equivalence result of equation \eqref{eq:fw_t3_firststage}.
OLS and reduced form results are statistically different in 84\% of cases for log(wages) and in 64\% of cases for employment, such that any method that does not account for self-selection into university careers is highly misleading (to compare the returns one-to-one, refer to figure \ref{fig:t3_te_compared}). 
Secondly, substantial variation is present when we compare the effect of university careers with the same undergraduate choice, which underscores the importance of accounting for advanced degrees in the discussion on returns to higher education. For example, log wage returns to undergraduate programs in chemistry and pharmacy vary greatly depending on the advanced degree. By plotting the distribution of the labor market returns by undergraduate choice, it is apparent that in almost all instances, the interquartile range of the conditional distribution spans positive and negative values with respect to the excluded category. 
Thirdly, OLS estimates more positive effects for 29 out of 43 log wage returns and 33 out of 43 returns to employment. This suggests that students self-select into degrees based on comparative advantage. Under the OLS equivalence assumptions, OLS coefficients overestimate on average the returns to university careers by 7.2pp (employment) and 0.26 log points (log wages). It also emphasizes the validity of exclusion restrictions $Z_j$ and $Z_m$ to partial out individual sorting. Another interpretation of these effects is thus the average returns to degree combinations enjoyed by individuals if they were randomly allocated to them. With this interpretation, it is perhaps not surprising that the average return to a career in engineering (Arch.Eng., Arch.Eng.) shifts from strictly positive when not accounting for self-selection to slightly negative when I do.



\let\saveFloatBarrier\FloatBarrier
\let\FloatBarrier\relax
\section{Results on Academic Curricula}\label{sec:courses}
\let\FloatBarrier\saveFloatBarrier

Here I exploit the information on academic curricula to shed light on outcome-enhancing characteristics of university careers. I focus on how the composition of the curriculum affects returns with interest in market responses to multidisciplinary careers, quantitative courses, and the timing of degrees and courses. 
To facilitate the understanding of the results, I refer to careers with $j=m$ as \textit{specialized careers}, such as (Econ.Mgmt., Econ.Mgmt.), careers with $j\neq m$ and $m\neq 0$ as \textit{multidisciplinary careers}, for example (Econ.Mgmt., Sci.Stat.), and careers with $m=0$ as \textit{no master careers}, for example (Econ.Mgmt., No Master).

\subsection{Academic Curricula and Degree Composition}\label{sec:degreecomposition}

	

\begin{sidewaysfigure}
\caption{Comparison of log wage and employment returns for all careers}
\includegraphics[width=\linewidth]{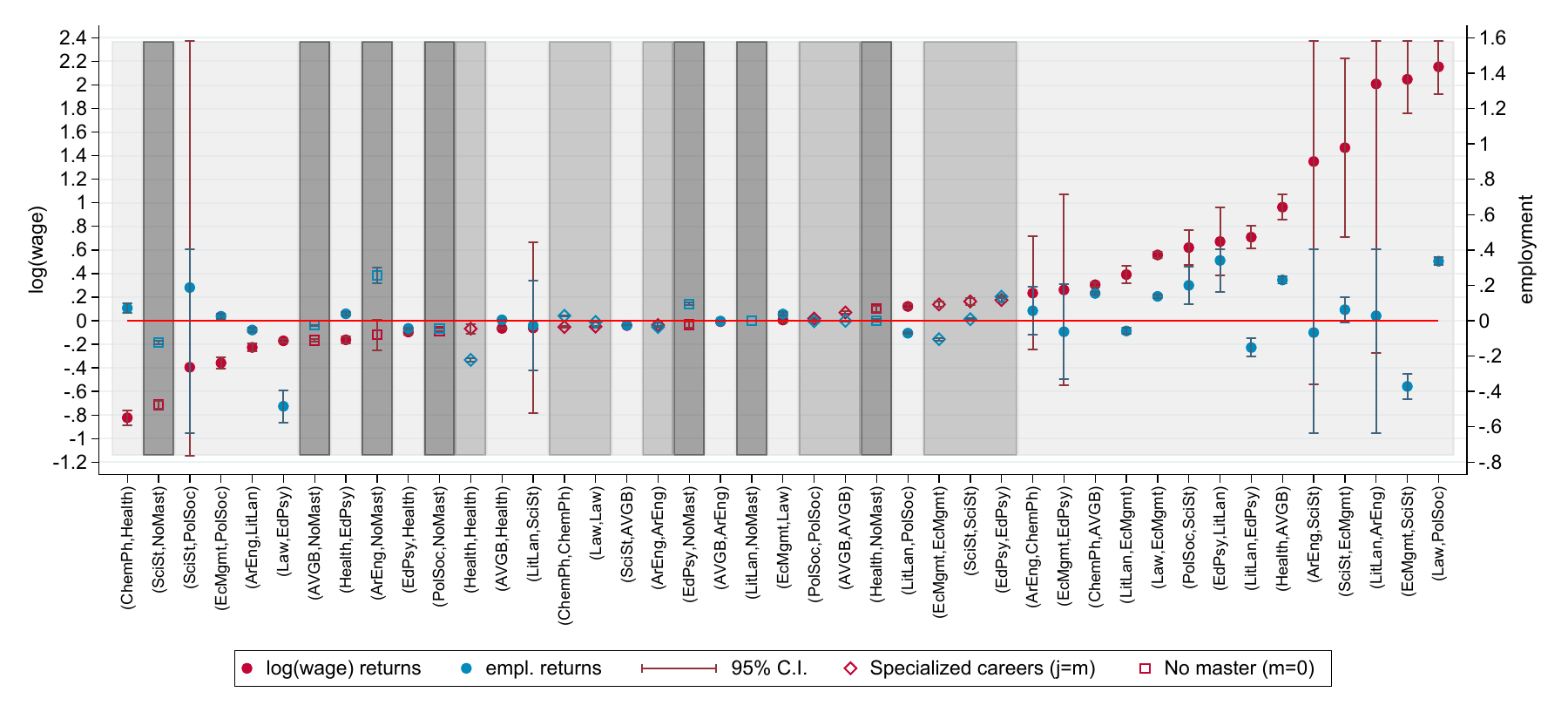}
\caption*{\footnotesize Comparison of log wage returns (red, left vertical axis) and returns to employment (blue, right vertical axis). Axes are centered around 0. Full-circle symbol markers indicate multidisciplinary careers $j\neq m$ (light gray shading), hollow diamond symbols indicate specialized careers $j=m$ (mid-gray shading), and hollow square symbols indicate careers with no master $m=0$ (dark gray shading). Careers are ordered by increasing returns to log wages. Only careers for which both labor market outcomes were estimated are displayed.}
\label{fig:TE_compared_all_v2}
\end{sidewaysfigure}

Figure \ref{fig:TE_compared_all_v2} directly compares the estimated returns to log wages and employment and orders careers by increasing returns to log wages. 
Both outcomes are significantly and positively correlated once we account for the precision of the estimates ($\rho(\tilde{\alpha}^{\text{lnwage}}, \tilde{\alpha}^{\text{empl}})=0.37$, $p=0.015$), although the relationship does not hold at the tails of the distribution of log wage returns. Especially for very high log wage returns, there seems to be a trade-off between higher pecuniary outcomes and a lower probability of employment. In the extreme case of (Ec.Mgmt., Sci.Stat.), the estimated return to log wages is 2.05 (average monthly wage of 5 784 Euros), however, the return to employment is extremely low (-0.37), resulting in a probability of employment of 24.5\%. The career with the best overall outcome is (Law, Pol.Soc.), with an estimated log wage return of 2.15 (6 393 Euros) and return to employment of 0.35 (0.95 probability of employment). More generally, only 7 of the 10 careers with the highest log wage returns display positive returns to employment with respect to the excluded career (Lit.Lang., No Master). On the opposite end of the distribution, the worst overall labor market returns are associated with career (Sci.St., No Master) which features a log wage return of -0.71  (366 Euros) and return to employment of -0.12 (0.49 probability of employment).\footnote{In terms of employment, the worst performing university career is (Law, Ed.Psy.), with an employment coefficient of -0.48 (13.5 probability of employment on average) and a -0.17 log wage coefficient (628 Euros).}
These results might be partially driven by different timelines that affect entry into the profession. The pathway to employment might be more complicated for individuals with peculiar university careers, for example, because of additional requirements regarding certification, training, or difficulty in building a client base. Certain careers require long apprenticeship periods after graduation (teachers, lawyers, doctors). In other instances, differences between wages and employment may reflect the riskiness of the career, whereby few individuals reap substantial benefits (creative careers, policy). Similarly, low-earning careers with relatively high levels of employment might reflect lower riskiness of the career, which is often the case for careers with no master.\footnote{The magnitude of the estimates is obtained by comparing the returns to the predicted outcomes for the excluded career (Lit.Lang., No Master) at sample averages of the observed characteristics. Value in levels (Euros) of log wage return $\tilde{\alpha}_{jm}^{\text{lnwage}}$ is $\exp(6.614+\tilde{\alpha}_{jm}^{\text{lnwage}})$, probability of employment for return to employment $\tilde{\alpha}_{jm}^{\text{empl}}$ is $0.615+\tilde{\alpha}_{jm}^{\text{empl}}$.} These low-earning careers also exhibit differences in the sign of the two labor market returns, with 5 out of the 10 lowest earning careers displaying positive returns to employment with respect to the excluded career. Figure \ref{fig:TE_compared1} in the appendix concentrates on careers with no master's and specialized careers that mostly populate the central part of figure \ref{fig:TE_compared_all_v2}. By considering the returns as a whole, I note that the returns to combinations with no master are ranked towards the bottom of the distribution of wage returns (dark gray shading), suggesting that in most instances there is a premium to having a master's degree. Specialized careers are bunched towards the middle of the distribution in mid-gray shading with sensible rankings (science ranks better than economics which ranks better than law), while the top of the distribution is exclusively populated by multidisciplinary careers (light gray shading).\footnote{Indeed, some specialized careers result in surprising results: the best-ranking specialized career is education and psychology, while the worst one is healthcare. The ones reported as sensible rankings remain stable throughout versions of this paper, while the ones cited in this note change (previous versions of this paper are available upon request). Furthermore, healthcare requires extensive training after the degree such that potential long term returns are not captured in this framework, and overall the returns to specialized careers are close to each other in magnitude, leading to variations in rankings even without substantial changes in the estimated.} 
Out of 43 estimated returns to careers, the 14 highest log wage returns are all multidisciplinary careers with $j\neq m$ and $m\neq 0$ (top third of the distribution), while the 10 lowest log wage returns are associated with no master careers in 3 cases and multidisciplinary careers in the other 7. 
These findings suggest that enrolling in a multidisciplinary career can substantially boost labor market outcomes if chosen well. Even though career (Econ.Mgmt., Econ.Mgmt.) yields the third highest log wage returns among specialized careers ($\tilde{\alpha}_{\text{(EcMg,EcMg)}}^{\text{lnwage}}=0.14$), returns can be up to fourteen times higher if combined with other degrees such as (Econ.Mgmt., Educ.Psy.), (Law, Econ.Mgmt.), or (Econ.Mgmt., Sci.Stat.), yielding $\tilde{\alpha}_{\text{(EcMg,EdPs)}}^{\text{lnwage}}=0.26$, $\tilde{\alpha}_{\text{(Law,EcMg)}}^{\text{lnwage}}=0.56$, $\tilde{\alpha}_{\text{(EcMg,Sci)}}^{\text{lnwage}}=2.05$, respectively. At the same time, multidisciplinarity can lead to drastically lower returns. For example, log wage returns to (Econ.Mgmt., Pol.Soc.) are equal to -0.36, or 1.4 times lower than the specialized career. 

While figure \ref{fig:TE_compared_all_v2} highlights the importance of the joint choice of bachelor's and master's beyond undergraduate majors, it does not reveal which characteristics of the careers are informative about outcomes. I investigate the composition of the curriculum of the best- and worst-performing careers to elicit any patterns in the type of knowledge that is covered. In order to avoid considerations on the trade off between employment and wages, I focus on the five best-performing careers -- compared to the benchmark -- which display the highest log wage returns as well as positive returns to employment. Similarly, the five worst-performing careers are selected such that they display negative returns to both outcomes.

\begin{figure}[htbp]
	\centering
	\caption{Comparison of academic curricula and log wage returns for selected careers}
	\includegraphics[width=\linewidth]{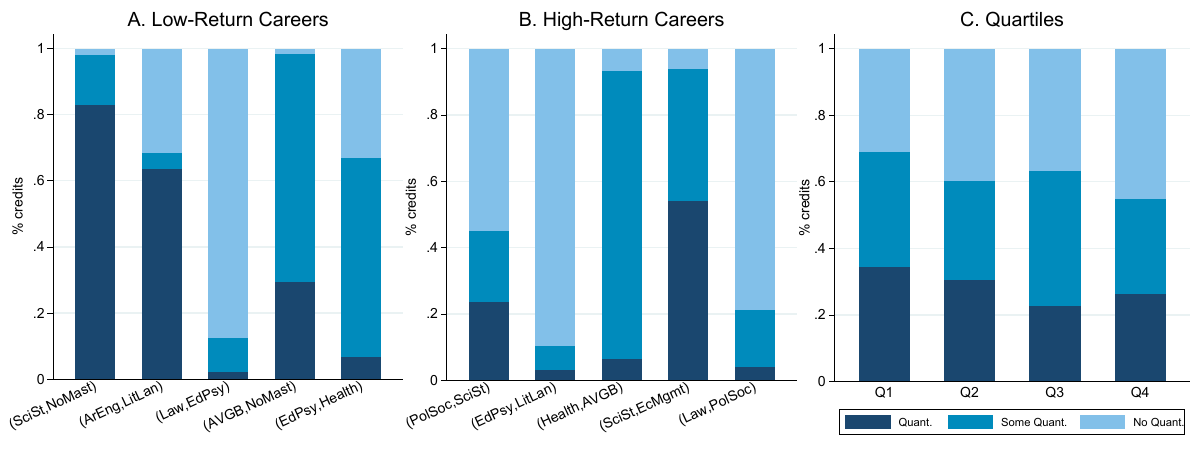}
	\caption*{\footnotesize Quantitative courses (dark blue): science and statistics, architecture and engineering, and chemistry and pharmacy. Some quantitative (technical) courses (blue): life sciences (agriculture, veterinary, geology and biology), economics and management, and healthcare. Non-quantitative courses (light blue): education and psychology, law, humanities (literature and languages), and political and social sciences. The total percentage of credits in each grouping is plotted on the vertical axis. The order of degrees follows the ranking of log-wage returns, increasing from left to right within each panel.}
	\label{fig:credit_composition}
	
\end{figure}

\begin{figure}[htbp]
	\centering
	\caption{Comparison of academic curricula along quartiles of the distribution of employment}
	\includegraphics[width=.8\linewidth]{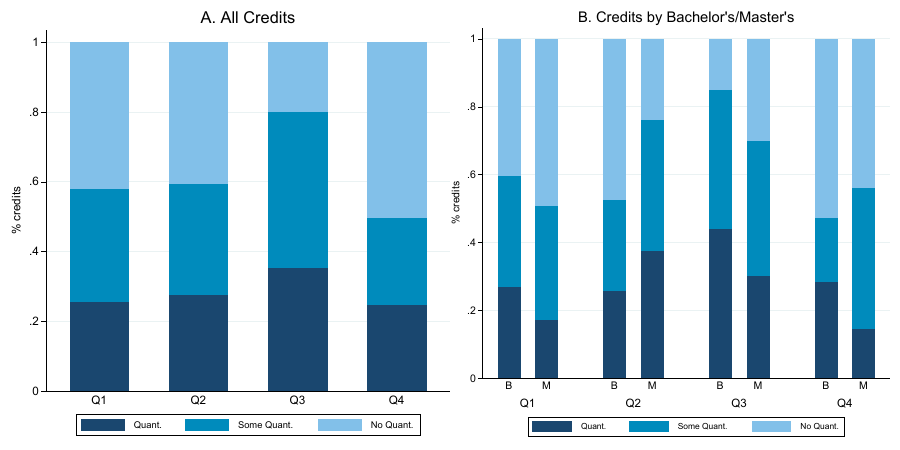}
	\caption*{\footnotesize Quantitative courses (dark blue): science and statistics, architecture and engineering, and chemistry and pharmacy. Some quantitative (technical) courses (blue): life sciences (agriculture, veterinary, geology and biology), economics and management, and healthcare. Non-quantitative courses (light blue): education and psychology, law, humanities (literature and languages), and political and social sciences. The total percentage of credits in each grouping is plotted on the vertical axis. The order of degrees follows the ranking of returns to employment, increasing from left to right within each panel. Panel B further decomposes by bachelor's (B) and master's (M).}
	\label{fig:credit_composition_empl}
	
\end{figure}

Panels A and B of figure \ref{fig:credit_composition} present the academic curricula of the selected high- and low-performing careers. The curricula are summarized as the share of credits in courses with different levels of quantitative content. Following the agreement among scholars in the categorization of STEM disciplines (table \ref{tab:stem_litreview}), I group university courses according to their quantitative content. Quantitative courses include science and statistics, architecture and engineering, and chemistry and pharmacy. These are the fields of study that most scholars agree can be defined as STEM. Courses with some quantitative component include life sciences (agriculture, veterinary, geology and biology), economics and management, and healthcare. These are more technical fields of study over which researchers disagree on whether they should belong to STEM education (I will alternatively refer to these disciplines as "technical"). Non-quantitative courses include education and psychology, law, humanities (literature and languages), and political and other social sciences. Most scholars agree that these fields of study do not fit the STEM definition. The ordering along the horizontal axis reflects increasing log wage returns. Figure \ref{fig:credit_composition} shows that quantitativeness alone does not explain the higher returns of certain careers. In fact, careers with high shares of credits in quantitative courses are represented both among the worst- and best-performing careers. Panel C presents the average composition of careers by quartiles of the distribution of log wage returns. This ensures that the lack of relationship between the share of quantitative credits and returns is not driven by the choice of low- and high-return careers. Indeed, the share of quantitative courses displays a slight U-shape relationship with log wage returns. The overall share of non-quantitative courses tends to increase along the distribution of log wage returns. Panel A in figure \ref{fig:credit_composition_empl} presents the same decomposition of academic curricula for the distribution of returns to employment. Increasing the share of quantitative courses only improves outcomes up to the third quartile, whereby the fourth quartile has the highest share of non-quantitative courses.

\begin{figure}[htbp]
	\centering
	\caption{Comparison of academic curricula for selected careers}
	\includegraphics[width=\linewidth]{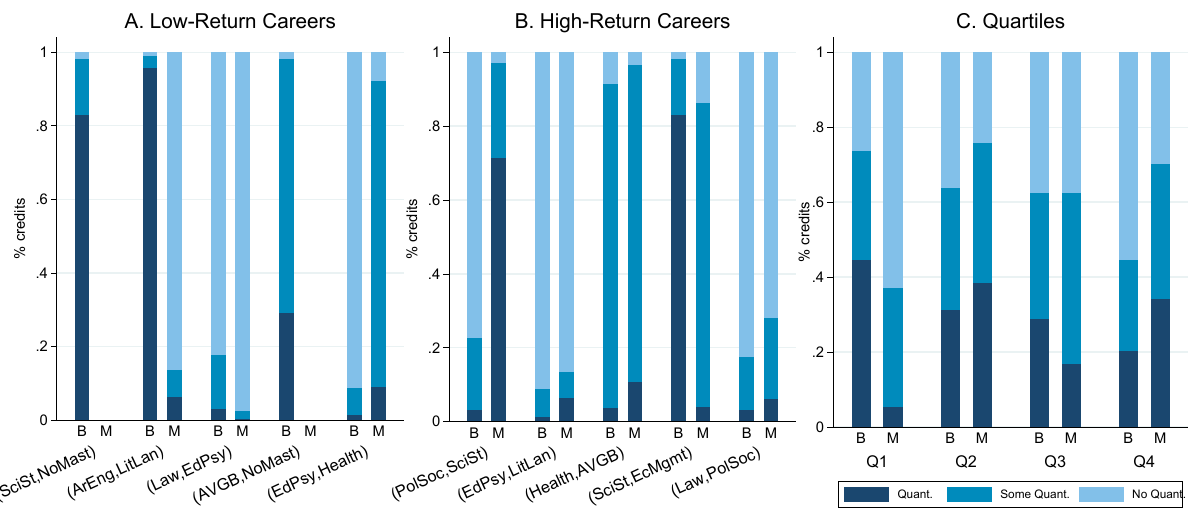}
	\caption*{\footnotesize Quantitative courses (dark blue): science and statistics, architecture and engineering, and chemistry and pharmacy. Some quantitative (technical) courses (blue): life sciences (agriculture, veterinary, geology and biology), economics and management, and healthcare. Non-quantitative courses (light blue): education and psychology, law, humanities (literature and languages), and political and social sciences. The total percentage of credits in each grouping is plotted on the vertical axis. The order of degrees follows the ranking of log-wage returns, increasing from left to right within each panel. Column labels B and M denote bachelor's and master's, respectively.}
	\label{fig:credit_composition_dis}
	
\end{figure}

I report the curriculum composition for the same groups of careers separately for the bachelor's and the master's degrees to elicit patterns in the timing of courses in figure \ref{fig:credit_composition_dis}. The most striking difference between low- and high-earning degrees in terms of curriculum that emerges once courses are plotted separately by bachelor's and master's is that degrees with low returns have a low share of technical courses in the bachelor's (panel A and quartile 1 of panel C). Conversely, high-return careers have a low share of non-quantitative credits in the master's (panel B and quartile 4 of panel C). Once again, a U-shaped relationship between the share of quantitative courses and log wage returns emerges, reiterating that quantitativeness alone does not explain higher returns, even when I account for timing. Panel B in figure \ref{fig:credit_composition_empl} presents the same decomposition of academic curricula for the distribution of returns to employment. In this case, high-performing degrees spend more time in more general type of courses in the bachelor's (non-quantitative and quantitative) while they invest substantially more in technical courses in the master's (quartile 4). These results are consistent with the paradigms of education, whereby more general education should be approached earlier and more vocational education later.\footnote{\cite{neal2018information} on optimal life cycle investments in skills, "\textit{learn to learn, learn to earn, earn}" (Appendix I.5).}

\begin{figure}[htbp]
	\centering
	\caption{Differences in returns for symmetric careers}
	\includegraphics[width=\linewidth]{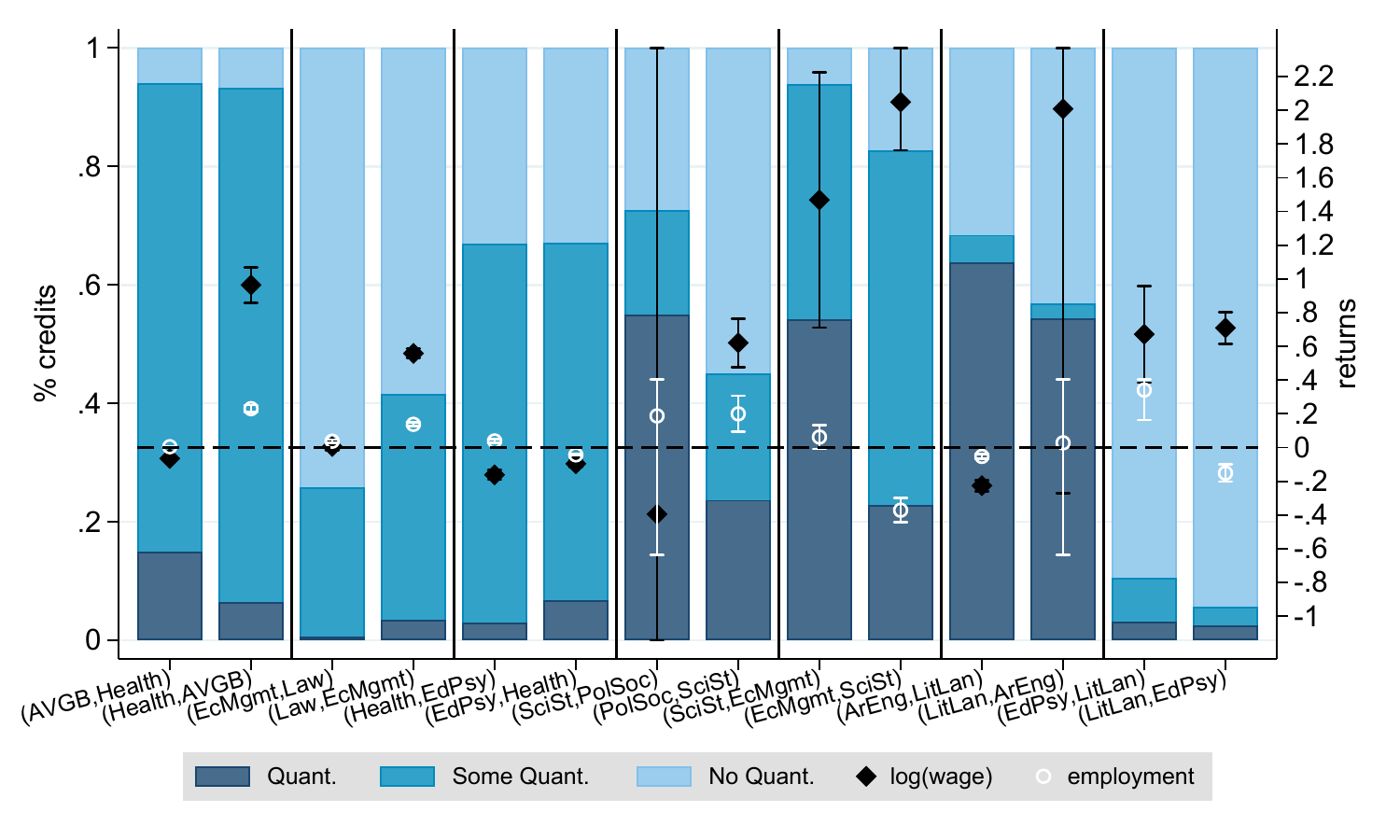}
	\caption*{\footnotesize Symmetric careers are grouped next to eachother. The share of courses by quantitativeness are plotted on the left vertical axis, while the estimated returns to log wages (black diamonds) and employment (white circles) follow the right vertical axis.}
	\label{fig:reciprocity}
\end{figure}

To further understand how the timing of degrees affects returns, I compare the returns and career composition for symmetric multidisciplinary careers, that is, given two fields of study $x$ and $y$, the returns to career $(x, y)$ compared with career $(y, x)$. Complete returns for both sets of outcomes are available for seven pairs of reciprocal degrees: (AVGB, Health), (Econ. Mgmt., Law), (Educ. Psyc., Health), (Pol. Soc., Sci. Stat.), (Ec.Mgmt., Sci.Stat.), (Arch.Eng., Lit.Lang.), (Ed.Psy., Lit.Lang.), and the reciprocals of these groups. Figure \ref{fig:reciprocity} presents the composition of these careers by degrees and the labor market returns, where each reciprocal is ordered such that the more quantitative group of the two is studied in the master's. Even though the composition of symmetric careers is comparable, the log wage returns vary substantially. In particular, log wage returns are higher when the more quantitative of the degrees is studied later, consistent with the findings of figure \ref{fig:credit_composition_dis}.\footnote{A degree in Agr.Vet.Geo.Bio. contains more quantitative courses (e.g. math, chemistry) than a degree in Health. Similarly, a degree in Health contains more quantitative courses than a degree in Education and Psychology and so on.} This trend in log wage returns is only partially carried across returns to employment. 

The analysis on the composition of curricula suggests that multidisciplinary careers can substantially increase or decrease labor market returns. While there is no clear recipe for a successful university career, several clues guide indicate best practices in the design of university programs. Quantitative courses are connected to log wage returns by a U-shaped relationship, whereby both low- and high-performing careers display relatively high shares of quantitative courses, and returns to employment increase with the share of quantitative courses only up to the third quartile. The timing of courses matters with higher shares of non-quantitative courses in the master's being related to lower returns. All high earning careers are characterized by relatively more general education early on (especially non-quantitative), and more technical courses in the master's. This is consistent with the comparison of symmetric multidisciplinary careers. It suggests that returns are different even when the overall structure of the curriculum is similar, with the returns being higher for careers with the most quantitative and technical degree studied later, even if globally it may result in less time spent in these subjects.

Even though these results should not be regarded as conclusive insights on the role of timing, multidisciplinarity and quantitativeness on labor market outcomes, they do suggest that these characteristics strongly affect outcomes and call for a deeper understanding of synergies across courses. When optimally designing a degree, additional constraints on total credits, substitution patterns and complementarities between courses should be considered, as well as measuring skill acquisition at university and skill use during the job, which are not observed in this setting. While this project does not allow for an in-depth discussion of how to increase the labor market returns of existing university careers, it does suggest that the combination of quantitative and technical courses is important for labor market outcomes, that well-thought multidisciplinary careers can lead to impressive labor market outcomes, and that timing of courses matters. In particular, it does seem that specializing in quantitative degrees in graduate school is positively associated with outcomes. Indeed, the signaling component of the degree might play a role in these results, so further research is needed to corroborate the role of timing. 

	\section{Conclusions}\label{sec:conclusions}

This article proposes a new method to causally estimate the returns to many combinations of bachelor's and master's degrees. It then leverages information on the course content of programs to investigate how multidisciplinarity, quantitativeness, and timing affect these returns. The findings highlight that considering the joint choices of bachelor's and master's degrees is crucial for accurately evaluating the effect of higher education on outcomes. Combining degrees in different fields can boost labor market returns, although there is no unique pattern of quantitative course content and timing that explains the success of certain university careers. A U-shaped relationship between labor market returns and the share of quantitative courses emerges. The breakdown of this relationship by bachelor's and master's highlight the importance of the timing of courses, suggesting that successful university careers often involve substantial non-quantitative education at the bachelor's level and less at the master's level. However, a deeper understanding of the complementarities between courses acquired early and late in the career is necessary. Finally, policy simulations that remove entry barriers at the bachelor's level suggest that students have preferences for non-quantitative degrees.

These results suggest that policies that incentivize enrollment in STEM education without considering the nonlinear relationship between quantitative education and outcomes might not benefit students. Furthermore, policies that incentivize STEM education by reducing entry barriers might be ineffective due to individual preferences and unable to affect the composition of the student body, such as increasing female enrollment. 
The results emphasize the importance of covering multiple disciplines throughout higher education with surprising positive effects on wages, challenging the notion that extreme specialization is profitable. This suggests that policies facilitating switching from one field to another may be highly beneficial to students. However, caution is advised, as certain combinations encompassing multiple fields can have negative effects.

One limitation of this setup is that it does not account for students' reactions to enrollment policies at the extensive margin. The negative effects of policies incentivizing STEM enrollment through reduced entry barriers might be mitigated if they generate a sufficient influx of students who would otherwise not obtain a degree. Additionally, the setup does not incorporate the signaling component of degrees. If employers primarily observe the highest level of education (as assumed by \cite{altonji1993demand}), master's degrees might be disproportionately weighted by employers, partially explaining the results on timing. Lastly, while the policy simulations hint at preferences for non-quantitative studies outweighing quantitative preferences, the model does not isolate this effect. Non-pecuniary returns not captured by the model might explain some features of the sorting, suggesting that policies affecting enrollment could have a greater impact if they incorporate these amenities.

This paper reveals two potential avenues for future research that would improve our understanding of how university-acquired knowledge translates into the labor market. Skills acquired during university vary across fields, but they are not observed in this setting. In particular, the results on the content of degrees signal the importance of the time spent in technical courses, such as medicine or management, which typically involve the acquisition of practical knowledge. We can speculate that part of the commonly observed success of STEM can be explained by the successful integration of quantitative and technical education that interplay with skills. Similarly, the concept of quantitativeness remains elusive, and high returns can be expected from specific types of quantitative education. Understanding how specialized knowledge in quantitative fields and how the mathematical language spills over into different courses -- for example through enhanced problem solving ability -- might be critical for optimally designing university degrees.

	\bibliographystyle{apalike}
	\bibliography{bibliography} 

@article{andrews2022returns,
  title={The returns to college major choice: Average and distributional effects, career trajectories, and earnings variability},
  author={Andrews, Rodney J and Imberman, Scott A and Lovenheim, Michael F and Stange, Kevin M},
  year={2022},
  journal={NBER Working Paper 30331},
  publisher={National Bureau of Economic Research}
}

@article{arcidiacono2016college,
	title={College attrition and the dynamics of information revelation},
	author={Arcidiacono, Peter and Aucejo, Esteban and Maurel, Arnaud and Ransom, Tyler},
	year={2016},
        journal={NBER Working Paper 22325},
	publisher={National Bureau of Economic Research}
}

@article{arcidiacono2016university,
	title={University differences in the graduation of minorities in STEM fields: Evidence from California},
	author={Arcidiacono, Peter and Aucejo, Esteban M and Hotz, V Joseph},
	journal={American Economic Review},
	volume={106},
	number={3},
	pages={525--62},
	year={2016}
}

@article{ahn2019equilibrium,
	title={Equilibrium grade inflation with implications for female interest in stem majors},
	author={Ahn, Thomas and Arcidiacono, Peter and Hopson, Amy and Thomas, James R},
	year={2019},
        journal={NBER Working Paper 26556},
	publisher={National Bureau of Economic Research}
}

@article{autor2013growth,
	title={The growth of low-skill service jobs and the polarization of the US labor market},
	author={Autor, David and Dorn, David},
	journal={American economic review},
	volume={103},
	number={5},
	pages={1553--97},
	year={2013}
}

@article{delaney2021high,
	title={High School Rank in Math and English and the Gender Gap in STEM},
	author={Delaney, Judith M and Devereux, Paul J},
	journal={Labour Economics},
	volume={69},
	pages={101969},
	year={2021},
	publisher={Elsevier}
}

@article{delaney2019understanding,
	title={Understanding gender differences in STEM: Evidence from college applications},
	author={Delaney, Judith M and Devereux, Paul J},
	journal={Economics of Education Review},
	volume={72},
	pages={219--238},
	year={2019},
	publisher={Elsevier}
}

@article{mcfadden1974conditional,
  title={Conditional logit analysis of qualitative choice behavior},
  author={McFadden, D},
  journal={Frontiers in Econometrics},
  pages={105--142},
  year={1974},
  publisher={Academic Press}
}

@article{montmarquette2002young,
	title={How do young people choose college majors?},
	author={Montmarquette, Claude and Cannings, Kathy and Mahseredjian, Sophie},
	journal={Economics of Education Review},
	volume={21},
	number={6},
	pages={543--556},
	year={2002},
	publisher={Elsevier}
}

@article{hoffman1988multinomial,
	title={Multinomial and conditional logit discrete-choice models in demography},
	author={Hoffman, Saul D and Duncan, Greg J},
	journal={Demography},
	volume={25},
	number={3},
	pages={415--427},
	year={1988},
	publisher={Springer}
}

@article{kahn2017women,
	title={Women and STEM},
	author={Kahn, Shulamit and Ginther, Donna},
	year={2017},
        journal={NBER Working Paper 23525},
	publisher={National Bureau of Economic Research}
}

@article{deming2017growing,
	title={The growing importance of social skills in the labor market},
	author={Deming, David J},
	journal={The Quarterly Journal of Economics},
	volume={132},
	number={4},
	pages={1593--1640},
	year={2017},
	publisher={Oxford University Press}
}

@book{neal2018information,
	title={Information, incentives, and education policy},
	author={Neal, Derek A},
	year={2018},
	publisher={Harvard University Press}
}

@misc{istat2021,
	author = {ISTAT},
	title = {Iscritti all'Universit\`{a}},
	month = september,
	year = {2021},
	howpublished={\url{http://dati.istat.it/Index.aspx?DataSetCode=DCIS_ISCRITTI}}
}

@article{winters2014stem,
	title={STEM graduates, human capital externalities, and wages in the US},
	author={Winters, John V},
	journal={Regional Science and Urban Economics},
	volume={48},
	pages={190--198},
	year={2014},
	publisher={Elsevier}
}

@article{porter2020gender,
	title={Gender differences in the choice of major: The importance of female role models},
	author={Porter, Catherine and Serra, Danila},
	journal={American Economic Journal: Applied Economics},
	volume={12},
	number={3},
	pages={226--54},
	year={2020}
}

@article{xie2015stem,
	title={STEM education},
	author={Xie, Yu and Fang, Michael and Shauman, Kimberlee},
	journal={Annual review of sociology},
	volume={41},
	pages={331--357},
	year={2015},
	publisher={Annual Reviews}
}

@article{schmeiser2016student,
	title={Student loan information provision and academic choices},
	author={Schmeiser, Maximilian and Stoddard, Christiana and Urban, Carly},
	journal={American Economic Review},
	volume={106},
	number={5},
	pages={324--28},
	year={2016}
}

@article{buffington2016stem,
	title={STEM training and early career outcomes of female and male graduate students: Evidence from UMETRICS data linked to the 2010 census},
	author={Buffington, Catherine and Cerf, Benjamin and Jones, Christina and Weinberg, Bruce A},
	journal={American Economic Review},
	volume={106},
	number={5},
	pages={333--38},
	year={2016}
}

@article{uddin2021research,
	title={Research interdisciplinarity: STEM versus non-STEM},
	author={Uddin, Shahadat and Imam, Tasadduq and Mozumdar, Mohammad},
	journal={Scientometrics},
	volume={126},
	number={1},
	pages={603--618},
	year={2021},
	publisher={Springer}
}

@article{rask2010attrition,
	title={Attrition in STEM fields at a liberal arts college: The importance of grades and pre-collegiate preferences},
	author={Rask, Kevin},
	journal={Economics of Education Review},
	volume={29},
	number={6},
	pages={892--900},
	year={2010},
	publisher={Elsevier}
}

@article{maple1991influences,
	title={Influences on the choice of math/science major by gender and ethnicity},
	author={Maple, Sue A and Stage, Frances K},
	journal={American Educational Research Journal},
	volume={28},
	number={1},
	pages={37--60},
	year={1991},
	publisher={Sage Publications}
}

@article{adams2016women,
	title={Women on boards in finance and STEM industries},
	author={Adams, Ren{\'e}e B and Kirchmaier, Tom},
	journal={American Economic Review},
	volume={106},
	number={5},
	pages={277--81},
	year={2016}
}

@incollection{altonji2016analysis,
	title={The analysis of field choice in college and graduate school: Determinants and wage effects},
	author={Altonji, Joseph G and Arcidiacono, Peter and Maurel, Arnaud},
	booktitle={Handbook of the Economics of Education},
	volume={5},
	pages={305--396},
	year={2016},
	publisher={Elsevier}
}

@article{altonji2021labor,
	title={The labor market returns to advanced degrees},
	author={Altonji, Joseph G and Zhong, Ling},
	journal={Journal of Labor Economics},
	volume={39},
	number={2},
	pages={303--360},
	year={2021},
	publisher={The University of Chicago Press Chicago, IL}
}

@article{ng2020returns,
        Author = {Ng, Kevin and Riehl, Evan},
        Title = {The Returns to STEM Programs for Less-Prepared Students},
        Journal = {American Economic Journal: Economic Policy},
        Volume = {16},
        Number = {2},
        Year = {2024},
        Month = {May},
        Pages = {37–77},
        DOI = {10.1257/pol.20200694},
        URL = {https://www.aeaweb.org/articles?id=10.1257/pol.20200694}
        }

@article{canaan2018returns,
	title={Returns to education quality for low-skilled students: Evidence from a discontinuity},
	author={Canaan, Serena and Mouganie, Pierre},
	journal={Journal of Labor Economics},
	volume={36},
	number={2},
	pages={395--436},
	year={2018},
	publisher={University of Chicago Press Chicago, IL}
}

@misc{granato2018gender,
	title={Gender Inequalities and Scarring Effects in School to Work Transitions},
	author={Granato, Silvia},
	year={2018},
	school={Queen Mary University of London}
}

@article{chise2021intergenerational,
	title={On the intergenerational transmission of STEM education among graduate students},
	author={Chise, Diana and Fort, Margherita and Monfardini, Chiara},
	journal={The BE Journal of Economic Analysis \& Policy},
	volume={21},
	number={1},
	pages={115--145},
	year={2021},
	publisher={De Gruyter}
}

@article{altonji2012heterogeneity,
	title={Heterogeneity in human capital investments: High school curriculum, college major, and careers},
	author={Altonji, Joseph G and Blom, Erica and Meghir, Costas},
	journal={Annu. Rev. Econ.},
	volume={4},
	number={1},
	pages={185--223},
	year={2012},
	publisher={Annual Reviews}
}

@article{braga2016teaching,
	author = {Braga, Michela and Paccagnella, Marco and Pellizzari, Michele},
	title = {The Impact of College Teaching on Students’ Academic and Labor Market Outcomes},
	journal = {Journal of Labor Economics},
	volume = {34},
	number = {3},
	pages = {781-822},
	year = {2016},
	doi = {10.1086/684952},
	
	URL = { 
	https://doi.org/10.1086/684952
	
	},
	eprint = { 
	https://doi.org/10.1086/684952
	
	}
}

@phdthesis{bamberger1987occupational,
	title={Occupational choice: The role of undergraduate education},
	author={Bamberger, Gustavo Ernesto},
	year={1987},
	school={The University of Chicago}
}

@techreport{eurydice2020fees,
	author = {European Commission/EACEA/Eurydice},
	title = {National Student Fee and Support Systems in
	European Higher Education – 2020/21},
	institution = {Eurydice -- Fact and Figures},
	publisher = {Luxembourg: Publications Office of the European Union},
	year = {2020}}

@article{kirkeboen2016field,
	title={Field of study, earnings, and self-selection},
	author={Kirkeboen, Lars J and Leuven, Edwin and Mogstad, Magne},
	journal={The Quarterly Journal of Economics},
	volume={131},
	number={3},
	pages={1057--1111},
	year={2016},
	publisher={Oxford University Press}
}

@article{heckman2006understanding,
	title={Understanding instrumental variables in models with essential heterogeneity},
	author={Heckman, James J and Urzua, Sergio and Vytlacil, Edward},
	journal={The Review of Economics and Statistics},
	volume={88},
	number={3},
	pages={389--432},
	year={2006},
	publisher={The MIT Press}
}

@article{heckman2010comparing,
	title={Comparing IV with structural models: What simple IV can and cannot identify},
	author={Heckman, James J and Urzua, Sergio},
	journal={Journal of Econometrics},
	volume={156},
	number={1},
	pages={27--37},
	year={2010},
	publisher={Elsevier}
}

@article{webber2016college,
	title={Are college costs worth it? How ability, major, and debt affect the returns to schooling},
	author={Webber, Douglas A.},
	journal={Economics of Education Review},
	volume={53},
	pages={296--310},
	year={2016},
	publisher={Elsevier}
}

@article{hastings2013some,
	title={Are some degrees worth more than others? Evidence from college admission cutoffs in Chile},
	author={Hastings, Justine S and Neilson, Christopher A and Zimmerman, Seth D},
	year={2013},
        journal={NBER Working Paper 19241},
	publisher={National Bureau of Economic Research}
}

@incollection{altonji2018costs,
  title={The costs of and net returns to college major},
  author={Altonji, Joseph G and Zimmerman, Seth D},
  booktitle={Productivity in higher education},
  pages={133--176},
  year={2018},
  publisher={University of Chicago Press}
}

@article{mountjoy2021community,
	title={Community colleges and upward mobility},
	author={Mountjoy, Jack},
	year={2022},
	journal={forthcoming American Economic Review}
}

@article{arcidiacono2004ability,
	title={Ability sorting and the returns to college major},
	author={Arcidiacono, Peter},
	journal={Journal of Econometrics},
	volume={121},
	number={1-2},
	pages={343--375},
	year={2004},
	publisher={Elsevier}
}

@article{heckman2018unordered,
	title={Unordered monotonicity},
	author={Heckman, James J and Pinto, Rodrigo},
	journal={Econometrica},
	volume={86},
	number={1},
	pages={1--35},
	year={2018},
	publisher={Wiley Online Library}
}

@article{bleemer2022will,
	title={Will studying economics make you rich? A regression discontinuity analysis of the returns to college major},
	author={Bleemer, Zachary and Mehta, Aashish},
	journal={American Economic Journal: Applied Economics},
	volume={14},
	number={2},
	pages={1--22},
	year={2022}
}

@article{deming2020earnings,
	title={Earnings dynamics, changing job skills, and STEM careers},
	author={Deming, David J and Noray, Kadeem},
	journal={The Quarterly Journal of Economics},
	volume={135},
	number={4},
	pages={1965--2005},
	year={2020},
	publisher={Oxford University Press}
}

@article{arcidiacono2011practical,
	title={Practical methods for estimation of dynamic discrete choice models},
	author={Arcidiacono, Peter and Ellickson, Paul B and others},
	journal={Annual Reviews of Economics},
	volume={3},
	number={1},
	pages={363-394},
	year={2011},
	publisher={Annual Reviews}
	
}

@article{bhuller20222sls,
	title={2SLS with multiple treatments},
	author={Bhuller, Manudeep and Sigstad, Henrik},
	journal={arXiv preprint arXiv:2205.07836},
	year={2022}
}

@article{biasi2022education,
	title={The education-innovation gap},
	author={Biasi, Barbara and Ma, Song},
	year={2022},
        journal={NBER Working Paper 29853},
	publisher={National Bureau of Economic Research}
}

@article{bianchi2020scientific,
	title={Scientific education and innovation: from technical diplomas to university STEM degrees},
	author={Bianchi, Nicola and Giorcelli, Michela},
	journal={Journal of the European Economic Association},
	volume={18},
	number={5},
	pages={2608--2646},
	year={2020},
	publisher={Oxford University Press}
}

@article{vytlacil2002independence,
	title={Independence, monotonicity, and latent index models: An equivalence result},
	author={Vytlacil, Edward},
	journal={Econometrica},
	volume={70},
	number={1},
	pages={331--341},
	year={2002},
	publisher={JSTOR}
}

@article{malamud2010breadth,
	title={Breadth versus depth: the timing of specialization in higher education},
	author={Malamud, Ofer},
	journal={Labour},
	volume={24},
	number={4},
	pages={359--390},
	year={2010},
	publisher={Wiley Online Library}
}

@article{malamud2011discovering,
	title={Discovering one's talent: learning from academic specialization},
	author={Malamud, Ofer},
	journal={ILR Review},
	volume={64},
	number={2},
	pages={375--405},
	year={2011},
	publisher={SAGE Publications Sage CA: Los Angeles, CA}
}

@online{CollegeEnrollmentUS,
	author = {Hanson, Melanie},
	title = {College Enrollment \& Student Demographic Statistics},
	year = 2022,
	month = July,
	day = 26,
	url = {https://educationdata.org/college-enrollment-statistics},
	urldate = {2022-09-28}
}

@online{CollegeEnrollmentEU,
	author = {EuroStat},
	title = {Tertiary Education Statistics},
	year = 2022,
	month = September,
	day = 14,
	url = {https://ec.europa.eu/eurostat/statistics-explained/index.php?title=Tertiary_education_statistics#Participation_by_level},
	urldate = {2022-09-28}
}

@article{patnaik2020college,
	title={College Majors},
	author={Patnaik, Arpita and Wiswall, Matthew and Zafar, Basit},
	journal={NBER Working Paper},
	number={w27645},
	year={2020}
}

@article{arcidiacono2008economic,
	title={The economic returns to an MBA},
	author={Arcidiacono, Peter and Cooley, Jane and Hussey, Andrew},
	journal={International Economic Review},
	volume={49},
	number={3},
	pages={873--899},
	year={2008},
	publisher={Wiley Online Library}
}

@article{altonji1993demand,
	title={The demand for and return to education when education outcomes are uncertain},
	author={Altonji, Joseph G},
	journal={Journal of Labor Economics},
	volume={11},
	number={1, Part 1},
	pages={48--83},
	year={1993},
	publisher={University of Chicago Press}
}

@article{black2003economic,
	title={The economic reward for studying economics},
	author={Black, Dan A and Sanders, Seth and Taylor, Lowell},
	journal={Economic Inquiry},
	volume={41},
	number={3},
	pages={365--377},
	year={2003},
	publisher={Wiley Online Library}
}

@article{bhattacharya2005specialty,
	title={Specialty selection and lifetime returns to specialization within medicine},
	author={Bhattacharya, Jayanta},
	journal={Journal of Human Resources},
	volume={40},
	number={1},
	pages={115--143},
	year={2005},
	publisher={University of Wisconsin Press}
}

@article{chen2012women,
	title={Are women overinvesting in education? Evidence from the medical profession},
	author={Chen, M Keith and Chevalier, Judith A},
	journal={Journal of Human Capital},
	volume={6},
	number={2},
	pages={124--149},
	year={2012},
	publisher={University of Chicago Press Chicago, IL}
}

@article{ketel2016returns,
	title={The returns to medical school: Evidence from admission lotteries},
	author={Ketel, Nadine and Leuven, Edwin and Oosterbeek, Hessel and van der Klaauw, Bas},
	journal={American Economic Journal: Applied Economics},
	volume={8},
	number={2},
	pages={225--54},
	year={2016}
}

@article{brewer1999does,
	title={Does it pay to attend an elite private college? Cross-cohort evidence on the effects of college type on earnings},
	author={Brewer, Dominic J and Eide, Eric R and Ehrenberg, Ronald G},
	journal={Journal of Human Resources},
	volume={34},
	number={1},
	pages={104--123},
	year={1999},
	publisher={University of Wisconsin Press}
}

@article{ashworth2021changes,
	title={Changes across cohorts in wage returns to schooling and early work experiences},
	author={Ashworth, Jared and Hotz, V Joseph and Maurel, Arnaud and Ransom, Tyler},
	journal={Journal of Labor Economics},
	volume={39},
	number={4},
	pages={931--964},
	year={2021},
	publisher={The University of Chicago Press Chicago, IL}
}

@article{dhaultfoeuille2013inference,
	title={Inference on an extended Roy model, with an application to schooling decisions in France},
	author={d’Haultfoeuille, Xavier and Maurel, Arnaud},
	journal={Journal of Econometrics},
	volume={174},
	number={2},
	pages={95--106},
	year={2013},
	publisher={Elsevier}
}

@article{beffy2012fieldofstudy,
	author = {Beffy, Magali and Fougère, Denis and Maurel, Arnaud},
	title = "{Choosing the Field of Study in Postsecondary Education: Do Expected Earnings Matter?}",
	journal = {The Review of Economics and Statistics},
	volume = {94},
	number = {1},
	pages = {334-347},
	year = {2012},
	month = {02},
	issn = {0034-6535},
	doi = {10.1162/REST_a_00212},
	url = {https://doi.org/10.1162/REST\_a\_00212},
	eprint = {https://direct.mit.edu/rest/article-pdf/94/1/334/1916876/rest\_a\_00212.pdf},
}

@article{oreopoulos2013making,
	title={Making College Worth It: A Review of the Returns to Higher Education.},
	author={Oreopoulos, Philip and Petronijevic, Uros},
	journal={Future of Children},
	volume={23},
	number={1},
	pages={41--65},
	year={2013},
	publisher={ERIC}
}

@article{acemoglu2022eclipse,
	title={Eclipse of Rent-Sharing: The Effects of Managers' Business Education on Wages and the Labor Share in the US and Denmark},
	author={Acemoglu, Daron and He, Alex and le Maire, Daniel},
	year={2022},
        journal={NBER Working Paper 29874},
	publisher={National Bureau of Economic Research}
}

@techreport{almalaurea2020,
	title={{XXIII} indagine -- Profilo dei laureati 2020},
	author={{AlmaLaurea}},
	year={2021},
	publisher={MUR}
}

@techreport{almalaurea2020note,
	title={Note Metodologiche: {XXIII} indagine -- Profilo dei laureati 2020},
	author={{AlmaLaurea}},
	year={2021},
	publisher={MUR}
}

@techreport{almalaurea2019note,
	title={Note Metodologiche: {XXII} indagine -- Profilo dei laureati 2019},
	author={{AlmaLaurea}},
	year={2020},
	publisher={MUR}
}

@article{chernozhukov2008reduced,
	title={The reduced form: A simple approach to inference with weak instruments},
	author={Chernozhukov, Victor and Hansen, Christian},
	journal={Economics Letters},
	volume={100},
	number={1},
	pages={68--71},
	year={2008},
	publisher={Elsevier}
}

@article{phillips2017structural,
	title={Structural inference from reduced forms with many instruments},
	author={Phillips, Peter CB and Gao, Wayne Yuan},
	journal={Journal of Econometrics},
	volume={199},
	number={2},
	pages={96--116},
	year={2017},
	publisher={Elsevier}
}

@misc{ssd,
	author = {CUN, Italian National University Council},
	title = {Academic Disciplines' List for Italian University Research and Teaching},
	year = 2000,
	month = October,
	day = 4,
	url = {https://www.cun.it/uploads/storico/settori_scientifico_disciplinari_english.pdf},
	urldate = {2022-10-15}
}

@misc{braccioli2022unordered,
	title={Education Expansion, Skills and Labour Market Success},
	author={Braccioli, Federica and Ghinetti, Paolo and Moriconi, Simone and Naguib, Costanza and Pellizzari, Michele},
	journal={Unpublished Manuscript},
	year={2022}
}

@article{crudu2021inference,
	title={Inference in instrumental variable models with heteroskedasticity and many instruments},
	author={Crudu, Federico and Mellace, Giovanni and S{\'a}ndor, Zsolt},
	journal={Econometric Theory},
	volume={37},
	number={2},
	pages={281--310},
	year={2021},
	publisher={Cambridge University Press}
}

@article{mikusheva2022inference,
	title={Inference with many weak instruments},
	author={Mikusheva, Anna and Sun, Liyang},
	journal={The Review of Economic Studies},
	volume={89},
	number={5},
	pages={2663--2686},
	year={2022},
	publisher={Oxford University Press}
}

@article{bianchi2020expansion,
	title={The indirect effects of educational expansions: Evidence from a large enrollment increase in university majors},
	author={Bianchi, Nicola},
	journal={Journal of Labor Economics},
	volume={38},
	number={3},
	pages={767--804},
	year={2020},
	publisher={The University of Chicago Press Chicago, IL}
}

@article{mann2013,
	title = {Trends in gender segregation in the choice of science and engineering majors},
	journal = {Social Science Research},
	volume = {42},
	number = {6},
	pages = {1519-1541},
	year = {2013},
	issn = {0049-089X},
	doi = {https://doi.org/10.1016/j.ssresearch.2013.07.002},
	url = {https://www.sciencedirect.com/science/article/pii/S0049089X13001051},
	author = {Allison Mann and Thomas A. DiPrete}
}

@article{fricke2018,
	title = {Exposure to academic fields and college major choice},
	journal = {Economics of Education Review},
	volume = {64},
	pages = {199-213},
	year = {2018},
	issn = {0272-7757},
	doi = {https://doi.org/10.1016/j.econedurev.2018.04.007},
	url = {https://www.sciencedirect.com/science/article/pii/S0272775717303795},
	author = {Hans Fricke and Jeffrey Grogger and Andreas Steinmayr}
}

@article{hotz1993conditional,
  title={Conditional choice probabilities and the estimation of dynamic models},
  author={Hotz, V Joseph and Miller, Robert A},
  journal={The Review of Economic Studies},
  volume={60},
  number={3},
  pages={497--529},
  year={1993},
  publisher={Wiley-Blackwell}
}

@misc{uscensus2019,
	author = {{US Census Bureau}},
	title = {Number of People With Master’s and Doctoral Degrees Doubles Since 2000},
	year = {2019},
	addendum = {(accessed: 28.10.2022)},
	howpublished={\url{https://www.census.gov/library/stories/2019/02/number-of-people-with-masters-and-phd-degrees-double-since-2000.html}}
}

@article{bleemer2021top,
  title={Top percent policies and the return to postsecondary selectivity},
  author={Bleemer, Zachary},
  journal={Research \& Occasional Paper Series: CSHE},
  volume={1},
  year={2021}
}

@incollection{MCFADDEN19841395,
title = {Chapter 24 Econometric analysis of qualitative response models},
booktitle = {Handbook of Econometrics},
publisher = {Elsevier},
volume = {2},
pages = {1395-1457},
year = {1984},
issn = {1573-4412},
doi = {https://doi.org/10.1016/S1573-4412(84)02016-X},
url = {https://www.sciencedirect.com/science/article/pii/S157344128402016X},
author = {McFadden, Daniel L.}
}

@book{amemiya1985advanced,
  title={Advanced econometrics},
  author={Amemiya, Takeshi},
  year={1985},
  publisher={Harvard university press}
}

	\newpage
	\appendix
\section{Appendix}\label{sec:appendix}
\subsection{Additional Descriptive Results}\label{sec:appendix_descriptives}

\begin{table}[htbp]
	\centering
	\caption{Differences in $X$ for the sample of employed and unemployed.}
	\begin{tabular}{lccc}
		\toprule
		& All   & Employed & Unemployed \\
		\midrule
		&       &       &  \\
		High School: grade (st.) & 0.00  & 0.03  & -0.11 \\
		& (1.000) & (0.998) & (0.998) \\
		&       &       &  \\
		High School: humanities & 0.15  & 0.14  & 0.18 \\
		& (0.359) & (0.352) & (0.384) \\
		&       &       &  \\
		High School: science & 0.39  & 0.40  & 0.36 \\
		& (0.487) & (0.489) & (0.479) \\
		&       &       &  \\
		Gender (1=female) & 0.62  & 0.60  & 0.68 \\
		& (0.485) & (0.489) & (0.466) \\
		&       &       &  \\
		Parents: graduate & 0.26  & 0.26  & 0.26 \\
		& (0.438) & (0.437) & (0.439) \\
		&       &       &  \\
		
		Parents: high-ranked occ. & 0.21  & 0.22  & 0.21 \\
		& (0.410) & (0.412) & (0.406) \\
		&       &       &  \\
		Employment & 0.77  & 1.00  & 0.00 \\
		& (0.418) & (0)   & (0) \\
		&       &       &  \\
		Observations & 655 847 & 508 242 & 147 605 \\
		\bottomrule
	\end{tabular}%
	\label{tab:diff_empl_unempl}%
\end{table}%

\begin{figure}
	\centering
	\caption{Occupation sectors by master degree's (ISTAT codes)}
	\label{fig:occupations}
	\begin{subfigure}{.5\textwidth}
		\centering
		\includegraphics[width=.95\linewidth]{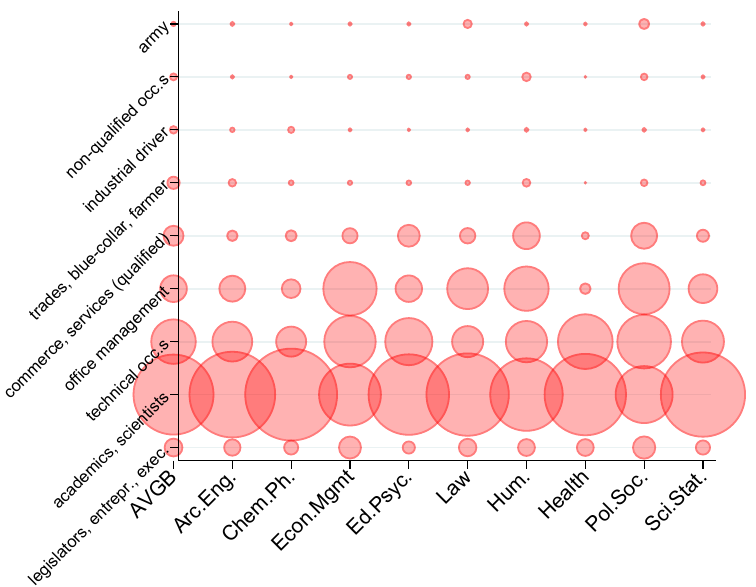}
		\caption{Share of master graduates in one-digit occupations.}
		\label{fig:occupation1}
	\end{subfigure}%
	\begin{subfigure}{.5\textwidth}
		\centering
		\includegraphics[width=.95\linewidth]{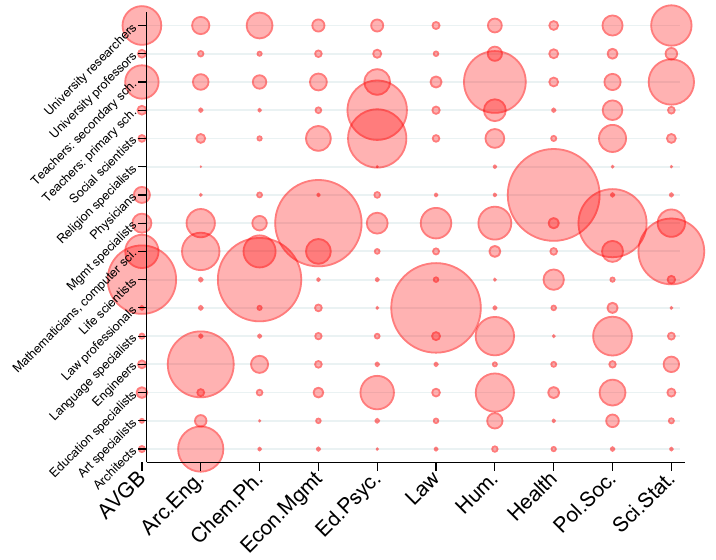}
		\caption{Two-digit occupations for master graduates employed in sector "academics and scientists".}
		\label{fig:occupation2}
	\end{subfigure}
	\caption*{\footnotesize Panel \ref{fig:occupation1} presents one-digit occupation sectors for all master graduates as defined by ISTAT's 2011 classification of occupations (in turn based on ILO's 2008 \textit{International Standard Classification of Occupations}). This information is available for 209 906 individuals. Panel \ref{fig:occupation2} focuses on two-digit occupation sectors for master graduates employed in sector "academics and scientists" (intellectual and highly specialized occupations), for a total of 126 166 observations. In both instances, occupation codes are only available for individuals who complete a master degree and are not available for students who start working after their bachelor. Both panels show that labor markets are segregated along specialized skill sets.}
	
\end{figure}

\begin{landscape}
	\begin{table}[ht]
		\centering
		\setlength{\tabcolsep}{2pt}
		\resizebox{\linewidth}{!}{
		\begin{threeparttable}
			\caption{Comparison of STEM classification methods in Economics papers}
			\begin{tabular}{l l l l l l l l l l l l}
				\toprule
				\textbf{Paper} & \textbf{Classification} & \multicolumn{10}{c}{\textbf{Groups of Degrees Classified as STEM}} \\
				&       & \textbf{Science} & \textbf{Arch.} & \textbf{Chem.} & \textbf{Agr.Vet. } & \textbf{Econ.} & \textbf{Health} & \textbf{Educ.} & \textbf{Law} & \textbf{Lit. } & \textbf{Pol.} \\
				&       &  & \textbf{Eng.} & \textbf{Pharm.} & \textbf{Geo.Bio.} & \textbf{Stat.} &  & \textbf{Psy.} &  & \textbf{Lang.} & \textbf{Soc.} \\
				\midrule
				\cite{adams2016women} & O*NET, authors & \cellcolor[rgb]{ .329,  .51,  .208}All & \cellcolor[rgb]{ .663,  .816,  .557}Most & \cellcolor[rgb]{ .663,  .816,  .557}Most & \cellcolor[rgb]{ .776,  .878,  .706}Some & \cellcolor[rgb]{ .663,  .816,  .557}Most & None  & None  & None  & None  & None \\
				\cite{ahn2019equilibrium} & author & \cellcolor[rgb]{ .329,  .51,  .208}All & \cellcolor[rgb]{ .663,  .816,  .557}Most & \cellcolor[rgb]{ .663,  .816,  .557}Most & \cellcolor[rgb]{ .776,  .878,  .706}Some & \cellcolor[rgb]{ .329,  .51,  .208}All & None  & None  & None  & None  & None \\
				\cite{altonji2016analysis} & author & \cellcolor[rgb]{ .329,  .51,  .208}All & \cellcolor[rgb]{ .329,  .51,  .208}All & \cellcolor[rgb]{ .329,  .51,  .208}All & \cellcolor[rgb]{ .329,  .51,  .208}All & \cellcolor[rgb]{ .329,  .51,  .208}All & \cellcolor[rgb]{ .329,  .51,  .208}All & None  & None  & None  & None \\
				\cite{altonji2021labor} & author, NSCG, NSRCG & \cellcolor[rgb]{ .329,  .51,  .208}All & \cellcolor[rgb]{ .663,  .816,  .557}Most & None  & \cellcolor[rgb]{ .329,  .51,  .208}All & None  & None  & None  & None  & None  & None \\
				\cite{arcidiacono2016university} & author & \cellcolor[rgb]{ .329,  .51,  .208}All & \cellcolor[rgb]{ .329,  .51,  .208}All & \cellcolor[rgb]{ .329,  .51,  .208}All & \cellcolor[rgb]{ .663,  .816,  .557}Most & \cellcolor[rgb]{ .776,  .878,  .706}Some & None  & \cellcolor[rgb]{ .776,  .878,  .706}Some & None  & None  & None \\
				\cite{arcidiacono2016college} & author & \cellcolor[rgb]{ .329,  .51,  .208}All & \cellcolor[rgb]{ .329,  .51,  .208}All & \cellcolor[rgb]{ .663,  .816,  .557}Most & \cellcolor[rgb]{ .329,  .51,  .208}All & None  & \cellcolor[rgb]{ .776,  .878,  .706}Some & None  & None  & None  & None \\
				\cite{bianchi2020scientific} & author & \cellcolor[rgb]{ .329,  .51,  .208}All & \cellcolor[rgb]{ .663,  .816,  .557}Most & \cellcolor[rgb]{ .663,  .816,  .557}Most & \cellcolor[rgb]{ .663,  .816,  .557}Most & \cellcolor[rgb]{ .776,  .878,  .706}Some & None  & None  & None  & None  & None \\
				\cite{biasi2022education} & ARC 2016 & \cellcolor[rgb]{ .329,  .51,  .208}All & \cellcolor[rgb]{ .329,  .51,  .208}All & \cellcolor[rgb]{ .329,  .51,  .208}All & \cellcolor[rgb]{ .329,  .51,  .208}All & None  & \cellcolor[rgb]{ .329,  .51,  .208}All & None  & None  & None  & None \\
				\cite{buffington2016stem} & author & \cellcolor[rgb]{ .329,  .51,  .208}All & \cellcolor[rgb]{ .663,  .816,  .557}Most & \cellcolor[rgb]{ .329,  .51,  .208}All & \cellcolor[rgb]{ .663,  .816,  .557}Most & None  & \cellcolor[rgb]{ .329,  .51,  .208}All & None  & None  & None  & None \\
				\cite{canaan2018returns} & author & \cellcolor[rgb]{ .329,  .51,  .208}All & \cellcolor[rgb]{ .329,  .51,  .208}All & \cellcolor[rgb]{ .329,  .51,  .208}All & \cellcolor[rgb]{ .329,  .51,  .208}All & \cellcolor[rgb]{ .663,  .816,  .557}Most & \cellcolor[rgb]{ .663,  .816,  .557}Most & None  & None  & None  & None \\
				\cite{chise2021intergenerational} & ISCED, MIUR, author & \cellcolor[rgb]{ .329,  .51,  .208}All & \cellcolor[rgb]{ .329,  .51,  .208}All & \cellcolor[rgb]{ .663,  .816,  .557}Most & \cellcolor[rgb]{ .776,  .878,  .706}Some & \cellcolor[rgb]{ .776,  .878,  .706}Some & None  & None  & None  & None  & None \\
				\cite{delaney2019understanding} & ISCED, authors & \cellcolor[rgb]{ .329,  .51,  .208}All & \cellcolor[rgb]{ .329,  .51,  .208}All & \cellcolor[rgb]{ .329,  .51,  .208}All & \cellcolor[rgb]{ .776,  .878,  .706}Some & \cellcolor[rgb]{ .776,  .878,  .706}Some & \cellcolor[rgb]{ .329,  .51,  .208}All & None  & None  & None  & None \\
				\cite{delaney2021high} & ISCED, authors & \cellcolor[rgb]{ .329,  .51,  .208}All & \cellcolor[rgb]{ .329,  .51,  .208}All & \cellcolor[rgb]{ .329,  .51,  .208}All & \cellcolor[rgb]{ .776,  .878,  .706}Some & \cellcolor[rgb]{ .776,  .878,  .706}Some & \cellcolor[rgb]{ .329,  .51,  .208}All & None  & None  & None  & None \\
				\cite{deming2017growing} & \cite{autor2013growth} & \cellcolor[rgb]{ .329,  .51,  .208}All & \cellcolor[rgb]{ .329,  .51,  .208}All & \cellcolor[rgb]{ .776,  .878,  .706}Some & \cellcolor[rgb]{ .776,  .878,  .706}Some & \cellcolor[rgb]{ .776,  .878,  .706}Some & \cellcolor[rgb]{ .776,  .878,  .706}Some & None  & None  & None  & None \\
				\cite{granato2018gender} & MIUR  & \cellcolor[rgb]{ .329,  .51,  .208}All & \cellcolor[rgb]{ .329,  .51,  .208}All & \cellcolor[rgb]{ .329,  .51,  .208}All & \cellcolor[rgb]{ .776,  .878,  .706}Some & \cellcolor[rgb]{ .776,  .878,  .706}Some & None  & None  & None  & None  & None \\
				\cite{kahn2017women} & author & \cellcolor[rgb]{ .329,  .51,  .208}All & \cellcolor[rgb]{ .663,  .816,  .557}Most & \cellcolor[rgb]{ .329,  .51,  .208}All & \cellcolor[rgb]{ .663,  .816,  .557}Most & \cellcolor[rgb]{ .663,  .816,  .557}Most & None  & \cellcolor[rgb]{ .776,  .878,  .706}Some & None  & None  & None \\
				\cite{maple1991influences} & author & \cellcolor[rgb]{ .329,  .51,  .208}All & \cellcolor[rgb]{ .663,  .816,  .557}Most & None  & None  & None  & None  & None  & None  & None  & None \\
				\cite{ng2020returns} & author & \cellcolor[rgb]{ .329,  .51,  .208}All & \cellcolor[rgb]{ .663,  .816,  .557}Most & \cellcolor[rgb]{ .663,  .816,  .557}Most & \cellcolor[rgb]{ .663,  .816,  .557}Most & \cellcolor[rgb]{ .776,  .878,  .706}Some & None  & None  & None  & None  & None \\
				\cite{porter2020gender} & author & \cellcolor[rgb]{ .329,  .51,  .208}All & \cellcolor[rgb]{ .329,  .51,  .208}All & \cellcolor[rgb]{ .329,  .51,  .208}All & None  & \cellcolor[rgb]{ .329,  .51,  .208}All & None  & None  & None  & None  & None \\
				\cite{rask2010attrition} & Anon. data provider & \cellcolor[rgb]{ .329,  .51,  .208}All & \cellcolor[rgb]{ .440,  .440,  .440}NA & \cellcolor[rgb]{ .329,  .51,  .208}All & \cellcolor[rgb]{ .663,  .816,  .557}Most & \cellcolor[rgb]{ .776,  .878,  .706}Some & None  & \cellcolor[rgb]{ .776,  .878,  .706}Some & None  & None  & None \\
				\cite{schmeiser2016student} & author & \cellcolor[rgb]{ .329,  .51,  .208}All & \cellcolor[rgb]{ .329,  .51,  .208}All & \cellcolor[rgb]{ .329,  .51,  .208}All & \cellcolor[rgb]{ .329,  .51,  .208}All & \cellcolor[rgb]{ .776,  .878,  .706}Some & \cellcolor[rgb]{ .663,  .816,  .557}Most & None  & None  & None  & None \\
				\cite{uddin2021research} & ARC 2016 & \cellcolor[rgb]{ .329,  .51,  .208}All & \cellcolor[rgb]{ .329,  .51,  .208}All & \cellcolor[rgb]{ .329,  .51,  .208}All & \cellcolor[rgb]{ .329,  .51,  .208}All & None  & \cellcolor[rgb]{ .329,  .51,  .208}All & None  & None  & None  & None \\
				\cite{webber2016college} & author, NLSY, NSCG, ACS  & \cellcolor[rgb]{ .329,  .51,  .208}All & \cellcolor[rgb]{ .663,  .816,  .557}Most & \cellcolor[rgb]{ .663,  .816,  .557}Most  & \cellcolor[rgb]{ .776,  .878,  .706}Some & None  & None  & None  & None  & None & None \\
				\cite{winters2014stem} & U.S. ICE & \cellcolor[rgb]{ .329,  .51,  .208}All & \cellcolor[rgb]{ .329,  .51,  .208}All & \cellcolor[rgb]{ .329,  .51,  .208}All & \cellcolor[rgb]{ .329,  .51,  .208}All & \cellcolor[rgb]{ .776,  .878,  .706}Some & \cellcolor[rgb]{ .776,  .878,  .706}Some & \cellcolor[rgb]{ .776,  .878,  .706}Some & None  & None  & None \\
				\bottomrule
			\end{tabular}%
			\label{tab:stem_litreview}%
			\begin{tablenotes}
				\small 
				\item The table presents a non-exhaustive review of the methods used to classify STEM fields in the literature. All possible degrees are grouped into ten categories: Science, Architecture and Engineering, Chemistry and Pharmacy, Agriculture Veterinary Geology and Biology, Economics and Statistics, Education and Psychology, Law, Literature and Languages, Political sciences and social sciences. The full list of degrees belonging to each group according to the Italian classification can be found in appendix \ref{sec:appendix_degrees}. The grouping and the comparison across papers is inherently lax as not all degrees are available across countries. The label "Most" indicates that almost all the degrees in the group according to my grouping are defined as STEM in the authors' paper. "Some" indicates that only a few of them are defined as STEM. "All" and "None" should be self-explanatory. 
			\end{tablenotes}
		\end{threeparttable}
		}
	\end{table}%
\end{landscape}

\FloatBarrier
\subsection{Marginal Effects of Main Regressions}\label{sec:app_mem}

\begin{table}[htbp]
  \centering
  \caption{$t=1$: Marginal Effects at Means of exclusion restrictions on choice of bachelor}
  \resizebox*{\textwidth}{!}{%
  \setlength{\tabcolsep}{1pt}
    \begin{tabular}{lcccccccccc}
    \toprule
    $Z_j$:  & AVGB  & Ar.Eng. & Ch.Pharm. & Ec.Mgmt. & Ed.Psy. & Law   & Lit.Lan. & Health & Pol.Soc. & Sci.Stat. \\
    \textit{Entry Exams}     & (1)   & (2)   & (3)   & (4)   & (5)   & (6)   & (7)   & (8)   & (9)   & (10) \\
    \midrule
          &       &       &       &       &       &       &       &       &       &  \\
    \textit{Outcomes} &       &       &       &       &       &       &       &       &       &  \\
    Pr(AVGB) & -0.031*** & 0.004 & 0.048*** & 0.016*** & -0.014*** & 0.049*** & -0.170*** & 0.196*** & -0.065*** & 0.046*** \\
          & (0.003) & (0.004) & (0.002) & (0.002) & (0.002) & (0.003) & (0.005) & (0.007) & (0.005) & (0.004) \\
    Pr(ArEng) & -0.061*** & 0.053*** & 0.138*** & 0.025*** & 0.049*** & 0.046*** & 0.173*** & -0.127*** & -0.280*** & -0.022*** \\
          & (0.004) & (0.005) & (0.003) & (0.002) & (0.003) & (0.004) & (0.006) & (0.009) & (0.007) & (0.006) \\
    Pr(ChPh) & -0.005** & 0.044*** & -0.009*** & -0.011*** & -0.026*** & 0.000 & -0.043*** & 0.137*** & 0.041*** & 0.016*** \\
          & (0.002) & (0.003) & (0.002) & (0.001) & (0.002) & (0.002) & (0.003) & (0.005) & (0.004) & (0.003) \\
    Pr(EcMg) & -0.171*** & -0.141*** & 0.064*** & 0.013*** & -0.031*** & 0.026*** & 0.305*** & 0.080*** & 0.046*** & -0.047*** \\
          & (0.004) & (0.005) & (0.003) & (0.002) & (0.003) & (0.004) & (0.006) & (0.009) & (0.006) & (0.005) \\
    Pr(EdPsy) & 0.091*** & -0.097*** & -0.042*** & 0.021*** & 0.035*** & -0.025*** & -0.145*** & 0.042*** & -0.073*** & 0.125*** \\
          & (0.003) & (0.004) & (0.003) & (0.002) & (0.003) & (0.004) & (0.006) & (0.008) & (0.006) & (0.005) \\
    Pr(Law) & 0.049*** & -0.036*** & 0.018*** & 0.005** & -0.059*** & 0.018*** & -0.050*** & -0.056*** & 0.003 & 0.044*** \\
          & (0.003) & (0.004) & (0.003) & (0.002) & (0.002) & (0.003) & (0.005) & (0.008) & (0.006) & (0.005) \\
    Pr(LitLan) & 0.003 & -0.037*** & 0.076*** & 0.055*** & -0.085*** & -0.156*** & 0.264*** & -0.543*** & 0.034*** & -0.016*** \\
          & (0.004) & (0.005) & (0.003) & (0.003) & (0.003) & (0.004) & (0.006) & (0.010) & (0.007) & (0.006) \\
    Pr(Health) & 0.078*** & 0.168*** & -0.286*** & -0.121*** & 0.184*** & 0.119*** & -0.302*** & 0.462*** & 0.171*** & -0.182*** \\
          & (0.004) & (0.005) & (0.003) & (0.002) & (0.003) & (0.004) & (0.007) & (0.010) & (0.007) & (0.006) \\
    Pr(PolSoc) & 0.037*** & 0.014*** & 0.020*** & -0.009*** & -0.063*** & -0.067*** & -0.035*** & -0.210*** & 0.104*** & 0.038*** \\
          & (0.003) & (0.005) & (0.003) & (0.002) & (0.003) & (0.004) & (0.006) & (0.009) & (0.007) & (0.005) \\
    Pr(SciSt) & 0.010*** & 0.026*** & -0.027*** & 0.007*** & 0.008*** & -0.009*** & 0.003 & 0.019*** & 0.017*** & -0.002 \\
          & (0.002) & (0.002) & (0.001) & (0.001) & (0.001) & (0.002) & (0.003) & (0.005) & (0.003) & (0.003) \\
          &       &       &       &       &       &       &       &       &       &  \\
    Observations & 655,847 & 655,847 & 655,847 & 655,847 & 655,847 & 655,847 & 655,847 & 655,847 & 655,847 & 655,847 \\
    \midrule
    \multicolumn{11}{l}{Standard errors in parentheses} \\
    \multicolumn{11}{l}{*** p<0.01, ** p<0.05, * p<0.1} \\
    \end{tabular}%
    }%
  \label{tab:t1_mem}%
\end{table}%

\FloatBarrier
\subsection{Choice of Master Degree -- Regression Tables}\label{sec:appendix_t2} 

\begin{table}[H]
	\centering
	\caption{Choice of master conditional on bachelor in Agriculture, Veterinary, Geology, Biology}
	\begin{tabular}{lccccc}
    \toprule
          & (1)   & (2)   & (3)   & (4)   & (5) \\
    VARIABLES & AVGB  & Arc.Eng. & Ed.Psy. & Health & Sci.Stat. \\
    \midrule
          &       &       &       &       &  \\
    $Z_m$ &       &       &       &       &  \\
    Credit req. (st.): AVGB & 1.141*** & -1.191** & 1.727 & -6.138*** & 6.461*** \\
          & (0.114) & (0.530) & (1.472) & (0.777) & (0.251) \\
    Credit req. (st.): ArEn & 3.897*** & 2.266* & 13.779*** & 17.824*** & 1.226 \\
          & (0.323) & (1.242) & (4.057) & (1.164) & (1.031) \\
    Credit req. (st.): Med & -2.667*** & 7.017*** & -8.738*** & 2.887*** & -1.746*** \\
          & (0.132) & (0.932) & (1.467) & (0.472) & (0.450) \\
    Credit req. (st.): Sci & -1.668*** & 0.931 & -15.627*** & -6.154*** & -9.848*** \\
          & (0.140) & (0.863) & (3.481) & (0.596) & (1.024) \\
    log(distance) & 0.006 & -0.166*** & 0.043 & -0.012 & -0.016 \\
          & (0.007) & (0.029) & (0.041) & (0.021) & (0.018) \\
    $X$   &       &       &       &       &  \\
    HS: grade (st.) & 0.538*** & 0.237*** & 0.124 & 0.314*** & 0.454*** \\
          & (0.016) & (0.076) & (0.090) & (0.048) & (0.041) \\
    HS: humanities & 0.872*** & 0.402 & 0.638*** & 0.547*** & 1.121*** \\
          & (0.055) & (0.277) & (0.225) & (0.141) & (0.135) \\
    HS: science & 0.867*** & 0.140 & -0.390** & 0.488*** & 0.860*** \\
          & (0.032) & (0.163) & (0.191) & (0.102) & (0.088) \\
    Gender (1=female) & -0.451*** & -0.097 & 0.667*** & 0.077 & 0.252** \\
          & (0.041) & (0.205) & (0.252) & (0.134) & (0.107) \\
    Parents: graduate & 0.366*** & 0.387** & 0.361* & 0.406*** & 0.468*** \\
          & (0.040) & (0.185) & (0.202) & (0.114) & (0.095) \\
    Parents: high-rank occ. & -0.033 & 0.198 & 0.209 & 0.089 & -0.126 \\
          & (0.042) & (0.191) & (0.208) & (0.124) & (0.106) \\
          &       &       &       &       &  \\
    Additional Controls & \multicolumn{5}{c}{Yes} \\
    $\Theta$ & \multicolumn{5}{c}{Yes} \\
          &       &       &       &       &  \\
    Constant & 10.364*** & -34.833*** & 17.671** & -41.140*** & 13.965*** \\
          & (0.909) & (5.033) & (7.548) & (3.938) & (2.570) \\
          &       &       &       &       &  \\
    Observations & 32,494 & 32,494 & 32,494 & 32,494 & 32,494 \\
    Pseudo R2 & 0.211 & 0.211 & 0.211 & 0.211 & 0.211 \\
    \midrule
    \multicolumn{6}{l}{Standard errors in parentheses} \\
    \multicolumn{6}{l}{*** p<0.01, ** p<0.05, * p<0.1} \\
    \multicolumn{6}{l}{Excluded category: no master. } \\
\end{tabular}%
	\label{tab:t2_1}%
\end{table}%

\begin{table}[htbp]
	\centering
	\caption{Choice of master conditional on bachelor in Architecture and Engineering}
	\begin{tabular}{lcccc}
		    \toprule
          & (1)   & (2)   & (3)   & (4) \\
    VARIABLES & Arc.Eng. & Chem.Pharm. & Lit.Lang. & Sci.Stat. \\
    \midrule
          &       &       &       &  \\
    $Z_m$ &       &       &       &  \\
    Credit req. (st.): ArEn & -0.053** & -2.726*** & -6.861*** & -0.268* \\
          & (0.024) & (0.144) & (0.687) & (0.143) \\
    Credit req. (st.): ChPh & -6.053*** & -1.743* & 23.136*** & -5.426*** \\
          & (0.127) & (0.977) & (2.300) & (1.132) \\
    Credit req. (st.): Hum & -0.734*** & 15.535*** & 14.477*** & 8.470*** \\
          & (0.127) & (2.934) & (2.252) & (2.453) \\
    Credit req. (st.): Sci & 0.649*** & 0.239 & 5.577*** & -1.262* \\
          & (0.053) & (0.729) & (0.521) & (0.754) \\
    log(distance) & -0.015*** & 0.005 & 0.004 & 0.005 \\
          & (0.005) & (0.022) & (0.028) & (0.037) \\
    $X$   &       &       &       &  \\
    HS: grade (st.) & 0.737*** & 0.742*** & 0.634*** & 0.808*** \\
          & (0.010) & (0.045) & (0.072) & (0.079) \\
    HS: humanities & 0.900*** & 1.340*** & 1.124*** & 0.849*** \\
          & (0.044) & (0.170) & (0.192) & (0.302) \\
    HS: science & 1.023*** & 1.425*** & 0.564*** & 1.273*** \\
          & (0.021) & (0.098) & (0.148) & (0.164) \\
    Gender (1=female) & -0.253*** & 0.208* & 0.549*** & 0.604*** \\
          & (0.030) & (0.113) & (0.194) & (0.179) \\
    Parents: graduate & 0.368*** & 0.385*** & 0.517*** & 0.544*** \\
          & (0.025) & (0.089) & (0.151) & (0.150) \\
    Parents: high-rank occ. & 0.074*** & -0.033 & -0.029 & 0.015 \\
          & (0.026) & (0.097) & (0.160) & (0.162) \\
          &       &       &       &  \\
    Additional Controls & \multicolumn{4}{c}{Yes} \\
    $\Theta$ & \multicolumn{4}{c}{Yes} \\
          &       &       &       &  \\
    Constant & 4.078*** & -2.262 & -18.458*** & -0.826 \\
          & (0.158) & (1.656) & (1.929) & (1.670) \\
          &       &       &       &  \\
    Observations & 79,817 & 79,817 & 79,817 & 79,817 \\
    Pseudo R2 & 0.241 & 0.241 & 0.241 & 0.241 \\
    \midrule
    \multicolumn{5}{l}{Standard errors in parentheses} \\
    \multicolumn{5}{l}{*** p<0.01, ** p<0.05, * p<0.1} \\
    \multicolumn{5}{l}{Excluded category: no master. } \\
	\end{tabular}%
	\label{tab:t2_2}%
\end{table}%

\begin{table}[htbp]
	\centering
	\setlength{\tabcolsep}{12pt}
	\caption{Probability of choosing a master degree given a bachelor in Chemistry and Pharmacy}
	
	\begin{tabular}{lccc}
	   \toprule
          & (1)   & (2)   & (3) \\
    VARIABLES & AVGB  & Chem.Pharm. & Health \\
    \midrule
          &       &       &  \\
    $Z_m$ &       &       &  \\
    Credit req. (st.): ChPh & 0.423*** & -2.994*** & 7.038 \\
          & (0.125) & (0.186) & (293.390) \\
    log(distance) & -0.072 & -0.013 & -0.037 \\
          & (0.044) & (0.021) & (0.037) \\
    $X$   &       &       &  \\
    HS: grade (st.) & 0.688*** & 0.770*** & 0.779*** \\
          & (0.112) & (0.047) & (0.088) \\
    HS: humanities & 0.346 & 0.877*** & 0.864*** \\
          & (0.315) & (0.176) & (0.239) \\
    HS: science & -0.049 & 0.866*** & 0.622*** \\
          & (0.217) & (0.093) & (0.167) \\
    Gender (1=female) & 0.300 & -0.203 & -0.781*** \\
          & (0.283) & (0.126) & (0.244) \\
    Parents: graduate & 0.464* & 0.403*** & -0.013 \\
          & (0.248) & (0.115) & (0.230) \\
    Parents: high-rank occ. & 0.197 & 0.080 & -0.002 \\
          & (0.272) & (0.130) & (0.261) \\
          &       &       &  \\
    Additional Controls & \multicolumn{3}{c}{Yes} \\
    $\Theta$ & \multicolumn{3}{c}{Yes} \\
          &       &       &  \\
    Constant & -3.444*** & -5.804*** & -23.672 \\
          & (0.911) & (0.580) & (1,430.729) \\
          &       &       &  \\
    Observations & 7,398 & 7,398 & 7,398 \\
    Pseudo R2 & 0.571 & 0.571 & 0.571 \\
    \midrule
    \multicolumn{4}{l}{Standard errors in parentheses} \\
    \multicolumn{4}{l}{*** p<0.01, ** p<0.05, * p<0.1} \\
    \multicolumn{4}{l}{Excluded category: no master. } \\
\end{tabular}%
	\label{tab:t2_3}%
\end{table}%

\begin{table}[htbp]
	\centering
	\setlength{\tabcolsep}{5pt}
	\caption{Probability of choosing a master degree given a bachelor in Economics and Management}
	\begin{tabular}{lccccc}
    \toprule
          & (1)   & (2)   & (3)   & (4)   & (5) \\
    VARIABLES & Econ.Mgmt. & Educ.Psy. & Law   & Pol.Soc. & Sci.Stat. \\
    \midrule
          &       &       &       &       &  \\
    $Z_m$ &       &       &       &       &  \\
    Credit req. (st.): PlSc & -18.471*** & -15.756*** & -13.175*** & -18.097*** & -13.711*** \\
          & (0.157) & (1.497) & (1.264) & (0.554) & (0.815) \\
    log(distance) & 0.012** & -0.125*** & 0.040** & -0.054*** & -0.025 \\
          & (0.005) & (0.046) & (0.019) & (0.016) & (0.028) \\
    $X$   &       &       &       &       &  \\
    HS: grade (st.) & 0.509*** & 0.027 & -0.009 & 0.237*** & 0.515*** \\
          & (0.010) & (0.101) & (0.075) & (0.034) & (0.053) \\
    HS: humanities & 0.908*** & 1.391*** & 1.506*** & 1.160*** & 0.875*** \\
          & (0.041) & (0.282) & (0.211) & (0.105) & (0.206) \\
    HS: science & 0.763*** & 0.771*** & -0.160 & 0.538*** & 1.128*** \\
          & (0.021) & (0.213) & (0.182) & (0.072) & (0.106) \\
    Gender (1=female) & -0.393*** & 1.282*** & -1.303*** & -0.065 & -0.432*** \\
          & (0.026) & (0.258) & (0.221) & (0.088) & (0.133) \\
    Parents: graduate & 0.377*** & 0.698*** & -0.091 & 0.608*** & 0.442*** \\
          & (0.026) & (0.225) & (0.216) & (0.080) & (0.122) \\
    Parents: high-rank occ. & 0.076*** & -0.392 & -0.868*** & -0.105 & -0.194 \\
          & (0.025) & (0.248) & (0.261) & (0.085) & (0.130) \\
          &       &       &       &       &  \\
    Additional Controls & \multicolumn{5}{c}{Yes} \\
    $\Theta$ & \multicolumn{5}{c}{Yes} \\
          &       &       &       &       &  \\
    Constant & -40.752*** & -44.205*** & -31.250*** & -41.948*** & -34.294*** \\
          & (0.365) & (3.554) & (2.950) & (1.284) & (1.900) \\
          &       &       &       &       &  \\
    Observations & 75,993 & 75,993 & 75,993 & 75,993 & 75,993 \\
    Pseudo R2 & 0.220 & 0.220 & 0.220 & 0.220 & 0.220 \\
    \midrule
    \multicolumn{6}{l}{Standard errors in parentheses} \\
    \multicolumn{6}{l}{*** p<0.01, ** p<0.05, * p<0.1} \\
    \multicolumn{6}{l}{Excluded category: no master. } \\
\end{tabular}%
	\label{tab:t2_4}%
\end{table}%

\begin{table}[htbp]
	\centering
	\setlength{\tabcolsep}{10pt}
	\caption{Probability of choosing a master degree given a bachelor in Physical Education, Teaching and Psychology}
	\begin{tabular}{lcccc}
    \toprule
          & (1)   & (2)   & (3)   & (4) \\
    VARIABLES & Educ.Psy. & Lit.Lang. & Health & Pol.Soc. \\
    \midrule
          &       &       &       &  \\
    $Z_m$ &       &       &       &  \\
    Credit req. (st.): EdPsy & 4.587*** & -0.610*** & 0.610 & -0.581*** \\
          & (0.081) & (0.137) & (0.573) & (0.094) \\
    Credit req. (st.): PlSc & -1.547*** & -0.731*** & -0.842*** & -0.512*** \\
          & (0.021) & (0.162) & (0.210) & (0.111) \\
    log(distance) & -0.016*** & -0.081*** & -0.055 & -0.031 \\
          & (0.004) & (0.028) & (0.035) & (0.019) \\
    $X$   &       &       &       &  \\
    HS: grade (st.) & 0.365*** & 0.194*** & 0.018 & 0.312*** \\
          & (0.011) & (0.070) & (0.102) & (0.048) \\
    HS: humanities & 0.786*** & 0.878*** & 0.362 & 0.574*** \\
          & (0.032) & (0.182) & (0.298) & (0.136) \\
    HS: science & 0.670*** & -0.055 & -0.173 & 0.313*** \\
          & (0.023) & (0.170) & (0.237) & (0.107) \\
    Gender (1=female) & -0.446*** & -0.454** & -0.661** & -0.548*** \\
          & (0.032) & (0.199) & (0.270) & (0.140) \\
    Parents: graduate & 0.371*** & 0.344* & -0.123 & 0.417*** \\
          & (0.029) & (0.186) & (0.315) & (0.123) \\
    Parents: high-rank occ. & 0.105*** & -0.168 & -0.711* & 0.185 \\
          & (0.029) & (0.203) & (0.367) & (0.128) \\
          &       &       &       &  \\
    Additional Controls & \multicolumn{4}{c}{Yes} \\
    $\Theta$ & \multicolumn{4}{c}{Yes} \\
          &       &       &       &  \\
    Constant & 3.314*** & -4.234*** & -5.255*** & -2.900*** \\
          & (0.134) & (0.702) & (1.131) & (0.457) \\
          &       &       &       &  \\
    Observations & 62,741 & 62,741 & 62,741 & 62,741 \\
    Pseudo R2 & 0.223 & 0.223 & 0.223 & 0.223 \\
    \midrule
    \multicolumn{5}{l}{Standard errors in parentheses} \\
    \multicolumn{5}{l}{*** p<0.01, ** p<0.05, * p<0.1} \\
    \multicolumn{5}{l}{Excluded category: no master. } \\
	\end{tabular}%
	\label{tab:t2_5}%
\end{table}%

\begin{table}[htbp]
	\centering
	\setlength{\tabcolsep}{10pt}
	\caption{Probability of choosing a master degree given a bachelor in Law}
	\begin{tabular}{lcccc}
    \toprule
          & (1)   & (2)   & (3)   & (4) \\
    VARIABLES & Econ.Mgmt. & Educ.Psy. & Law   & Pol.Soc. \\
    \midrule
          &       &       &       &  \\
    $Z_m$ &       &       &       &  \\
    Credit req. (st.): PlSc & -3.800*** & -2.129*** & -5.008*** & -2.087*** \\
          & (0.149) & (0.418) & (0.154) & (0.160) \\
    log(distance) & -0.040*** & 0.032 & 0.012 & -0.032* \\
          & (0.013) & (0.057) & (0.010) & (0.017) \\
    $X$   &       &       &       &  \\
    HS: grade (st.) & 0.540*** & 0.233** & 0.270*** & 0.311*** \\
          & (0.034) & (0.098) & (0.033) & (0.037) \\
    HS: humanities & 0.166 & 1.392*** & 0.814*** & 0.893*** \\
          & (0.115) & (0.240) & (0.097) & (0.104) \\
    HS: science & 0.796*** & 0.755*** & 0.571*** & 0.592*** \\
          & (0.075) & (0.227) & (0.076) & (0.084) \\
    Gender (1=female) & -0.137 & 0.078 & -0.640*** & -0.509*** \\
          & (0.088) & (0.257) & (0.088) & (0.095) \\
    Parents: graduate & 0.592*** & 0.669*** & 0.335*** & 0.404*** \\
          & (0.091) & (0.242) & (0.096) & (0.102) \\
    Parents: high-rank occ. & 0.411*** & 0.121 & -0.042 & -0.001 \\
          & (0.091) & (0.258) & (0.100) & (0.106) \\
          &       &       &       &  \\
    Additional Controls & \multicolumn{4}{c}{Yes} \\
    $\Theta$ & \multicolumn{4}{c}{Yes} \\
          &       &       &       &  \\
    Constant & -13.281*** & -28.260 & -12.949*** & -8.243*** \\
          & (0.567) & (374.262) & (0.494) & (0.551) \\
          &       &       &       &  \\
    Observations & 10,882 & 10,882 & 10,882 & 10,882 \\
    Pseudo R2 & 0.182 & 0.182 & 0.182 & 0.182 \\
    \midrule
    \multicolumn{5}{l}{Standard errors in parentheses} \\
    \multicolumn{5}{l}{*** p<0.01, ** p<0.05, * p<0.1} \\
    \multicolumn{5}{l}{Excluded category: no master. } \\
	\end{tabular}%
	\label{tab:t2_6}%
\end{table}%

\begin{table}[htbp]
	\centering
	\setlength{\tabcolsep}{5pt}
	\caption{Probability of choosing a master degree given a bachelor in Literature and Languages}
	\begin{tabular}{lcccccc}
    \toprule
          & (1)   & (2)   & (3)   & (4)   & (5)   & (6) \\
    VARIABLES & Arc.Eng. & Econ.Mgmt. & Educ.Psy. & Lit.Lang. & Pol.Soc. & Sci.Stat. \\
    \midrule
          &       &       &       &       &       &  \\
    $Z_m$ &       &       &       &       &       &  \\
    Credit req. (st.): EdPs & 3.252*** & 1.105*** & -1.907*** & -1.582*** & 0.219* & 0.222 \\
          & (1.001) & (0.371) & (0.208) & (0.052) & (0.123) & (0.888) \\
    Credit req. (st.): Hum & 7.735*** & -3.854*** & 0.118 & -1.500*** & -4.206*** & 0.931 \\
          & (0.970) & (0.611) & (0.513) & (0.089) & (0.207) & (0.845) \\
    Credit req. (st.): PlSc & -1.525** & 2.315*** & -0.244 & -0.122** & 1.966*** & -2.063*** \\
          & (0.633) & (0.412) & (0.330) & (0.057) & (0.138) & (0.543) \\
    Credit req. (st.): Sci & -0.176 & 2.466*** & -0.433 & -2.117*** & 0.411** & -0.963 \\
          & (1.986) & (0.521) & (0.520) & (0.089) & (0.173) & (0.900) \\
    log(distance) & -0.041 & 0.045** & -0.005 & -0.008** & -0.030*** & -0.037 \\
          & (0.042) & (0.022) & (0.015) & (0.004) & (0.006) & (0.038) \\
    $X$   &       &       &       &       &       &  \\
    HS: grade (st.) & 0.580*** & 0.428*** & -0.123*** & 0.565*** & 0.381*** & 0.251*** \\
          & (0.096) & (0.042) & (0.044) & (0.008) & (0.015) & (0.083) \\
    HS: humanities & 0.315 & 0.839*** & 0.367*** & 1.115*** & 0.846*** & 0.593*** \\
          & (0.288) & (0.108) & (0.115) & (0.021) & (0.040) & (0.224) \\
    HS: science & 0.615*** & 0.784*** & 0.250** & 0.753*** & 0.836*** & 0.605*** \\
          & (0.213) & (0.093) & (0.102) & (0.019) & (0.035) & (0.187) \\
    Gender (1=female) & -0.833*** & -0.068 & 0.073 & -0.393*** & -0.740*** & -1.032*** \\
          & (0.251) & (0.125) & (0.135) & (0.024) & (0.043) & (0.252) \\
    Parents: graduate & 0.769*** & 0.373*** & 0.136 & 0.403*** & 0.244*** & 0.439** \\
          & (0.213) & (0.096) & (0.111) & (0.020) & (0.036) & (0.199) \\
    Parents: high-rank occ. & 0.189 & 0.020 & -0.209* & -0.058*** & 0.052 & -0.271 \\
          & (0.219) & (0.102) & (0.121) & (0.021) & (0.038) & (0.233) \\
          &       &       &       &       &       &  \\
    Additional Controls & \multicolumn{6}{c}{Yes} \\
    $\Theta$ & \multicolumn{6}{c}{Yes} \\
          &       &       &       &       &       &  \\
    Constant & -3.131** & -6.094*** & -1.798*** & -0.913*** & 0.201 & -7.927*** \\
          & (1.546) & (0.588) & (0.528) & (0.105) & (0.189) & (1.456) \\
          &       &       &       &       &       &  \\
    Observations & 90,681 & 90,681 & 90,681 & 90,681 & 90,681 & 90,681 \\
    Pseudo R2 & 0.108 & 0.108 & 0.108 & 0.108 & 0.108 & 0.108 \\
    \midrule
    \multicolumn{7}{l}{Standard errors in parentheses} \\
    \multicolumn{7}{l}{*** p<0.01, ** p<0.05, * p<0.1} \\
    \multicolumn{7}{l}{Excluded category: no master. } \\
	\end{tabular}%

	\label{tab:t2_7}%
\end{table}%


\begin{table}[htbp]
	\centering
	\setlength{\tabcolsep}{15pt}
	\caption{Probability of choosing a master degree given a bachelor in Health}
	\begin{tabular}{lccc}
		    \toprule
          & (1)   & (2)   & (3) \\
    VARIABLES & AVGB  & Educ.Psy. & Health \\
    \midrule
          &       &       &  \\
    $Z_m$ &       &       &  \\
    Credit req. (st.): AVGB & -4.685*** & 9.015*** & -4.584*** \\
          & (0.189) & (1.686) & (0.160) \\
    log(distance) & -0.017 & -0.033 & 0.011* \\
          & (0.019) & (0.023) & (0.006) \\
    $X$   &       &       &  \\
    HS: grade (st.) & 0.403*** & -0.032 & 0.115*** \\
          & (0.054) & (0.062) & (0.015) \\
    HS: humanities & 0.205 & -0.376* & 0.147*** \\
          & (0.180) & (0.219) & (0.049) \\
    HS: science & 0.401*** & -0.402*** & -0.193*** \\
          & (0.109) & (0.132) & (0.032) \\
    Gender (1=female) & -0.258* & -0.006 & -0.027 \\
          & (0.142) & (0.164) & (0.040) \\
    Parents: graduate & 0.186 & 0.309* & -0.098** \\
          & (0.143) & (0.174) & (0.047) \\
    Parents: high-rank occ. & 0.063 & 0.200 & -0.070 \\
          & (0.157) & (0.183) & (0.050) \\
          &       &       &  \\
    Additional Controls & \multicolumn{3}{c}{Yes} \\
    $\Theta$ & \multicolumn{3}{c}{Yes} \\
          &       &       &  \\
    Constant & -5.587*** & -1.271 & -4.954*** \\
          & (0.453) & (0.806) & (0.158) \\
          &       &       &  \\
    Observations & 81,883 & 81,883 & 81,883 \\
    Pseudo R2 & 0.0591 & 0.0591 & 0.0591 \\
    \midrule
    \multicolumn{4}{l}{Standard errors in parentheses} \\
    \multicolumn{4}{l}{*** p<0.01, ** p<0.05, * p<0.1} \\
    \multicolumn{4}{l}{Excluded category: no master. } \\
	\end{tabular}%
	\label{tab:t2_8}%
\end{table}%

\begin{table}[htbp] 
	\centering
	\setlength{\tabcolsep}{5pt}
	\caption{Probability of choosing a master degree given a bachelor in Political and Social Sciences}
	\begin{tabular}{lcccccc}
    \toprule
          & (1)   & (2)   & (3)   & (4)   & (5)   & (6) \\
    VARIABLES & Econ.Mgmt. & Educ.Psy. & Law   & Lit.Lang. & Pol.Soc. & Sci.Stat. \\
    \midrule
          &       &       &       &       &       &  \\
    $Z_m$ &       &       &       &       &       &  \\
    Credit req. (st.): EcMg & -1.417** & 4.024*** & 8.987*** & 1.368*** & 3.404*** & 4.183** \\
          & (0.661) & (0.557) & (1.241) & (0.453) & (0.154) & (1.859) \\
    Credit req. (st.): EdPs & -2.027*** & -4.539*** & -7.398*** & -3.631*** & -2.605*** & -3.432*** \\
          & (0.733) & (0.345) & (0.867) & (0.681) & (0.086) & (1.207) \\
    Credit req. (st.): Law & -1.630*** & -3.461*** & -7.974*** & -2.296*** & -2.837*** & -3.141** \\
          & (0.613) & (0.355) & (0.829) & (0.489) & (0.103) & (1.262) \\
    Credit req. (st.): Hum & 13.178*** & -15.963*** & -15.441*** & -4.140** & -3.580*** & -10.459 \\
          & (2.992) & (1.868) & (4.253) & (1.981) & (0.608) & (7.022) \\
    Credit req. (st.): PlSc & 6.158*** & -8.130*** & -7.197*** & 2.305*** & -3.379*** & -4.010 \\
          & (1.142) & (0.674) & (1.688) & (0.725) & (0.233) & (2.711) \\
    log(distance) & -0.024** & -0.000 & 0.029*** & -0.013 & -0.003 & 0.104 \\
          & (0.011) & (0.018) & (0.010) & (0.009) & (0.004) & (0.066) \\
    $X$   &       &       &       &       &       &  \\
    HS: grade (st.) & 0.433*** & 0.385*** & 0.228*** & 0.604*** & 0.468*** & 0.795*** \\
          & (0.029) & (0.049) & (0.031) & (0.026) & (0.010) & (0.104) \\
    HS: humanities & 0.342*** & 0.615*** & 0.053 & 0.864*** & 0.790*** & -0.079 \\
          & (0.086) & (0.134) & (0.089) & (0.066) & (0.028) & (0.369) \\
    HS: science & 0.794*** & 0.646*** & 0.249*** & 0.483*** & 0.668*** & 0.610*** \\
          & (0.061) & (0.108) & (0.067) & (0.058) & (0.022) & (0.210) \\
    Gender (1=female) & -0.192*** & 0.616*** & -0.497*** & -0.135** & -0.165*** & -1.282*** \\
          & (0.073) & (0.145) & (0.083) & (0.067) & (0.027) & (0.266) \\
    Parents: graduate & 0.222*** & 0.398*** & -0.015 & 0.467*** & 0.274*** & 0.210 \\
          & (0.068) & (0.118) & (0.082) & (0.059) & (0.025) & (0.247) \\
    Parents: high-rank occ. & -0.070 & 0.039 & -0.240*** & -0.079 & -0.078*** & -0.260 \\
          & (0.072) & (0.124) & (0.089) & (0.063) & (0.026) & (0.267) \\
          &       &       &       &       &       &  \\
    Additional Controls & \multicolumn{6}{c}{Yes} \\
    $\Theta$ & \multicolumn{6}{c}{Yes} \\
          &       &       &       &       &       &  \\
    Constant & 20.289*** & -29.793*** & -11.597** & 5.244** & -4.120*** & -14.040* \\
          & (3.374) & (1.974) & (4.706) & (2.182) & (0.667) & (7.746) \\
          &       &       &       &       &       &  \\
    Observations & 65,798 & 65,798 & 65,798 & 65,798 & 65,798 & 65,798 \\
    Pseudo R2 & 0.162 & 0.162 & 0.162 & 0.162 & 0.162 & 0.162 \\
    \midrule
    \multicolumn{7}{l}{Standard errors in parentheses} \\
    \multicolumn{7}{l}{*** p<0.01, ** p<0.05, * p<0.1} \\
    \multicolumn{7}{l}{Excluded category: no master. } \\
	\end{tabular}%
	\label{tab:t2_9}%
\end{table}%

\begin{table}[htbp] 
	\centering
	\small
	\setlength{\tabcolsep}{5pt}
	\caption{Probability of choosing a master degree given a bachelor in Science and Statistics}
	\begin{tabular}{lcccccc}
    \toprule
          & (1)   & (2)   & (3)   & (4)   & (5)   & (6) \\
    VARIABLES & AVGB  & Arc.Eng. & Chem.Phar. & Econ.Mgmt. & Pol.Soc. & Sci.Stat. \\
    \midrule
          &       &       &       &       &       &  \\
    $Z_m$ &       &       &       &       &       &  \\
    Credit req. (st.): ChPh & -14.708*** & 2.612*** & -81.978 & 7.585*** & 1.568* & -4.359*** \\
          & (1.878) & (0.842) & (2,381.101) & (1.886) & (0.911) & (0.173) \\
    Credit req. (st.): Sci & -0.320 & 1.618*** & -21.711 & 0.989*** & 0.567* & -0.879*** \\
          & (1.022) & (0.272) & (709.538) & (0.279) & (0.335) & (0.062) \\
    log(distance) & -0.042* & -0.021 & -0.067 & 0.072 & -0.006 & -0.005 \\
          & (0.024) & (0.049) & (0.042) & (0.076) & (0.037) & (0.009) \\
    $X$   &       &       &       &       &       &  \\
    HS: grade (st.) & 0.031 & 0.433*** & 0.386*** & 0.606*** & 0.183** & 0.736*** \\
          & (0.052) & (0.102) & (0.090) & (0.100) & (0.086) & (0.019) \\
    HS: humanities & 1.326*** & 0.698* & 0.271 & 0.928** & 0.736** & 0.829*** \\
          & (0.187) & (0.363) & (0.339) & (0.432) & (0.364) & (0.084) \\
    HS: science & 1.020*** & -0.047 & 0.370** & 1.046*** & 0.377** & 0.955*** \\
          & (0.117) & (0.219) & (0.186) & (0.205) & (0.179) & (0.038) \\
    Gender (1=female) & 1.020*** & -0.312 & 0.484* & 0.197 & -0.662*** & 0.140*** \\
          & (0.143) & (0.276) & (0.252) & (0.259) & (0.233) & (0.053) \\
    Parents: graduate & -0.181 & 0.530** & 0.316 & 0.140 & 0.110 & 0.365*** \\
          & (0.119) & (0.235) & (0.198) & (0.232) & (0.216) & (0.045) \\
    Parents: high-rank occ. & 0.021 & 0.610** & -0.110 & 0.252 & 0.019 & 0.008 \\
          & (0.131) & (0.239) & (0.227) & (0.243) & (0.231) & (0.050) \\
          &       &       &       &       &       &  \\
    Additional Controls & \multicolumn{6}{c}{Yes} \\
    $\Theta$ & \multicolumn{6}{c}{Yes} \\
          &       &       &       &       &       &  \\
    Constant & 9.007*** & -4.773*** & 32.378 & -13.920*** & -1.120 & 2.247*** \\
          & (0.791) & (1.286) & (896.781) & (2.239) & (0.994) & (0.222) \\
          &       &       &       &       &       &  \\
    Observations & 20,721 & 20,721 & 20,721 & 20,721 & 20,721 & 20,721 \\
    Pseudo R2 & 0.300 & 0.300 & 0.300 & 0.300 & 0.300 & 0.300 \\
    \midrule
    \multicolumn{7}{l}{Standard errors in parentheses} \\
    \multicolumn{7}{l}{*** p<0.01, ** p<0.05, * p<0.1} \\
    \multicolumn{7}{l}{Excluded category: no master. } \\
	\end{tabular}%
	\label{tab:t2_10}%
\end{table}%

\FloatBarrier
\subsection{Additional Results}\label{sec:appendix_addresults}

\begin{table}[htbp]
	\centering
	\footnotesize
	\setlength{\tabcolsep}{5pt}
	\caption{Summary of treatments $D_{jm}$ and probabilities $P_{jm}$}
    \begin{tabular}{llcccccc}
	\toprule
	\#    & $(j, m)$ & \multicolumn{2}{c}{$D_{jm}$} & \multicolumn{3}{c}{$P_{jm}$} & $P_{jm}-D_{jm}$ \\
	\cmidrule(lr){3-4} \cmidrule(lr){5-7}
	&       & Mean  & Std. Dev. & Mean  & Std. Dev. & Max   &  \\
	\midrule
	1     & (AVGB, No Master) & 0.0128 & (0.1124) & 0.0019 & (0.0057) & 0.1613 & -0.0109 \\
	2     & (AVGB, AVGB) & 0.0401 & (0.1963) & 0.0141 & (0.023) & 0.1944 & -0.026 \\
	3     & (AVGB, Arch.Eng.) & 0.0003 & (0.0183) & 0     & (0.0002) & 0.0236 & -0.0003 \\
	4     & (AVGB, Educ.Psy.) & 0.0003 & (0.0166) & 0.0071 & (0.0205) & 0.1911 & 0.0069 \\
	5     & (AVGB, Pol.Soc.) & 0.0009 & (0.0308) & 0.0001 & (0.0008) & 0.0586 & -0.0009 \\
	6     & (AVGB, Sci.Stat.) & 0.0014 & (0.0377) & 0.0326 & (0.0317) & 0.2041 & 0.0312 \\
	7     & (Arch.Eng., No Master) & 0.034 & (0.1812) & 0.0631 & (0.0722) & 0.5837 & 0.0291 \\
	8     & (Arch.Eng., Arch.Eng.) & 0.1217 & (0.327) & 0.0939 & (0.1237) & 0.7117 & -0.0278 \\
	9     & (Arch.Eng., Chem.Pharm.) & 0.0012 & (0.0344) & 0.0004 & (0.0023) & 0.066 & -0.0008 \\
	10    & (Arch.Eng., Lit.Lang.) & 0.0004 & (0.0209) & 0.0001 & (0.0009) & 0.1028 & -0.0004 \\
	11    & (Arch.Eng., Sci.Stat.) & 0.0004 & (0.0196) & 0.0002 & (0.0012) & 0.0995 & -0.0002 \\
	12    & (Chem.Pharm., No Master) & 0.0059 & (0.0769) & 0.0264 & (0.023) & 0.2088 & 0.0204 \\
	13    & (Chem.Pharm., AVGB) & 0.0002 & (0.0134) & 0.0014 & (0.0031) & 0.0839 & 0.0012 \\
	14    & (Chem.Pharm., Chem.Pharm.) & 0.0315 & (0.1746) & 0.0085 & (0.0216) & 0.1834 & -0.023 \\
	15    & (Chem.Pharm., Health) & 0.0004 & (0.0199) & 0.0017 & (0.0057) & 0.1438 & 0.0013 \\
	16    & (Econ.Mgmt., No Master) & 0.0424 & (0.2015) & 0.0724 & (0.068) & 0.5057 & 0.03 \\
	17    & (Econ.Mgmt., Econ.Mgmt.) & 0.0705 & (0.256) & 0.042 & (0.0611) & 0.462 & -0.0285 \\
	18    & (Econ.Mgmt., Educ.Psy.) & 0.0002 & (0.0137) & 0.0001 & (0.0002) & 0.0115 & -0.0001 \\
	19    & (Econ.Mgmt., Law) & 0.0003 & (0.0178) & 0.0001 & (0.0006) & 0.0641 & -0.0002 \\
	20    & (Econ.Mgmt., Pol.Soc.) & 0.0018 & (0.0419) & 0.0011 & (0.002) & 0.0523 & -0.0007 \\
	21    & (Econ.Mgmt., Sci.Stat.) & 0.0007 & (0.0264) & 0.0002 & (0.0006) & 0.0281 & -0.0005 \\
	22    & (Educ.Psy., No Master) & 0.0435 & (0.204) & 0.0449 & (0.0559) & 0.5477 & 0.0014 \\
	23    & (Educ.Psy., Educ.Psy.) & 0.0703 & (0.2556) & 0.0547 & (0.061) & 0.5451 & -0.0156 \\
	24    & (Educ.Psy., Lit.Lang.) & 0.0004 & (0.0195) & 0.0035 & (0.0116) & 0.313 & 0.0031 \\
	25    & (Educ.Psy., Health) & 0.0002 & (0.0138) & 0.0002 & (0.0008) & 0.0644 & 0 \\
	26    & (Educ.Psy., Pol.Soc.) & 0.0008 & (0.0286) & 0.0119 & (0.0359) & 0.4998 & 0.0111 \\
	27    & (Law, No Master) & 0.0123 & (0.1101) & 0.052 & (0.0512) & 0.5102 & 0.0398 \\
	28    & (Law, Econ.Mgmt.) & 0.0022 & (0.0472) & 0.0097 & (0.0168) & 0.3091 & 0.0074 \\
	29    & (Law, Educ.Psy.) & 0.0002 & (0.0139) & 0.0009 & (0.0025) & 0.0933 & 0.0007 \\
	30    & (Law, Law) & 0.0713 & (0.2573) & 0.0186 & (0.0435) & 0.4688 & -0.0527 \\
	31    & (Law, Pol.Soc.) & 0.0017 & (0.0409) & 0.0065 & (0.0121) & 0.2348 & 0.0048 \\
	32    & (Lit.Lang., No Master) & 0.0585 & (0.2346) & 0.0649 & (0.0539) & 0.5661 & 0.0064 \\
	33    & (Lit.Lang., Arch.Eng.) & 0.0002 & (0.0136) & 0.0037 & (0.0145) & 0.5213 & 0.0035 \\
	34    & (Lit.Lang., Econ.Mgmt.) & 0.0011 & (0.0325) & 0.0007 & (0.0019) & 0.0706 & -0.0004 \\
	35    & (Lit.Lang., Educ.Psy.) & 0.0009 & (0.0301) & 0.0027 & (0.0058) & 0.1723 & 0.0018 \\
	36    & (Lit.Lang., Lit.Lang.) & 0.0686 & (0.2527) & 0.0614 & (0.0604) & 0.6429 & -0.0071 \\
	37    & (Lit.Lang., Pol.Soc.) & 0.0088 & (0.0935) & 0.0046 & (0.0071) & 0.1531 & -0.0042 \\
	38    & (Lit.Lang., Sci.Stat.) & 0.0003 & (0.0159) & 0.0002 & (0.0016) & 0.1725 & 0 \\
	39    & (Health, No Master) & 0.1155 & (0.3196) & 0.1337 & (0.1242) & 0.7477 & 0.0183 \\
	40    & (Health, AVGB) & 0.0006 & (0.0248) & 0.0025 & (0.0073) & 0.3174 & 0.0018 \\
	41    & (Health, Educ.Psy.) & 0.0005 & (0.0218) & 0.0004 & (0.0011) & 0.0337 & 0 \\
	42    & (Health, Health) & 0.0428 & (0.2024) & 0.0227 & (0.041) & 0.6067 & -0.0201 \\
	43    & (Pol.Soc., No Master) & 0.0534 & (0.2248) & 0.0124 & (0.0309) & 0.3077 & -0.041 \\
	44    & (Pol.Soc., Econ.Mgmt.) & 0.0024 & (0.0487) & 0.0596 & (0.0545) & 0.3717 & 0.0572 \\
	45    & (Pol.Soc., Educ.Psy.) & 0.0009 & (0.0302) & 0.008 & (0.0259) & 0.2688 & 0.0071 \\
	46    & (Pol.Soc., Law) & 0.002 & (0.0452) & 0.0104 & (0.0276) & 0.3019 & 0.0083 \\
	47    & (Pol.Soc., Lit.Lang.) & 0.003 & (0.0544) & 0.0044 & (0.0141) & 0.2164 & 0.0014 \\
	48    & (Pol.Soc., Pol.Soc.) & 0.0386 & (0.1927) & 0.0054 & (0.0155) & 0.2249 & -0.0332 \\
	49    & (Pol.Soc., Sci.Stat.) & 0.0002 & (0.0131) & 0.0003 & (0.0015) & 0.0539 & 0.0001 \\
	50    & (Sci.Stat., No Master) & 0.0131 & (0.1137) & 0.0132 & (0.0174) & 0.2344 & 0.0001 \\
	51    & (Sci.Stat., AVGB) & 0.0015 & (0.0393) & 0.0001 & (0.0018) & 0.0881 & -0.0014 \\
	52    & (Sci.Stat., Arch.Eng.) & 0.0002 & (0.0132) & 0.0011 & (0.0026) & 0.0651 & 0.0009 \\
	53    & (Sci.Stat., Chem.Pharm.) & 0.0003 & (0.0167) & 0.0097 & (0.0217) & 0.2926 & 0.0094 \\
	54    & (Sci.Stat., Econ.Mgmt.) & 0.0002 & (0.0137) & 0.0006 & (0.0014) & 0.0462 & 0.0004 \\
	55    & (Sci.Stat., Pol.Soc.) & 0.0002 & (0.0156) & 0.0005 & (0.0011) & 0.055 & 0.0002 \\
	56    & (Sci.Stat., Sci.Stat.) & 0.0161 & (0.1257) & 0.0064 & (0.0115) & 0.218 & -0.0096 \\
	\bottomrule
\end{tabular}%
	\label{tab:diff_D_P}%
	\caption*{\footnotesize Summary statistics for the vector of treatments $D_{jm}$ and probabilities $P_{jm}$ for 56 combinations of bachelor's and master's degrees. Treatments $D_{jm}$ take values 0 and 1. The minimum value for instruments $P_{jm}$ is, hence the omission. Sums calculated on 655 847 observations.}
\end{table}%

\begin{figure}[ht]
	\caption{Comparison of OLS coefficients $\gamma$ and reduced form treatment effects}
	\includegraphics[width=\linewidth]{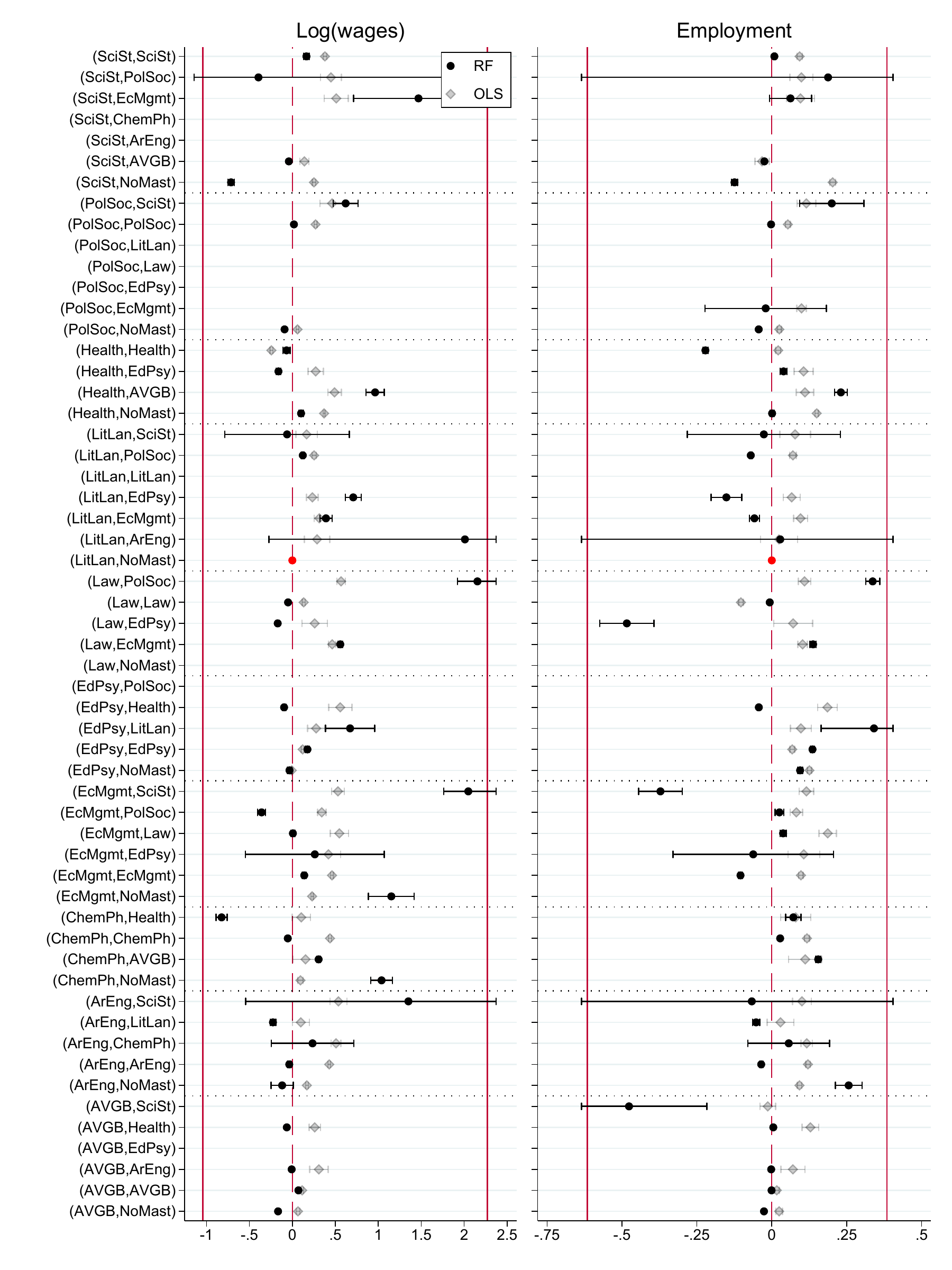}
	\centering
	\label{fig:t3_te_compared}
	\vspace{-.5cm}
	\caption*{\footnotesize Black markers indicate reduced form (RF) results \eqref{eq:fw_t3_reducedform}, grey markers indicate OLS results \eqref{eq:fw_t3}. Whiskers denote 95\% CIs and the red dot denotes the baseline (Lit.Lang., No Master). Red lines denote credible boundaries for the treatment effects. $TE(\text{ln(wage)})\in[-1.04, 2.27]$, employment: $TE(\text{employment})\in[-0.62, 0.38]$.}
\end{figure}

	

\begin{figure}[htbp]
	\centering
	\caption{Comparison of log wage and employment returns for all multidisciplinary careers}
	\includegraphics[width=\linewidth]{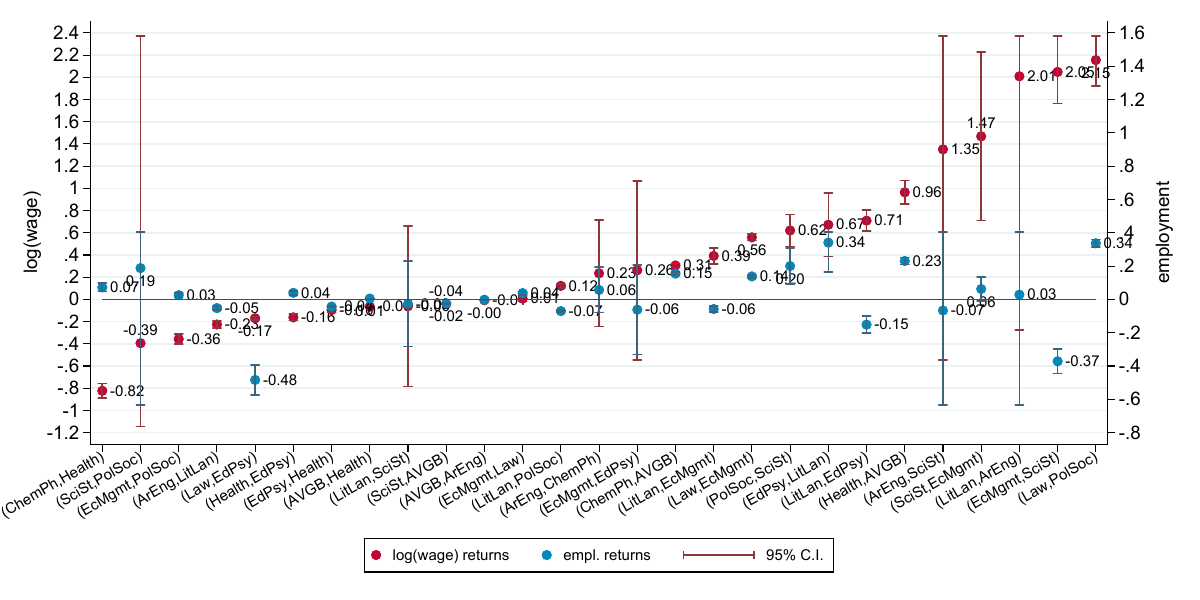}
	\caption*{\footnotesize Comparison of log wage returns (red, left vertical axis) and returns to employment (blue, right vertical axis). Axes are centered around 0. The excluded category is (Lit.Lang., No Master) centered at 0. Any missing returns could not be estimated for both outcomes. Panel B presents returns to specialized careers with the same bachelor's and master's. The order follows the ranking of log wage returns from lowest to highest.}
	\label{fig:TE_compared_all}
	
\end{figure}

\begin{figure}[htbp]
	\centering
	\caption{Comparison of log wage and employment returns for non-multidisciplinary careers}
	\includegraphics[width=\linewidth]{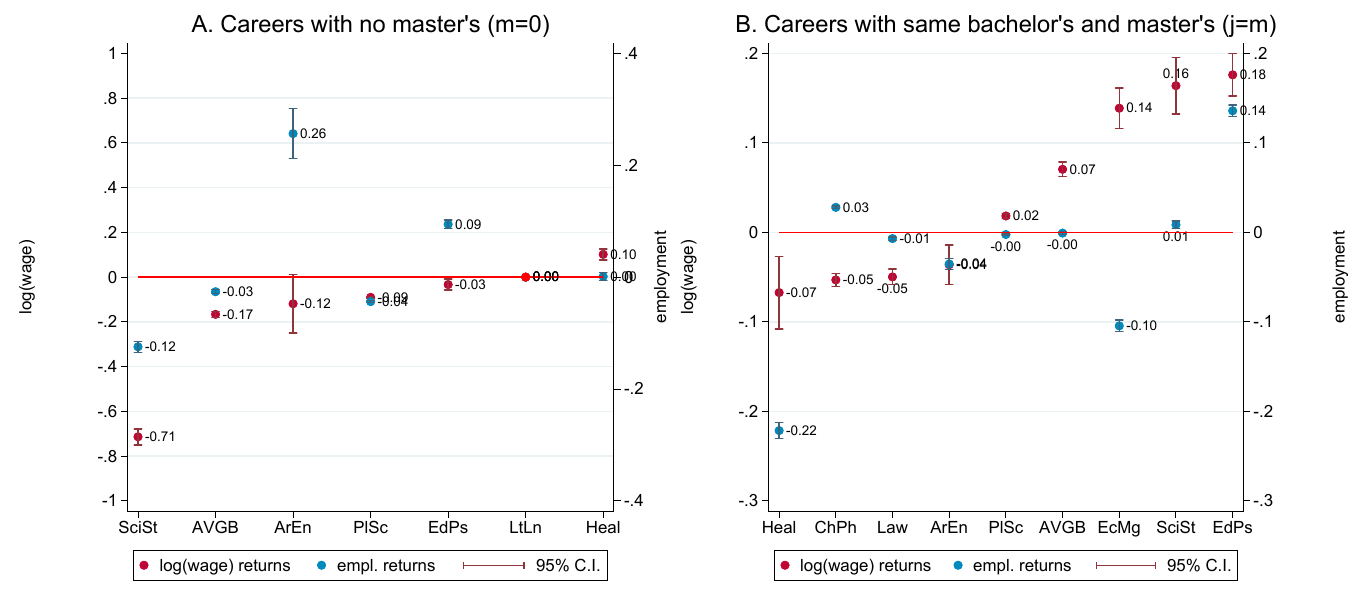}
	\caption*{\footnotesize Comparison of log wage returns (red, left vertical axis) and returns to employment (blue, right vertical axis). Axes are centered around 0. Panel A presents labor market returns for careers with no master, where (Lit.Lang., No Master) denotes the excluded category centered at 0. Any missing returns could not be estimated for both outcomes. Panel B presents returns to specialized careers with the same bachelor's and masters. In both panels, the order follows the ranking of log wage returns from lowest to highest.}
	\label{fig:TE_compared1}
	
\end{figure}

\begin{figure}[ht]
	\caption{Simulation 1 -- Decomposition by individual characteristics}
	\includegraphics[width=\linewidth]{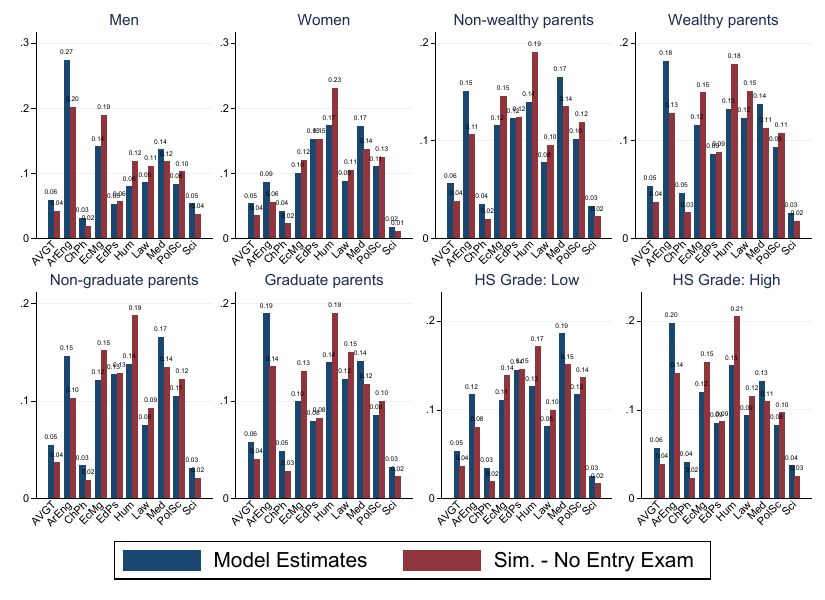}
	\centering
	\label{fig:sim1_decomp_alldemo}
\end{figure}

\FloatBarrier

\newpage	

\subsection{Methodological notes}\label{sec:appendix_method}

\subsubsection{Universities}
The universities that I consider are the following: Politecnico di Ancona, Bari, Politecnico di Bari, Basilicata, Bergamo, Bologna, Bolzano, Brescia, Cagliari, Calabria, Camerino, Campania - Luigi Vanvitelli, Cassino e Lazio Meridionale, Catania, Catanzaro, Chieti e Pescara, Enna Kore, Ferrara, Firenze, Foggia, Genova, Insubira, L'Aquila, LIUC Castellanza, Macerata, Messina, Milano Bicocca, Milano IULM, Milano Statale, Milano Vita-Salute S. Raffaele, Modena e Reggio Emilia, Molise, Napoli - Federico II, Napoli - Seconda Università, Napoli - L'Orientale, Napoli - Parthenope, Padova, Palermo, Parma, Pavia, Perugia, Università per Stranieri di Perugia, Piemonte Orientale, Pisa, Reggio Calabria Mediterranea, Roma - Campus Bio-Medico, Roma LUMSA, Roma Foro Italico, Roma Tre, Roma - La Sapienza, Roma - Tor Vergata, Salento, Salerno, Sannio e Benevento, Sassari, Siena, Università per Stranieri di Siena, Teramo, Torino, Politecnico di Torino, Trento, Trieste, Udine, Urbino, Viterbo Tuscia, Valle D'Aosta Venezia - Ca' Foscari, Venezia - IUAV, Verona.\\

Some universities which are not in this list may nonetheless appear in the dataset (e.g. Milano Bocconi). The reason is that students appear in the dataset if they graduated (master) from a university in the consortium, yet information is collected also for their bachelor which may differ. Only about 5\% of students in the sample switches institution throughout their career.

\subsubsection{Degrees and groups}\label{sec:appendix_degrees}
Here, I present the exact pooling of degrees into groups. The allocation has been done by the AlmaLaurea consortium. A few groups of degrees were further grouped to improve estimation: agriculture and veterinary was grouped with geology and biology, architecture with engineering, teaching with physical education and psychology, and literature with languages. Information on an additional group -- defense and security -- was dropped as access into these degrees is managed differently from standard university degrees. \\

\begin{longtable}{p{0.15\linewidth} p{0.8\linewidth}}
	\caption{Degree grouping}\label{tab:gruppi_cdl_description}\\
	\hline
	\textbf{Code} & \textbf{Description} \\ \hline
	\endfirsthead
	\textbf{Code} & \textbf{Description} \\ \hline
	\endhead
	\multicolumn{2}{l}{\indent \textit{1. Agriculture, veterinarian sciences, geology, biology}} \\
	L-2  & BIOTECNOLOGIE \\
	L-13  & SCIENZE BIOLOGICHE \\
	L-25  & SCIENZE E TECNOLOGIE AGRARIE E FORESTALI \\
	L-26  & SCIENZE E TECNOLOGIE AGRO‐ALIMENTARI \\
	L-32  & SCIENZE E TECNOLOGIE PER L'AMBIENTE E LA NATURA \\
	L-34  & SCIENZE GEOLOGICHE \\
	L-38  & SCIENZE ZOOTECNICHE E TECNOLOGIE DELLE PRODUZIONI ANIMALI \\
	LM-6  & BIOLOGIA \\
	LM-7  & BIOTECNOLOGIE AGRARIE \\
	LM-8  & BIOTECNOLOGIE INDUSTRIALI \\
	LM-9  & BIOTECNOLOGIE MEDICHE, VETERINARIE E FARMACEUTICHE \\
	LM-42  & MEDICINA VETERINARIA \\
	LM-69  & SCIENZE E TECNOLOGIE AGRARIE   \\
	LM-70  & SCIENZE E TECNOLOGIE ALIMENTARI \\
	LM-73 & SCIENZE E TECNOLOGIE FORESTALI ED AMBIENTALI \\
	LM-74 & SCIENZE E TECNOLOGIE GEOLOGICHE \\
	LM-75  & SCIENZE E TECNOLOGIE PER L'AMBIENTE E IL TERRITORIO \\
	LM-86  & SCIENZE ZOOTECNICHE E TECNOLOGIE ANIMALI \\
	&  \\
	\multicolumn{2}{l}{\indent \textit{2. Architecture and Engineering}} \\
	L-4  & DISEGNO INDUSTRIALE \\
	L-7  & INGEGNERIA CIVILE E AMBIENTALE \\
	L-8  & INGEGNERIA DELL'INFORMAZIONE \\
	L-9  & INGEGNERIA INDUSTRIALE \\
	L-17  & SCIENZE DELL'ARCHITETTURA \\
	L-21  & SCIENZE DELLA PIANIFICAZIONE TERRITORIALE, URBANISTICA, PAESAGGISTICA E AMBIENTALE \\
	L-23  & SCIENZE E TECNICHE DELL'EDILIZIA \\
	LM-3  & ARCHITETTURA DEL PAESAGGIO  \\
	LM-4  & ARCHITETTURA E INGEGNERIA EDILE‐ARCHITETTURA \\
	LM-12  & DESIGN \\
	LM-20  & INGEGNERIA AEROSPAZIALE E ASTRONAUTICA  \\
	LM-21  & INGEGNERIA BIOMEDICA \\
	LM-22  & INGEGNERIA CHIMICA \\
	LM-23 & INGEGNERIA CIVILE \\
	LM-24  & INGEGNERIA DEI SISTEMI EDILIZI \\
	LM-25  & INGEGNERIA DELL'AUTOMAZIONE \\
	LM-26  & INGEGNERIA DELLA SICUREZZA \\
	LM-27  & INGEGNERIA DELLE TELECOMUNICAZIONI \\
	LM-28  & INGEGNERIA ELETTRICA \\
	LM-29  & INGEGNERIA ELETTRONICA \\
	LM-30  & INGEGNERIA ENERGETICA E NUCLEARE \\
	LM-31  & INGEGNERIA GESTIONALE \\
	LM-32  & INGEGNERIA INFORMATICA \\
	LM-33 & INGEGNERIA MECCANICA \\
	LM-34  & INGEGNERIA NAVALE \\
	LM-35 & INGEGNERIA PER L'AMBIENTE E IL TERRITORIO \\
	LM-44  & MODELLISTICA MATEMATICO‐FISICA PER L'INGEGNERIA \\
	LM-48  & PIANIFICAZIONE TERRITORIALE URBANISTICA E AMBIENTALE \\
	LM-53  & SCIENZA E INGEGNERIA DEI MATERIALI \\
	&  \\
	\multicolumn{2}{l}{\indent \textit{3. Chemistry and Pharmacy}} \\
	L-27  & SCIENZE E TECNOLOGIE CHIMICHE \\
	L-29  & SCIENZE E TECNOLOGIE FARMACEUTICHE \\
	LM-13  & FARMACIA E FARMACIA INDUSTRIALE \\
	LM-54  & SCIENZE CHIMICHE \\
	LM-71  & SCIENZE E TECNOLOGIE DELLA CHIMICA INDUSTRIALE \\
	&  \\
	\multicolumn{2}{l}{\indent \textit{4. Economics and Management}} \\
	L-15  & SCIENZE DEL TURISMO \\
	L-16  & SCIENZE DELL'AMMINISTRAZIONE E DELL'ORGANIZZAZIONE \\
	L-18  & SCIENZE DELL'ECONOMIA E DELLA GESTIONE AZIENDALE \\
	L-33  & SCIENZE ECONOMICHE \\
	LM-16  & FINANZA \\
	LM-56  & SCIENZE DELL'ECONOMIA \\
	LM-76  & SCIENZE ECONOMICHE PER L'AMBIENTE E LA CULTURA \\
	LM-77 & SCIENZE ECONOMICO‐AZIENDALI \\
	&  \\
	\multicolumn{2}{l}{\indent \textit{5. Teaching, Physical Education and Psychology}} \\
	L-19  & SCIENZE DELL'EDUCAZIONE E DELLA FORMAZIONE \\
	L-22  & SCIENZE DELLE ATTIVITA MOTORIE E SPORTIVE \\
	L-24  & SCIENZE E TECNICHE PSICOLOGICHE \\
	LM-47  & ORGANIZZAZIONE E GESTIONE DEI SERVIZI PER LO SPORT E LE ATTIVITA MOTORIE \\
	LM-50  & PROGRAMMAZIONE E GESTIONE DEI SERVIZI EDUCATIVI \\
	LM-51  & PSICOLOGIA \\
	LM-55  & SCIENZE COGNITIVE \\
	LM-57  & SCIENZE DELL'EDUCAZIONE DEGLI ADULTI E DELLA FORMAZIONE CONTINUA \\
	LM-67  & SCIENZE E TECNICHE DELLE ATTIVITA MOTORIE PREVENTIVE E ADATTATE \\
	LM-68  & SCIENZE E TECNICHE DELLO SPORT \\
	LM-85  & SCIENZE PEDAGOGICHE \\
	LM-93  & TEORIE E METODOLOGIE DELL'E‐LEARNING E DELLA MEDIA EDUCATION \\
	&  \\
	\multicolumn{2}{l}{\indent \textit{6. Law}} \\
	L-14  & SCIENZE DEI SERVIZI GIURIDICI \\
	LMG-1  & GIURISPRUDENZA \\
	&  \\
	\multicolumn{2}{l}{\indent \textit{7. Literature and Languages}} \\
	L-1  & BENI CULTURALI \\
	L-3  & DISCIPLINE DELLE ARTI FIGURATIVE, DELLA MUSICA, DELLO SPETTACOLO E DELLA MODA (DAMS) \\
	L-5  & FILOSOFIA \\
	L-6  & GEOGRAFIA \\
	L-10  & LETTERE \\
	L-11  & LINGUE E CULTURE MODERNE \\
	L-12  & MEDIAZIONE LINGUISTICA \\
	L-42  & STORIA \\
	L-43  &  TECNOLOGIE PER LA CONSERVAZIONE E IL RESTAURO DEI BENI CULTURALI \\
	LM-1  & ANTROPOLOGIA CULTURALE ED ETNOLOGIA \\
	LM-2  & ARCHEOLOGIA \\
	LM-5  & ARCHIVISTICA E BIBLIOTECONOMIA \\
	LM-10  & CONSERVAZIONE DEI BENI ARCHITETTONICI E AMBIENTALI \\
	LM-11  & CONSERVAZIONE E RESTAURO DEI BENI CULTURALI \\
	LM-14  & FILOLOGIA MODERNA \\
	LM-15  & FILOLOGIA, LETTERATURE E STORIA DELL'ANTICHITA \\
	LM-36  & LINGUE E LETTERATURE DELL'AFRICA E DELL'ASIA \\
	LM-37  & LINGUE E LETTERATURE MODERNE EUROPEE E AMERICANE \\
	LM-38  & LINGUE MODERNE PER LA COMUNICAZIONE E LA COOPERAZIONE \\
	LM-39 & LINGUISTICA \\
	LM-45 &  MUSICOLOGIA E BENI MUSICALI \\
	LM-65  & SCIENZE DELLO SPETTACOLO E PRODUZIONE MULTIMEDIALE \\
	LM-78  & SCIENZE FILOSOFICHE \\
	LM-80  & SCIENZE GEOGRAFICHE \\
	LM-84  & SCIENZE STORICHE \\
	LM-89  & STORIA DELL'ARTE \\
	LM-94  & TRADUZIONE SPECIALISTICA E INTERPRETARIATO \\
	&  \\
	\multicolumn{2}{l}{\indent \textit{8. Health and Medicine}} \\
	L/SNT-1  & SCIENZE INFERMIERISTICHE E OSTETRICHE \\
	L/SNT-2  & SCIENZE RIABILITATIVE DELLE PROFESSIONI SANITARIE \\
	L/SNT-3  & SCIENZE DELLE PROFESSIONI SANITARIE TECNICHE \\
	L/SNT-4  & SCIENZE DELLE PROFESSIONI SANITARIE DELLA PREVENZIONE \\
	LM/SNT-1  & SCIENZE INFERMIERISTICHE E OSTETRICHE \\
	LM/SNT-2  & SCIENZE RIABILITATIVE DELLE PROFESSIONI SANITARIE \\
	LM/SNT-3  & SCIENZE DELLE PROFESSIONI SANITARIE TECNICHE \\
	LM/SNT-4  & SCIENZE DELLE PROFESSIONI SANITARIE DELLA PREVENZIONE \\
	LM-41  & MEDICINA E CHIRURGIA \\
	LM-46  & ODONTOIATRIA E PROTESI DENTARIA  \\
	LM-61  & SCIENZE DELLA NUTRIZIONE UMANA \\
	&  \\
	\multicolumn{2}{l}{\indent \textit{9. Political and social sciences}} \\
	L-20  & SCIENZE DELLA COMUNICAZIONE \\
	L-36  & SCIENZE POLITICHE E DELLE RELAZIONI INTERNAZIONALI \\
	L-37  & SCIENZE SOCIALI PER LA COOPERAZIONE, LO SVILUPPO E LA PACE \\
	L-39  & SERVIZIO SOCIALE \\
	L-40  & SOCIOLOGIA \\
	LM-19  & INFORMAZIONE E SISTEMI EDITORIALI \\
	LM-49  & PROGETTAZIONE E GESTIONE DEI SISTEMI TURISTICI \\
	LM-52  & RELAZIONI INTERNAZIONALI \\
	LM-59 & SCIENZE DELLA COMUNICAZIONE PUBBLICA, D'IMPRESA E PUBBLICITÀ  \\
	LM-62  & SCIENZE DELLA POLITICA \\
	LM-63 & SCIENZE DELLE PUBBLICHE AMMINISTRAZIONI \\
	LM-81  & SCIENZE PER LA COOPERAZIONE ALLO SVILUPPO \\
	LM-87  & SERVIZIO SOCIALE E POLITICHE SOCIALI \\
	LM-88  & SOCIOLOGIA E RICERCA SOCIALE \\
	LM-90  & STUDI EUROPEI \\
	LM-91  & TECNICHE E METODI PER LA SOCIETA DELL'INFORMAZIONE \\
	LM-92  & TEORIE DELLA COMUNICAZIONE \\
	&  \\
	\multicolumn{2}{l}{\indent \textit{10. Science and Statistics}} \\
	L-28  & SCIENZE E TECNOLOGIE DELLA NAVIGAZIONE \\
	L-30  & SCIENZE E TECNOLOGIE FISICHE \\
	L-31  & SCIENZE E TECNOLOGIE INFORMATICHE \\
	L-35  & SCIENZE MATEMATICHE \\
	LM-17  & FISICA \\
	LM-18  & INFORMATICA \\
	LM-40  & MATEMATICA \\
	L-41  & STATISTICA \\
	LM-43  & METODOLOGIE INFORMATICHE PER LE DISCIPLINE UMANISTICHE \\
	LM-58  & SCIENZE DELL'UNIVERSO \\
	LM-60 & SCIENZE DELLA NATURA \\
	LM-66  & SICUREZZA INFORMATICA \\
	LM-72 & SCIENZE E TECNOLOGIE DELLA NAVIGAZIONE \\
	LM-82  & SCIENZE STATISTICHE \\
	LM-83 & SCIENZE STATISTICHE ATTUARIALI E FINANZIARIE \\
	\hline
	\multicolumn{2}{l}{\textit{Note: Prefix L- refers to bachelor degrees, LM- to master degrees.}} \\
\end{longtable}%

\end{document}